\documentclass[letterpaper, 10pt, final, journal]{IEEEtran}

\IEEEoverridecommandlockouts  

\usepackage{array}
\usepackage{graphics} 
\usepackage{epsfig} 
\usepackage{mathptmx} 
\usepackage{times} 
\usepackage{amsmath} 
\usepackage{mathtools}
\usepackage{amssymb}  
\usepackage{color}
\usepackage[colorlinks]{hyperref}
\usepackage[ruled,linesnumbered]{algorithm2e} 
\usepackage{subfigure}
\usepackage{pbox}
\usepackage{multirow}

\usepackage{enumitem}

\hypersetup{citecolor=black, linkcolor=black, urlcolor=black}

\newcommand{\rcolor}{black}

\newcommand{\rrcolor}{black}

\title{\LARGE \bf
A Survey of Distributed Consensus Protocols for Blockchain Networks
}

\author{
\IEEEauthorblockN{
      {Yang Xiao}\IEEEauthorrefmark{1},
      {Ning Zhang}\IEEEauthorrefmark{2},
      {Wenjing Lou}\IEEEauthorrefmark{1},
      {Y. Thomas Hou}\IEEEauthorrefmark{1}
      } \\
    \IEEEauthorblockA{\IEEEauthorrefmark{1}Virginia Polytechnic Institute and State University, VA, USA} \\
    \IEEEauthorblockA{\IEEEauthorrefmark{2}Washington University in St. Louis, MO, USA}
\thanks{\copyright 2020 IEEE. Personal use of this material is permitted. Permission from IEEE must be obtained for all other uses, including reprinting/republishing this material for advertising or promotional purposes, collecting new collected works for resale or redistribution to servers or lists, or reuse of any copyrighted component of this work in other works}
}

\begin{document}

\maketitle

\begin{abstract} Since the inception of Bitcoin, cryptocurrencies and the underlying blockchain technology have attracted an increasing interest from both academia and industry. Among various core components, consensus protocol is the defining technology behind the security and performance of blockchain. From incremental modifications of Nakamoto consensus protocol to innovative alternative consensus mechanisms, many consensus protocols have been proposed to improve the performance of the blockchain network itself or to accommodate other specific application needs.

In this survey, we present a comprehensive review and analysis on the state-of-the-art blockchain consensus protocols. To facilitate the discussion of our analysis, we first introduce the key definitions and relevant results in the classic theory of fault tolerance which help to lay the foundation for further discussion. We identify five core components of a blockchain consensus protocol, namely, block proposal, block validation, information propagation, block finalization, and incentive mechanism. A wide spectrum of blockchain consensus protocols are then carefully reviewed accompanied by algorithmic abstractions and vulnerability analyses. The surveyed consensus protocols are analyzed using the five-component framework and compared with respect to different performance metrics. These analyses and comparisons provide us new insights in the fundamental differences of various proposals in terms of their suitable application scenarios, key assumptions, expected fault tolerance, scalability, drawbacks and trade-offs. We believe this survey will provide blockchain developers and researchers a comprehensive view on the state-of-the-art consensus protocols and facilitate the process of designing future protocols.

\begin{IEEEkeywords}
Blockchain, distributed consensus, fault tolerance, protocol design.
\end{IEEEkeywords}

\end{abstract}
\section{Introduction}

\IEEEPARstart{S}{ince}
Bitcoin's inception in late 2008, cryptocurrencies and the underlying blockchain technology have piqued great interest from the financial industry and society as a whole.
%
Blockchain is widely cited as a fully decentralized system and a secure-by-design technology. The blockchain itself is a database that keeps track of all transactions occurred in the network and is replicated at every participating node. It essentially realizes a distributed ledger without relying on a central authority to bootstrap the trust among participants or to clear the transactions. It also does not assume trust among the participating nodes. Blockchain is meant to enable trusted computation among a group of mutually distrustful participants. On the other hand, blockchain is also known for providing trustworthy immutable record keeping service. The block data structure adopted in a blockchain embeds the hash of the previous block in the next block generated. The use of hash chain ensures that data written on the blockchain can not be modified. In addition, a public blockchain system supports third-party auditing and some blockchain systems support a high level of anonymity, that is, a user can transact online using a pseudonym without revealing his/her true identity. 

The security properties promised by blockchain is unprecedented and truly inspiring. Pioneering blockchain systems such as Bitcoin have greatly impacted the digital payment world. It is envisioned that blockchain technology and applications built on top of it will revolutionize a broad array of financial service industries as well as non-financial sectors. 
Among the many technical components that a blockchain system is composed of, the distributed consensus protocol is the key technology that enables blockchain's decentralization, or more specifically, that ensures all participants agree on a unified transaction ledger without the help of a central authority. The distributed consensus protocol specifies message passing and local decision making at each node. Various design choices in the consensus protocol can greatly impact a blockchain system's performance, including its transaction capacity, scalablity, and fault tolerance. 

The Nakamoto consensus protocol \cite{nakamoto2008bitcoin} is the protocol implemented in the Bitcoin network. With the help of this consensus protocol, Bitcoin became the first digital currency system to resist double-spending attacks in a decentralized peer-to-peer network of little trust.
As the Bitcoin network continues to grow, Nakamoto consensus has encountered several performance bottlenecks and sustainability problems. Researchers in blockchain communities have raised the following concerns on Nakamoto consensus and particularly its proof-of-work (PoW) mining mechanism: 1) unsustainable energy consumption, 2) low transaction capacity and poor scalability, 3) long-term security concerns as mining rewards diminish. For instance, the Bitcoin network currently consists of roughly ten thousand nodes \cite{bitnodes2019bitcoin}, while the maximum transaction capacity of Bitcoin is 7 transactions per second (TPS) and can be increased to at most 25 TPS by tuning protocol parameters without jeopardizing consensus safety \cite{croman2016scaling}. In contrast, the VISA network consists of 50 million participants and can handle up to 65,000 TPS \cite{visa2018factsheet}. A single Bitcoin transaction (November 2019) consumes the equivalent amount of electricity that would power 21 average U.S. households for one day \cite{digiconomist2019bitcoin}. 

In response to the above performance limitations of PoW mining, blockchain researchers have been investigating new block proposing mechanisms such as proof of stake (PoS), proof of authority (PoA), and proof of elapsed time (PoET) which do not require computation-intensive mining, thus effectively reducing energy consumption. In some cases, cryptographic methods can be used to establish trust among nodes, enabling the use of more coordinated block proposing schemes such as round-robin and committee-based block generation. Appropriate incentives that will continue to encourage honest participation in the blockchain network is another key component of consensus protocol. Therefore, alternative block proposing schemes are often accompanied by a new incentive mechanism that promotes participation fairness and increases overall system sustainability. Popular blockchain consensus protocols encompassing these ideas include Peercoin \cite{king2012ppcoin}, Bitcoin-NG \cite{eyal2016bitcoin}, Ourosboros (Cardano) \cite{kiayias2017ouroboros}, Snow White \cite{daian2017snow}, and EOSIO \cite{eos2018whitepaper}, POA Network \cite{poa2018whitepaper}, etc.

Besides block proposing and incentive mechanisms, researchers have been seeking solutions from the prior wisdom---primarily {\color{\rcolor}classical Byzantine fault tolerant (BFT) consensus and secure multi-party computation (MPC)}---for efficient block finalization methods. For example, the state machine replication (SMR) based BFT consensus algorithms have great potential in permissioned blockchain networks operated with static and revealed identities, of which Tendermint \cite{kwon2014tendermint}, Algorand \cite{gilad2017algorand}, Casper FFG \cite{buterin2017casper}, and Hyperledger Fabric \cite{androulaki2018hyperledger} are well-known use cases. Moreover, asynchronous consensus protocols such as HoneyBadgerBFT \cite{miller2016honey} and BEAT \cite{duan2018beat} were proposed to provide robust block finalization under severe network conditions with uncertain message delays.




\textbf{Our contribution} 
With more blockchain consensus mechanisms being proposed, there is a pressing need to analyze and compare them in a formal and cohesive manner.
In this survey we present a comprehensive review and analysis of the state-of-the-art blockchain consensus protocols and their development history, 
with a special focus on their performance, fault tolerance, and security implications. Our information sources include academic papers, consensus protocol white papers, official documentation and statistics websites of cryptocurrencies.
Specifically, our survey features the following contributions:




\begin{enumerate}
    \item providing a background of classical distributed consensus research, including partially synchronous and asynchronous BFT protocols that are applicable to blockchain consensus;
    \item {\color{\rcolor}
     reviewing a broad array of blockchain consensus protocols with a proposed five-component framework and analyzing their design philosophy and security issues;}
    \item identifying four classes of proof of stake (PoS) based consensus protocols and providing algorithmic abstractions for them;
    \item comparing all mentioned consensus protocols with respect to the five-component framework, fault tolerance, and transaction processing capability.
    \item {\color{\rcolor}providing a succinct tutorial on blockchain consensus protocol design with respect to the security-decentralization-scalability trilemma.}
\end{enumerate}

 


The remaining part of this survey is organized as follows. 
Section \ref{sec:prev-survey} reviews related surveys and tutorials on blockchain consensus protocols.
Section \ref{sec:background} provides a background of classical fault tolerant consensus in distributed systems. Several legacy BFT consensus protocols designed for both partially synchronous and asynchronous networks are introduced. 
Section \ref{sec:blockchain-consensus} presents the basic framework of blockchain and the consensus goals, and introduces the five essential components of a blockchain consensus protocol. 
Section \ref{sec:nakamoto} focuses on the well-known Nakamoto consensus protocol, the defining technology of Bitcoin, and its vulnerabilities and improvement ideas.
Section \ref{sec:pos} provides a systematic view of the PoS protocols, the most promising competitors to the PoW-based Nakamoto consensus for public blockchain. 
Section \ref{sec:other-promising} discusses alternative consensus protocols usable under specific application scenarios. 
Section \ref{sec:comparison} compares all blockchain consensus protocols studied and summarizes their design philosophy.
Section \ref{sec:tutorial} discusses the paradigm shift in consensus protocol design and provides a succinct tutorial.
Section \ref{sec:conclude} concludes the paper.

\section{Previous Surveys and Tutorials} 
\label{sec:prev-survey}

The comparison study by Vukolic \cite{vukolic2015quest} treats two genres of blockchain consensus protocols, namely PoW-based and BFT-based, with respect to transaction throughput, scalability limits, consensus finality, and security implications. PoW-based protocols scale well with network size and are suited for permissionless blockchains, but yield very limited throughput and long confirmation latency. This is due to the security implication of their lack of consensus finality and limited capacity of raising block frequency and block size. In comparison, BFT protocols have built-in consensus finality and achieve much higher transaction capacity, but incur high messaging complexity per block ($O(N^2)$ versus PoW's $O(N)$, $N$ is network size) and need a permissioned network for identity management. As a result they do not scale well with network size. Besides remarking their differences, this paper also explores how the hybrid use of BFT and PoW can enhance the performance of established blockchain systems and provides tentative solutions to improve blockchain scalability.

Cachin et al. \cite{cachin2017blockchain} gives an overview of thirteen prominent consensus protocols designed for permissioned blockchain platforms along with their fault tolerance and security properties. The authors also make a powerful statement that the design of blockchain consensus protocols should follow the rigor established in prevailing wisdom of cryptography, security, and distributed systems, rather than in an ad hoc manner. This is particularly true when the consensus protocol regulates significant financial values and societal trust.
However, it does not provide a methodology to combine the prevailing wisdom in order to design a consensus protocol for specific needs.

The work by Bano et al. \cite{bano2017consensus} is the first well-structured survey of blockchain consensus protocols. It identifies three classes of consensus protocols based on committee formation and block proposing rules: 1) PoW, 2) proof of X (PoX) alternatives to PoW, and 3) hybrid consensus protocols that take advantage of classical distributed consensus techniques. 
This paper emphasizes the role of committee in hybrid consensus protocols. General discussions on the formation and configuration of committee and possible solutions to multi-committee (i.e. sharding-based) consensus are provided.
This paper also presents an evaluation framework that takes into account the protocol safety (censorship resistance, DoS resistance and fault tolerance) and performance (throughput and latency).
Notably, this paper is the closest work to our survey in terms of classification of block proposing mechanisms. However, analysis of new protocols in the area of PoS requires new methods of analysis. 

{\color{\rcolor}
Wang et al. \cite{wang2018survey} provides a comprehensive survey of blockchain consensus protocols and an in-depth review of incentive mechanism designs. The paper starts with a layered view of the blockchain network including the consensus part followed by a general discussion on the compatibility between consensus protocol and incentive mechanism. Then detailed consensus schemes are introduced, including Nakamoto consensus, proof-of-concept consensus schemes, and virtual mining techniques such as proof of stake (PoS) along with the usability of trusted hardware. Hybridization between PoX and classical BFT protocols is also discussed. This survey also features a game-theoretical characterization of Nakamoto consensus' incentive mechanism and its influence on system fairness and decentralization. The efficiency-scalability trade-off is also explored.
Despite of having rich details on consensus techniques and insights into blockchain protocol design, this survey does not provide a concise abstraction that captures different functional components of a consensus protocol or a cohesive characterization on fault-tolerant distributed consensus primitives that can be used for blockchain consensus. 
}

Xiao et al. \cite{xiao2019distributed} provides a succinct tutorial on distributed consensus protocols, from classical BFT protocols to the Nakamoto consensus protocol as well as recent breakthroughs in blockchain consensus. Specially this tutorial provides an abstraction of consensus goals for permissionless blockchains following the paradigm of classical BFT consensus. 
{\color{\rcolor}
The authors also remark that the network model and trust model should be jointly considered when designing blockchain consensus protocols for practical applications. However, this tutorial skips details on each consensus protocols and does not clarify how different components of a blockchain consensus protocol contribute to system performance and security.
}

{\color{\rcolor}
Belotti et al. \cite{belotti2019vademecum} provides a \emph{vademecum} on blockchain technology including development history, transaction and ledger structure, blockchain system abstraction, consensus mechanisms, and a detailed guide on when and how to use which blockchain technology. The consensus mechanisms covered include PoW, PoS, BFT algorithms, and hybrid BFT-based algorithms. Notably, this paper also gives a quantitative comparison of consensus mechanisms with respect to fault tolerance, node scalability, throughput, and transaction latency. Though with great details on building and managing a blockchain system, this vademecum can be perfected with an abstraction for each type of consensus mechanisms and a high-level tutorial on how to reach a compromise between different desired features.
}
\section{Fault-Tolerant Distributed Consensus}
\label{sec:background}


The fault-tolerant (FT) distributed consensus problem has been extensively studied in distributed systems since the late 1970s and recently gained popularity in the blockchain community, especially for permissioned blockchains where every consensus participant reveals its identity. Generally, consensus in a distributed system represents a state that all participants agree on the same data values. Depending on the medium for message exchange, distributed systems are classified into two types: \emph{message passing} and \emph{shared memory} \cite{attiya1995sharing}. In this section we are interested in message passing systems because of their resemblance to contemporary blockchain systems, wherein distributed consensus on a single network history is reached through peer-to-peer communication. We will use the terms process/node/server interchangeably, as they all refer to an individual participant of distributed consensus.

\subsection{System Model}

\subsubsection{Distributed system and task} We consider a distributed system that consists of $N$ independent processes. Each process $p_i$ begins with an individual initial value $x_i$ and communicates with others to update this value. Each local value can be used for a certain task, such as computation or just storage. If the processes are required to perform the same task, consensus on a single value is required before they proceed to the task.

\subsubsection{Process failure}
A process suffers a \emph{crash failure} if it abruptly stops working without resuming. The common causes of a crash failure include power shutdown, software errors, and DoS attacks. A \emph{Byzantine failure}, however, is much severer in that the process can act arbitrarily while appearing normal. It can send contradicting messages to other processes in hope of sabotaging the consensus. ``Byzantine" was coined by Lamport et al. \cite{lamport1982byzantine} in 1982 when describing the \emph{Byzantine Generals Problem}, an allegorical case for single-value consensus among distributed processes. The common cause of a Byzantine failure is adversarial influence, such as malware injection and physical device capture. Multiple Byzantine processes may collude to deal more damage.

\subsubsection{Network synchrony}
Network synchrony defines the level of coordination among all processes. Three levels of synchrony, namely \emph{synchronous}, \emph{partially synchronous}, and \emph{asynchronous}, are often assumed in the literature \cite{xiao2019distributed,attiya2004distributed}.
\begin{itemize}
    \item In a \emph{synchronous} network, operations of processes are coordinated in rounds with clear time constraints. In each round, all processes perform the same type of operations. This can be achieved by a centralized clock synchronization service and good network connectivity. Practically, a network is considered synchronous if message delivery is guaranteed within a fixed delay $\Delta$, for which the network is also called $\Delta$-synchronous.
    
    \item In a \emph{partially synchronous} network, operations of processes are loosely coordinated in a way that message delivery is guaranteed but with uncertain amount of delays. Within the scope of partial synchrony, \textit{weak synchrony} requires message delay not grow faster than the elapsing time indefinitely \cite{castro1999practical}, while \textit{eventual synchrony} ensures $\Delta$-synchrony only after some unknown instant \cite{dwork1988consensus}. In either case, operations of the networked processes can still follow that of a synchronous network if the time horizon is long enough.
    
    \item In an \emph{asynchronous} network, operations of processes are hardly coordinated. There is no delay guarantee on a message except for its eventual delivery. And the coordination of processes (if there is any) is solely driven by the message delivery events. This is often caused by the absence of clock synchronization (thus no notion of shared time) or the dominance of a mighty adversary over all communication channels.
\end{itemize}

It has been shown by Fischer, Lynch, and Paterson \cite{fischer1985impossibility} that under the asynchronous case, consensus cannot be guaranteed with even a single crash failure. This is commonly known as the FLP impossibility, for the authors' namesake. However, this impossibility can be practically circumvented using randomized decision making and a relaxed termination property, as we will show in \ref{subsec:async-bft}.

\subsection{Byzantine Fault Tolerant Consensus}

We call a consensus protocol \textit{crash fault tolerant} (CFT) or \textit{Byzantine fault tolerant} (BFT) if it can tolerate a certain amount of crash or Byzantine process failures while keeping normal functioning. Because of the inclusive relationship between a crash failure and Byzantine failure, a BFT consensus protocol is naturally CFT. BFT consensus is defined by the following four requirements \cite{xiao2019distributed,attiya2004distributed,dwork1988consensus}:
\begin{itemize}
    \item \textit{Termination:} Every non-faulty process decides an output.
    \item \textit{Agreement:} Every non-faulty process eventually decides the same output $\hat{y}$.
    \item \textit{Validity:} If every process begins with the same input $\hat{x}$, then $\hat{y}=\hat{x}$.
    \item \textit{Integrity:} Every non-faulty process' decision and the consensus value $\hat{y}$ must have been proposed by some non-faulty process.
\end{itemize}
The four requirements provide a general target for distributed consensus protocols. For any consensus protocol to attain these BFT requirements, the underlying distributed network should satisfy the following condition: $N\geq3f+1$ where $f$ is the number of Byzantine processes. This fundamental result was first proved by Pease et al. \cite{pease1980reaching} in 1980 and later adapted to the BFT consensus framework. The proof involves induction from $N=3$ and partitioning all processes into three equal-sized groups, with one containing the faulty ones. Interested readers are referred to \cite{pease1980reaching,attiya2004distributed} for detailed proofs.

\subsection{Consensus in Distributed Computing}

\begin{figure}
    \centering
    \includegraphics[width=3.4in]{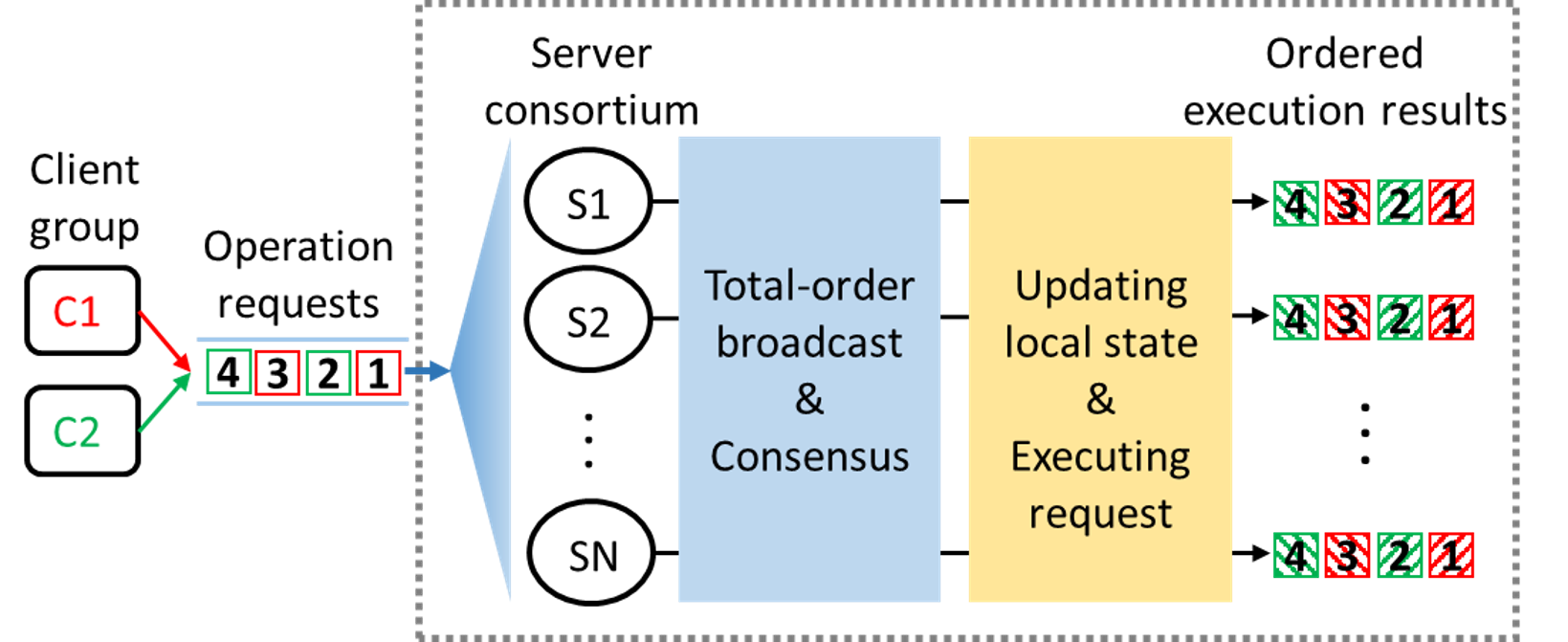}
    \caption{A high-level illustration of SMR-based distributed computing that serves four operation requests from two clients. 
    }
    \label{fig:smr-diagram}
\end{figure}

Consensus in distributed computing is a more sophisticated realization of the aforementioned distributed system. In a typical distributed computing system, one or more clients issue operation requests to the server consortium, which provides timely and correct computing service in response to the requests despite some of servers may fail. Here the correctness requirement is two-fold: correct execution results for all requests and correct ordering of them. According to Alpern and Schneider's work on liveness definition \cite{alpern1985defining} in 1985, the correctness of consensus can be formulated into two requirements: \textit{safety}---every server correctly executes the same sequence of requests, and \textit{liveness}---all requests should be served.

To fulfill these requirements even in the presence of faulty servers, server replication schemes especially \textit{state machine replication} (SMR) are often heralded as the de facto solution. SMR, originated from Lamport's early works on clock synchronization in distributed systems \cite{lamport1978time,lamport1984using}, was formally presented by Schneider \cite{schneider1990implementing} in 1990. Setting in the client-server framework,
SMR sets the following requirements:
\begin{enumerate}
    \item All servers start with the same initial state;
    \item Total-order broadcast: All servers receive the same sequence of requests as how they were generated from clients;
    \item All servers receiving the same request shall output the same execution result and end up in the same state.
\end{enumerate}

Total-order broadcast is also known as \emph{atomic broadcast} (ABC) \cite{cristian1986atomic}, which is in contrast to the \emph{reliable broadcast} (RBC) \cite{bracha1985asynchronous} primitive. The latter only requires all servers receive the same requests without enforcing the order. It is shown in \cite{defago2004total,chandra1996unreliable} that atomic broadcast and distributed consensus are equivalent problems.

A high-level diagram of SMR-based distributed computing is illustrated in Fig. \ref{fig:smr-diagram}. The $N$-server consortium accepts client requests and servers confirm each other's state before reaching consensus and executing requests. In many cases, especially randomized consensus protocols, there can be an alternating procedure of total-order broadcast and local state update until a certain consensus target is met, before moving on to execution. In practice, SMR is often implemented in a leader-based fashion. A primary server (say S1 in Fig. \ref{fig:smr-diagram}) receives client requests and starts the broadcast procedure so that the other $N-1$ replica servers receive the same requests and update their local states to that of the primary. 


In the rest of this section we summarize several well-known consensus protocols (some are based on SMR) designed under different network synchrony assumptions.





\subsection{Consensus Protocols for Partially Synchronous Network}

\begin{figure*}
    \centering
    \subfigure[VR]{
        \centering
        \includegraphics[width=2in]{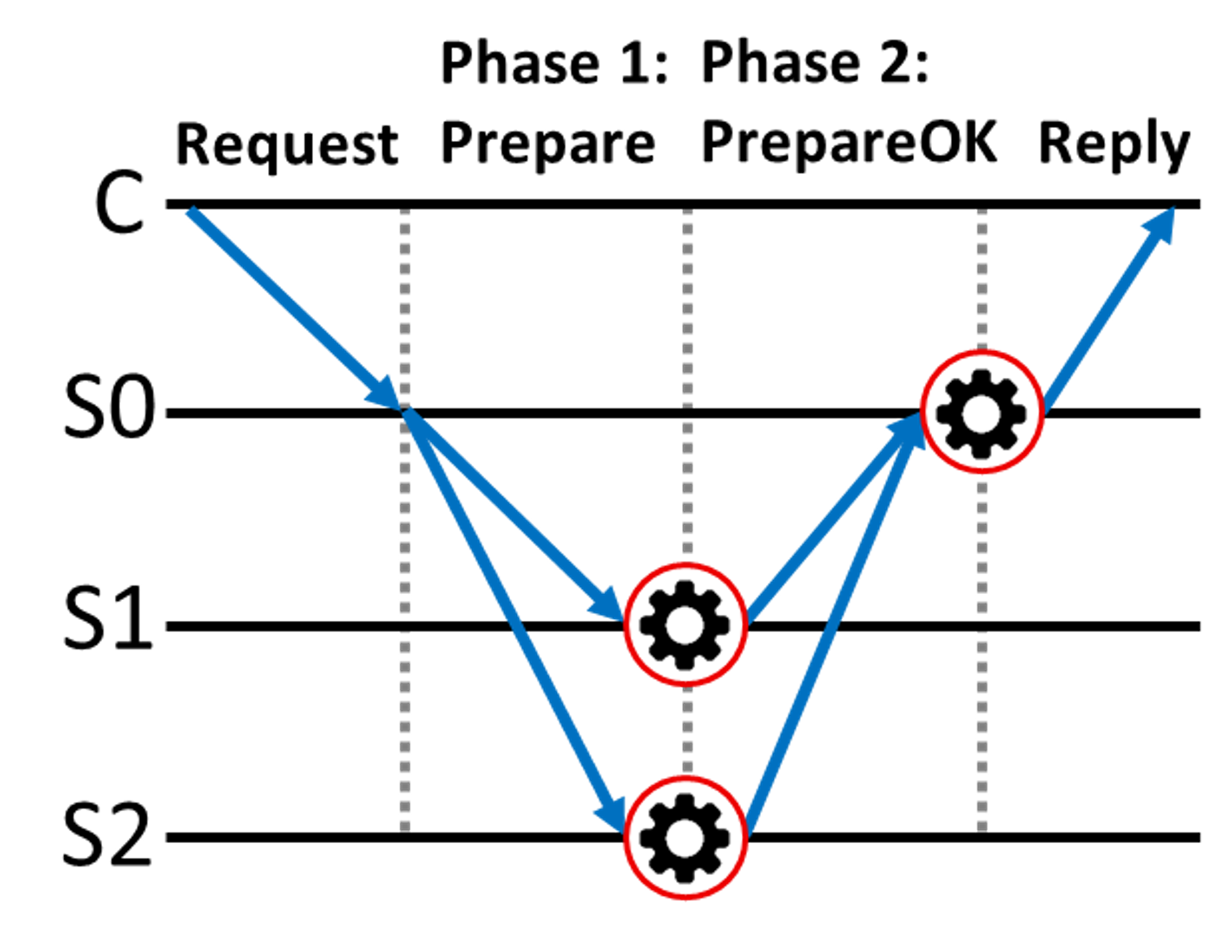}
        \label{fig:smr-diagram-a}
    }\hspace{.1in}
    \subfigure[Paxos]{
        \centering
        \includegraphics[width=2in]{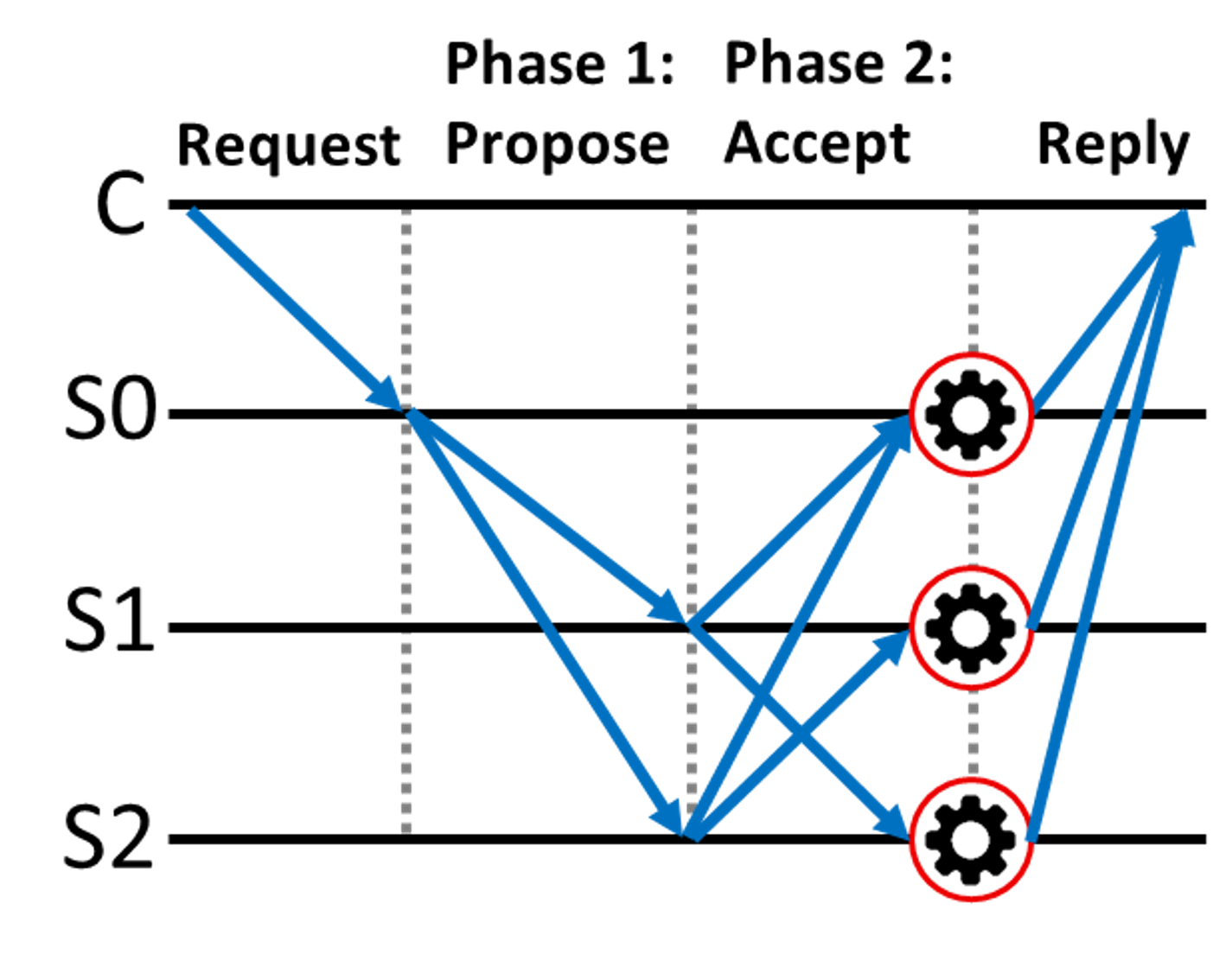}
        \label{fig:smr-diagram-b}
    }\hspace{.1in}
    \subfigure[PBFT]{
        \centering
        \includegraphics[width=2.47in]{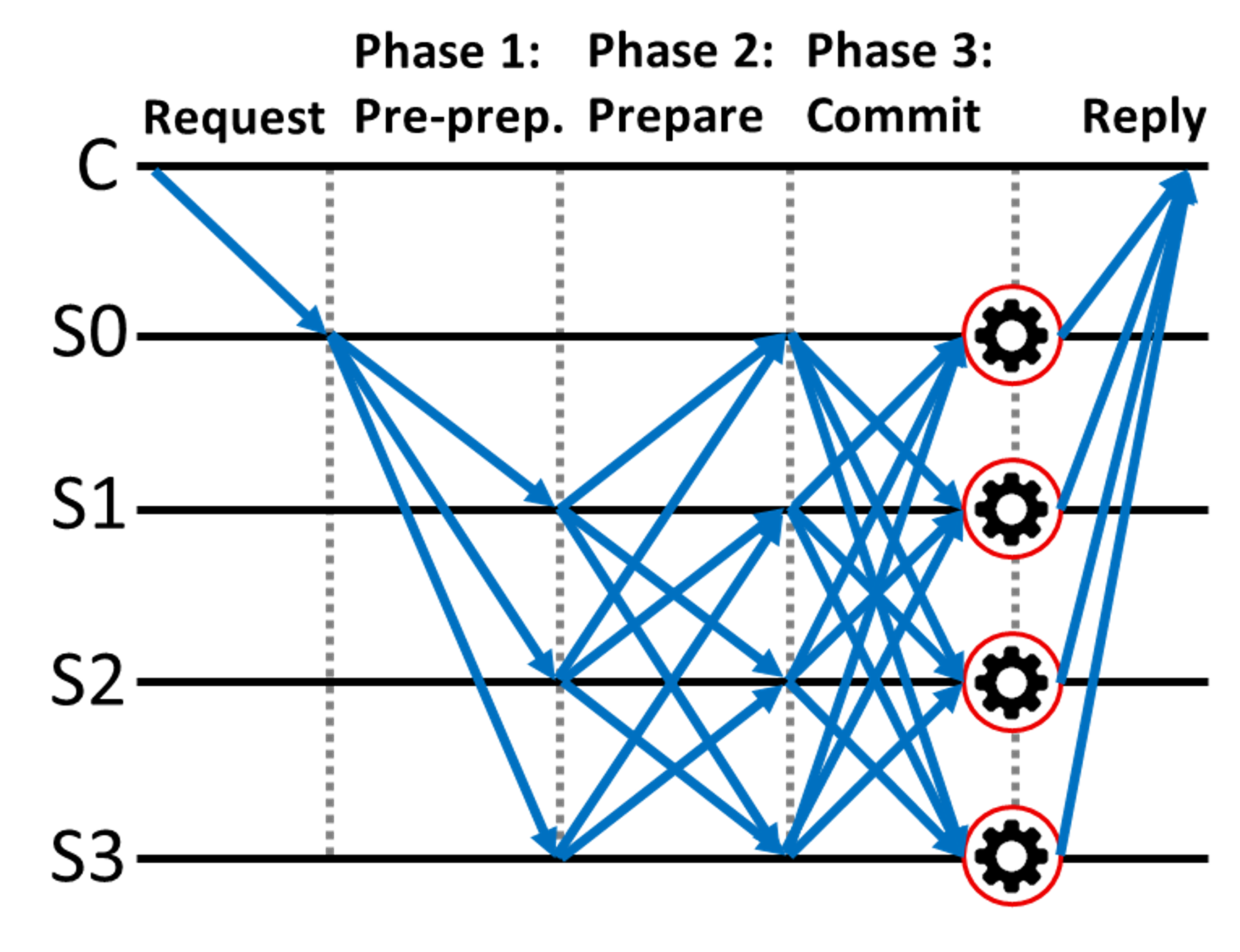}
        \label{fig:smr-diagram-c}
    }
    \caption{Messaging diagram during the normal operation of three SMR protocols. C is the client. S0 is the primary server (leader) who receives requests from the client and starts the consensus. S1+ are replica servers. Every server updates local state after receiving a message. Circled gear icon represents request execution. VR and Paxos can tolerate one crash failure when $N=3$. PBFT can tolerate one Byzantine failure when $N=4$.}
    \label{fig:smr-diagrams}
\end{figure*}

The ground-breaking work by Dwork, Lynch, and Stockmeyer \cite{dwork1988consensus} in 1988 laid the theoretical foundation of partially synchronous consensus. By dissecting the consensus objective into \emph{termination} and \emph{safey}, the authors were able to formally prove the feasibility of four consensus goals, including CFT, omission-tolerance, BFT, unauthenticated BFT, under the $\Delta$-synchrony/eventual synchrony condition. Notably, this work has inspired numerous proposals for partially synchronous consensus schemes, including the later known PBFT.

\subsubsection{DLS protocol}

The same paper \cite{dwork1988consensus} also proposes a prototype consensus protocol (called DLS for authors' namesake) featuring a broadcast primitive for each consensus cycle. Specifically, the broadcast primitive is started by an arbitrary process $p$ and consists of two initial rounds and subsequent iterative rounds. Through message exchanges in each round, the iterative procedure eventually drives the processes to reach agreement on a common value (either the one proposed by $p$ or a default value). At message complexity $O(N^2)$ ($N$ is the number of processes), the broadcast primitive essentially enables the DLS protocol to tolerate $f$ Byzantine processes if $N\geq3f+1$. The cryptocurrency Tendermint uses an adapted version of DLS for block finalization.

\subsubsection{Viewstamped Replication (VR)}

Proposed by Oki and Liskov \cite{oki1988viewstamped} in 1988, viewstamped replication is a server replication scheme for handling server crashes. It was later extended into a consensus protocol by Liskov and Cowling \cite{liskov2012viewstamped} in 2012, which we will refer to as VR. VR is a SMR scheme designed in the client-server framework and consists of three sub-protocols: 1) Normal-operation, 2) View-change, 3) Recovery. The primary server receives a client request and starts the normal operation, as is shown in Fig. \ref{fig:smr-diagram-a}.
In the case of a crash failure of the primary, the View-change protocol is triggered at every replica per the timeout of the \emph{Prepare} message. They broadcast \emph{View-change} messages to each other and count the receptions. After receiving \emph{View-change} messages from more than half of the replicas, the next-in-line replica becomes the new primary and informs the others to resume the normal operation. The Recovery protocol is used by any server to recover from a crash. 
VR can tolerate $f$ crashed replicas if the network population $N\geq2f+1$. However it does not tolerate any Byzantine failure, because the replicas simply follow the instructions from the primary without mutual state confirmation nor communication with the client. On the up side, this makes VR efficient, with $O(N)$ message complexity.

\subsubsection{Paxos}
Paxos is a SMR scheme proposed by Lamport \cite{lamport1998part} in 1989 that imitates the ancient Paxos part-time parliament and later elaborated in 2001 \cite{lamport2001paxos}. It was designed specifically for fault tolerant consensus while bearing many similarities to VR. Paxos classifies nodes into three roles: proposers, acceptors, and learners. A proposer suggests a value in the beginning and the system goal is to make acceptors agree on a single value, and learners learn this value from acceptors. In the client-server scenario depicted in Fig. \ref{fig:smr-diagram-b}, the client is the learner, the primary is the proposer, and the replicas are acceptors. After updating to the same state, all servers execute the request and send it to the client who then chooses the majority result.
When the proposer suffers a crash failure, the acceptors elects a new leader through a similar propose-accept procedure.
Akin to VR, Paxos can tolerate $f$ crashed acceptors when $N\geq2f+1$, but no Byzantine failures. Because of the mutual messaging during the accept phase, the message complexity of Paxos is $O(N^2)$. 

Embarking from its original design, Paxos has grown into a family of consensus protocols, including multi-Paxos, cheap-Paxos, and fast-Paxos, each features a specific goal. Raft, a SMR consensus protocol developed by Ongaro and Ousterhout \cite{ongaro2014search} in 2014 and popular in the blockchain community, is based off Paxos but with a more understandable design.

\subsubsection{Practical Byzantine Fault Tolerance (PBFT)}

Developed by Castro and Liskov \cite{castro1999practical} in 1999, PBFT is the first SMR-based BFT consensus protocol that has gained wide recognition for practicality. It has become almost synonymous to BFT consensus in the blockchain community. PBFT originated from the VR framework and took inspiration from Paxos. PBFT consists of three sub-protocols: 1) Normal-operation, 2) Checkpoint, 3) View-change. 
    
    
    

\begin{algorithm}
\label{algo:pbft-normal-op}
\caption{PBFT (Normal-operation protocol)}
\SetAlgoLined
\SetKwProg{Fn}{Function}{:}{end}
\setcounter{AlgoLine}{0}

\tcc{Request}
Client sends an operation request to the primary;

\tcc{Phase 1: Pre-prepare}
The primary relays this request to replicas via \emph{Pre-prepare} messages;

Replicas record the request and update local states;
 
\tcc{Phase 2: Prepare}
Replicas send \emph{Prepare} messages to all servers (replicas and the primary);

Once receiving $\geq2f+1$ \emph{Prepare} messages, a server updates local state and is ready to commit;

\tcc{Phase 3: Commit}
Servers send \emph{Commit} messages to each other;

Once receiving $\geq2f+1$ \emph{Commit} messages, a server starts to execute the client request and then updates local state;

\tcc{Reply}
Every server replies its result to the client.

\end{algorithm}


The Normal-operation protocol is shown in Algorithm \ref{algo:pbft-normal-op} and Fig. \ref{fig:smr-diagram-c}. Ideally, all results replied to the client should be the same; otherwise the client chooses the majority result. 
The Checkpoint protocol serves as a logging tool that keeps a sliding window (of which the lower bound is the stable checkpoint) to track active operation requests. The latest stable checkpoint is used for safely discarding older requests in the operation log and facilitating the view change protocol. 
In the case of a primary failure, the View-change protocol is triggered at every replica that detects the timeout of the primary's message. They oust the incumbent primary and broadcast view-change messages to each other and count receptions. After receiving \emph{View-change} messages from $2f$ peers, the next-in-line replica becomes the new primary and informs the rest to resume the normal operation.

{\color{\rcolor}
The message complexity of PBFT normal operation is $O(N^2)$ because of the mutual messaging in \textit{Prepare} and \textit{Commit} phase. As for the  fault tolerance, since a server needs to receive more than $2f+1$ \textit{Prepare} (\textit{Commit}) messages in Prepare (Commit) phase before proceeding to the next action, there will be at least $2f+1$ (as $N\geq 3f+1$) honest servers in the same state after \textit{Commit} phase and producing the same result; the $f$ Byzantine servers are not able to sway the majority consensus.
}
Therefore PBFT can tolerate $f$ Byzantine replicas when the server population $N\geq3f+1$, which is in accordance to the fundamental $1/3$ BFT threshold.
{\color{\rcolor}
Interested readers are referred to \cite{castro1999practical,xiao2019distributed} for detailed proofs.
}


PBFT has inspired numerous BFT consensus protocols with enhanced security and performance. Well-known proposals include Quorum/Update (QU) \cite{abd2005fault}, Hybrid Quorum (HQ) \cite{cowling2006hq}, Zyzzyva (using speculative execution) \cite{kotla2007zyzzyva}, FaB \cite{martin2006fast}, Spinning \cite{veronese2009spin}, Robust BFT SMR \cite{clement2009making}, and Aliph \cite{guerraoui2010next}. Interested readers are referred to Bessoni's tutorial \cite{bessani2012bft} for an overview of these protocols.

\subsection{Consensus Protocols for Asynchronous Network}
\label{subsec:async-bft}


For distributed systems that are predominantly built upon wired communication and reliable transport-layer protocols, partial synchrony is a practical assumption. However in scenarios such as mobile ad hoc network (MANET) and delay tolerant networks (DTN), the network is considered of near-to-none synchrony.
As is proved by the FLP impossibility \cite{fischer1985impossibility}, consensus can not be guaranteed in a fully asynchronous network with even one crash failure. Moreover, unreliable communication links have an equivalent effect of a Byzantine scheduler. Nonetheless, this impossibility result can be practically circumvented by two primitives: \emph{probabilistic termination} and \emph{randomization}.

First of all, according to \cite{bracha1987asynchronous} the termination property presented in Section \ref{sec:background} can be subdivided into two classes:
\begin{itemize}
    \item \textit{Deterministic termination:} Every non-faulty process decides an output by round $r$, a predetermined parameter.
    \item \textit{Probabilistic termination:} The probability that a non-faulty process is undecided after $r$ rounds approaches zero as $r$ grows to infinity.
\end{itemize}


For synchronous or partially synchronous networks where message delay and round period are bounded, protocols like PBFT can exploit a timeout mechanism to detect anomaly of the primary, which makes deterministic termination an achievable goal. For asynchronous networks where messages delivery has no timing guarantee, the consensus process can only be driven by the message delivery events themselves. This limitation demands probabilistic termination. 
To realize probabilistic termination, randomization (simultaneously proposed by Ben-Or \cite{ben1983another} and Rabin \cite{rabin1983randomized} in 1983) can be instantiated in the consensus protocol. The basic idea is that a process makes a random choice when there are not enough trusted messages received for making a final decision. 

Next we introduce four primitives/protocols that aim to solve asynchronous BFT consensus. Though in different contexts, they all feature probabilistic termination and make use of randomization.


\subsubsection{Bracha's RBC and asynchronous consensus protocol}

Bracha et al. \cite{bracha1984asynchronous} proposed the pioneering reliable broadcast (RBC) primitive and an asynchronous consensus protocol in 1984 
to solve the Byzantine Generals Problem \cite{lamport1982byzantine}, in which all non-faulty processes should eventually make the same binary decision.
Bracha's RBC guarantees that non-faulty processes will never accept contradicting messages from any process and forces the faulty ones to output either nothing (mimicking the crash failure) or the correct value.
Bracha's asynchronous consensus protocol, adapted from Ben-Or's 1983 work \cite{ben1983another}, runs by phases and each phase contains three RBC rounds for inter-process value exchange. We show the round-3 of each phase, which contains the randomization step. After receiving at least $N-f$ value messages ($f$ is the presumed Byzantine population), a process $P_i$ does:
\begin{enumerate}[label=\arabic*.]
    \item If receiving a value $v$ from more than $2f$ peers, decide $v$;
    \item Else if receiving a value $v$ from more than $f$ peers, hold $v$ as proposal value and go to the next phase;
    \item Else, toss a coin (1/2 chance for 0 or 1) for the proposal value and go to the next phase.
\end{enumerate}
When enough phases pass, the executions of step 2 and step 3 of RBC round-3 at all non-faulty processes will gradually filter out the influence of contradicting messages and eventually make the correct decision via step 1. Note this convergence only happens if $N\geq 3f+1$, which is the fundamental bound of BFT consensus. 

In terms of performance, the message complexity of RBC is $O(N^2)$ in each round and the expected number of rounds to reach consensus is $O(2^N)$ if $f=O(N)$, which gives a total message complexity of $O(N^2 2^N)$. If $f=O(\sqrt{N})$ (the benign case), it is shown in \cite{ben1983another,bracha1984asynchronous} that the expected number of rounds to reach consensus for the randomized protocol is a constant, yielding a total message complexity of $O(N^2)$.


\subsubsection{Ben-Or's ACS protocol for MPC}
Agreement on a common subset (ACS) was used by Ben-Or et al. \cite{ben1994asynchronous} in 1994 as a consensus primitive for secure and efficient multi-party computation (MPC) under asynchronous setting. In a network of $N$ players, each player holds a private input $x_i$ that was acquired secretly. The goal of MPC is to let the players collectively compute a function $\mathcal{F}(x_1,...,x_N)$ and obtain the same result.
Assuming $f$ players can be faulty, the ACS primitive requires the players to agree on a common subset $ComSubset$ of at least $N-f$ honest inputs, which are then used for computing $\mathcal{F}(\cdot)$. 

Ben-Or's ACS protocol builds on two primitives: RBC and binary asynchronous Byzantine agreement (ABA) which allows players to agree on the value of a single bit. Bracha's RBC \cite{bracha1984asynchronous} and Canetti et al.'s Fast ABA \cite{canetti1993fast} are suggested respectively in \cite{ben1994asynchronous} and used as black-boxes. Algorithm \ref{algo:benor-acs} shows the ACS protocol at each player. In the end, there will be at least $N-f$ completed ABA instances with output 1, yielding a $\mathcal{F}$-computable $ComSubset$.

\begin{algorithm}
\label{algo:benor-acs}
\caption{Ben-Or's ACS protocol (at player $P_i$)}
\SetAlgoLined
\SetKwProg{Fn}{Function}{:}{end}
\setcounter{AlgoLine}{0}

\tcc{Phase 1: Reliable Broadcast}
Start $RBC_i$ to propose my input $x_i$ to the network;

Participate in other $RBC$ instances;

\tcc{Phase 2: Asynchronous BA}
\While{round $\leq MaxRound$}{
    \If{receiving $x_j$ from $RBC_j$}{
        Join $ABA_j$ with input 1;
    }
    
    \If{completion of $N-f$ $ABA$ instances}{
        Join other $BA$ instances with input 0;
    }
    
    \If{completion of all $N$ $ABA$ instances}{
        $ComSubset=\{x_k|ABA_k \text{ outputs } 1\}$;
        
        \KwRet $ComSubset$;
    }
}

\end{algorithm}

Because both Bracha's RBC and Canetti's ABA can tolerate $f$ Byzantine players when $N\geq3f+1$, the same fault tolerance result is inherited by Ben-Or's ACS protocol. For complexity analysis, Bracha's RBC (in the benign case) and Canetti's ABA have message complexity of $O(N^2)$ and $O(N^3)$ \cite{canetti1993fast} respectively, and all ABA instances end in constant rounds. As a result, Ben-Or's ACS protocol has a bit-denominated communication complexity of $O(m N^2+N^3)$ at each player, where $m$ is the maximum bit-size of any input.

As we will see next, ACS can be conveniently adapted to asynchronous BFT consensus for blockchain systems, by substituting inputs with transaction sets.








\subsubsection{HoneyBadgerBFT}

Proposed by Miller et al. \cite{miller2016honey} in 2016, HoneyBadgerBFT is the first asynchronous BFT consensus protocol specifically designed for blockchain. It essentially realizes atomic broadcast: $N$ players with different sets of transactions work to agree on a common set of sorted transactions that will be included in a block.

\begin{figure}
    \centering
    \includegraphics[width=3.4in]{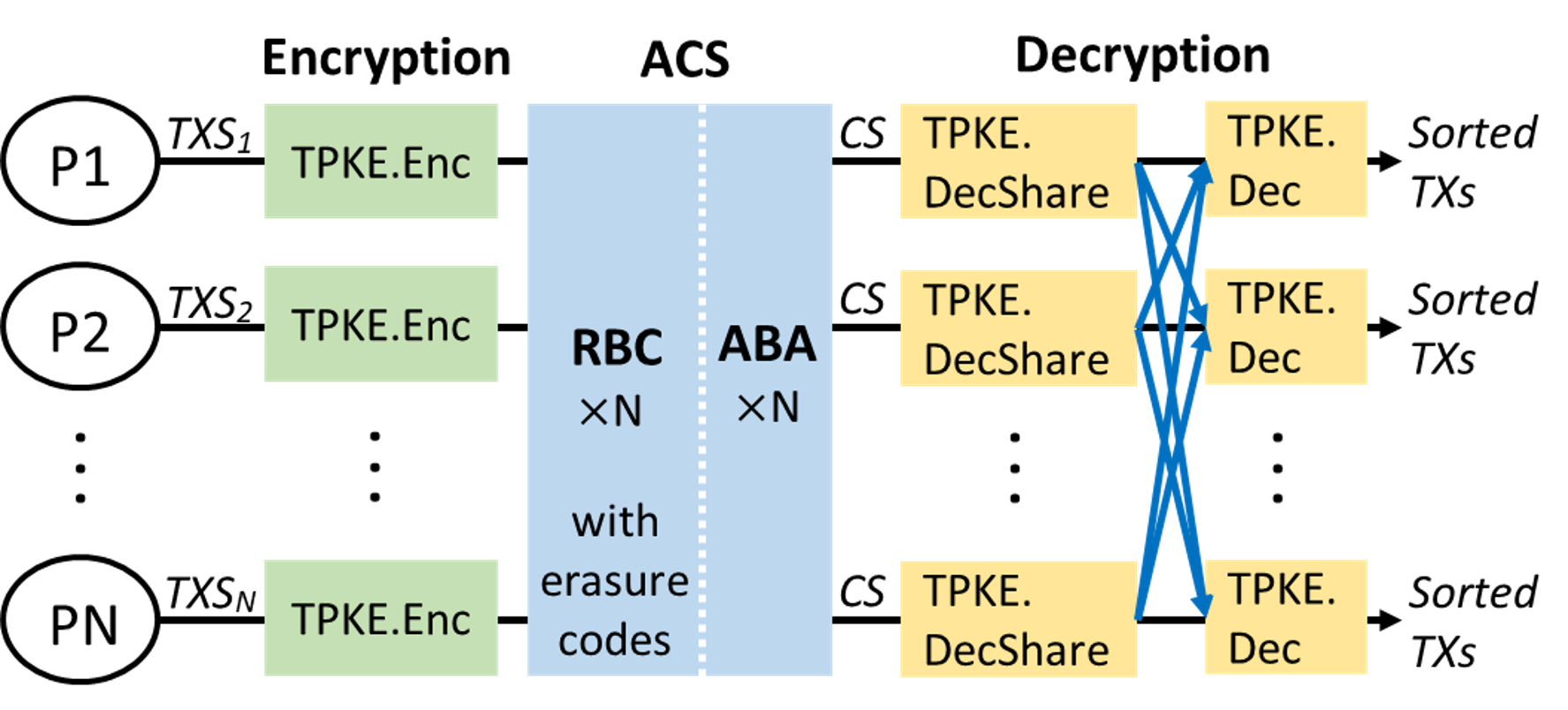}
    \caption{HoneyBadgerBFT workflow. $TXS_i$ is the set of transactions proposed by player $P_i$. $CS$ is the common subset output of ACS, comprising of at least $N-f$ encrypted transactions. The decryption process outputs sorted transactions that will be finalized in the block.}
    \label{fig:honeybadgerbft}
\end{figure}

Though using the multi-value Byzantine agreement primitive (MVBA) from Cachin et al. \cite{cachin2001secure} as the benchmark, HoneyBadgerBFT actually follows Ben-Or's ACS construction \cite{ben1994asynchronous} for better communication efficiency.
HoneyBadgerBFT's ACS cherry-picks the design of its sub-components: Cachin and Tessaro's erasure-coded RBC \cite{cachin2005asynchronous} and Most\'{e}faouil et al.'s common-coin based ABA \cite{mostefaoui2014signature}. They together incur $O(mN+N^2\log N)$ communication complexity at each player. 
To prevent an adversary from censoring particular transactions, threshold public key encryption (TPKE) \cite{baek2003simple} is used before ACS so that consensus is performed on ciphertexts.
In the decryption phase when a player receives enough shares from peers (generated by \emph{TPKE.DecShare}) that exceed a threshold, it proceeds to the actual decryption task (\emph{TPKE.Dec}) and sorts the transactions. The communication complexity of the decryption process is $O(N^2)$. Fig. \ref{fig:honeybadgerbft} illustrates the workflow of these components for one block cycle.

HoneyBadgerBFT processes transactions in batches. Let $B$ be the predefined batch size, denoting the maximum number of transactions that a block may enclose. For every block cycle, each player proposes a set of $B/N$ transactions, which are randomly chosen from recorded transactions. This is to ensure transaction sets proposed by different players are mostly disjoint so as to maximize blockchain throughput. Assuming the average bit-size of a transaction set $m:=\frac{|t|B}{N}\gg N$ where $|t|$ is the average transaction bit-size. Then the protocol's communication overhead will be dominated by the RBC, yielding overall communication complexity of $O(|t|B)$ at one player, or $O(|t|N)$ for one transaction.

Compared to popular partially synchronous consensus protocols such as PBFT, HoneyBadgerBFT has a higher cryptography overhead but features two advantages. First, as an asynchronous protocol HoneyBadgerBFT does not rely on a timeout mechanism for detecting malfunctioning players. This makes HoneyBadgerBFT less sensitive to unpredictable network delays that might stall consensus. Second, HoneyBadgerBFT does not need a leader rotation scheme. In PBFT every round of consensus is started by a leader (the primary), while in HoneyBadgerBFT every node starts its own broadcast and Byzantine agreement instance for proposed transactions; the concurrent execution of these instances effectively saves the need of a leader. 
As a result, the bandwidth of any individual leader will not become the bottleneck of overall network's capacity.
Currently the blockchain initiative POA Network \cite{poa2018whitepaper} is considering to adopt HoneyBadgerBFT.


On the other hand, due to HoneyBadgerBFT's asynchronous design philosophy that consensus progress is driven by message deliveries, transaction confirmation latency is externally influenced and uncontrollable. This leads HoneyBadgerBFT to overly emphasize high transaction throughput and decentralization.
In various applications such as industrial control and supply chain management, low transaction latency is often times a more important metric than throughput. 

{\color{\rcolor}
In response to the said inflexibility of HoneyBadgerBFT, Duan et al. \cite{duan2018beat} proposed BEAT in 2018, which is a collection of five asynchronous BFT protocols based off HoneyBadgerBFT but with carefully picked components that are optimized for different objectives.
Among the five constituent protocols, the baseline BEAT0 uses a more efficient threshold encryption scheme \cite{shoup1998securing} and outperforms HoneyBadgerBFT in throughput, latency and access overhead. BEAT1 and BEAT2 adopt a more efficient broadcast scheme, Bracha's RBC \cite{bracha1987asynchronous}, and are optimized for transaction latency. BEAT3 is optimized for throughput and storage and bandwidth saving while BEAT4 further reduces the bandwidth usage for clients that read particular stored transactions.
Interested readers are referred to the BEAT paper \cite{duan2018beat} for detailed discussion on the authors' design choices. 
}


\subsection{Blockchain Compatibility of Classical BFT-SMR Protocols} 


In a blockchain network, every consensus participant can validate transactions and propose new blocks.
For BFT-SMR consensus protocols that rely on a dedicated primary server to receive client requests and start the consensus, the following adaptation is needed: allowing all servers to act as a primary to propose transactions/blocks and reaching consensus on the finality of multiple transactions/blocks concurrently. 
For example, Casper FFG \cite{buterin2017casper}, a BFT-style blockchain protocol, allows every eligible participant to propose a block during a checkpoint cycle. The network finalizes only one block out of multiple proposed blocks for each checkpoint.

{\color{\rcolor}

For blockchain networks with a complex application layer such as smart contract, transaction execution often incurs significant computation. While in popular BFT-SMR schemes such as PBFT execution is integrated into the consensus process. An early work by Yin et al. \cite{yin2003separating} presents an alternative BFT-SMR framework that separates consensus (i.e. agreement on execution order) from execution, as the latter conveniently requires only an honest majority of execution nodes instead of an honest two-thirds of consensus nodes required by the former. This separation scheme is adopted by Zyzzyva \cite{kotla2007zyzzyva} and Tendermint \cite{kwon2014tendermint} wherein a small group of nodes are dedicated to the consensus task. Hyperledger Fabric \cite{androulaki2018hyperledger} further separates the consensus task into ordering service and validation service for better modularity.
}

From the performance perspective, BFT protocols are notorious for their limited scalability in network size.
Epitomized by PBFT, the message complexity of partially synchronous BFT protocols grows quadratically with the network size $N$. This means that given a fixed network bandwidth at each node, a growing network size leads to exploding communication overhead and diminishing transaction capacity. 
According to the performance evaluation in \cite{miller2016honey}, PBFT achieves a maximum throughput of 16,000 TPS when $N=8$; this figure drops to around 3,000 when $N=64$. 
On the other hand, for asynchronous protocols like HoneyBadgerBFT where erasure coding and threshold encryption are used to reduce communication complexity and enhance security, the extensive use of cryptography also brings non-negligible computation overhead, adding to local processing delays.
On the bright side, a typical BFT protocol achieves \emph{deterministic finality}, which is also known as \emph{forward security} \cite{decker2016bitcoin} in that a settled transaction will never be altered. As we will discuss in Section \ref{sec:nakamoto}, this allows BFT protocols to take advantage of shorter block intervals and attain high transaction throughput. 


Other blockchain compatibility challenges for BFT-SMR protocols include: 1) allowing nodes to join and leave flexibly without interrupting consensus while countering Sybil attacks; 2) adapting to real-world peer-to-peer networks that are sparsely connected. In later sections we will revisit these issues for blockchain protocols that incorporate BFT consensus.
\section{An Overview of Blockchain Consensus}
\label{sec:blockchain-consensus}


Compared to traditional distributed computing with a clear client-server model, a blockchain network allows every participant to be both a client (to issue transactions) and a server (to validate and finalize transactions). The underlying ledger data structure, the blockchain, is the consensus target and consists of chronologically ordered and hash-chained blocks. Each block contains a bundle of valid transactions and transactions across the blockchain should be consistent with each other (i.e. no double-/over-spending nor appropriation). 
Meanwhile, a blockchain system is often associated with a financial application and bears the responsibility of transaction processing and clearing. As a result, the responsibility of a blockchain consensus protocol is further-reaching than traditional distributed consensus protocols. In this section we provide a background of the blockchain network and data structure, introduce the blockchain consensus goal adapted from the BFT consensus paradigm and the five-component framework that we use to analyze blockchain consensus protocols.

\subsection{Blockchain Infrastructure}
\label{subsec:block-infra}

\textbf{Network~} 
The foundational infrastructure of blockchain, as is adopted by most public cryptocurrencies and distributed ledger systems, is a peer-to-peer overlay network on top of the Internet. 
Every node (or peer) in the network operates autonomously with respect to the same set of rules that cover peering protocol, consensus protocol, transaction processing, ledger management, and in some cases a wire protocol for transport-layer communication \cite{bitcoinwire2019,ethereumwire2019}.

\begin{table}
    \centering
    \color{\rcolor}
    \caption{A Comparison of Permissionless and Permissioned Blockchain.}
    \renewcommand{\arraystretch}{1.2}
    \small
    \begin{tabular} {|p{0.7in}||p{1.1in}|p{1.1in}|}
        \hline
        & \textbf{Permissionless blockchain} & \textbf{Permissioned blockchain} \\ \hline \hline
        Governance & Public & Private / Consortium \\ \hline
        Participation & Free join and leave & Authorized \\ \hline
        Node identity & Pseudonymous & Revealed \\ \hline
        Transparency & Open & Closed / Open \\ \hline
        Network \hspace{1in} size & Large \hspace{1in} (thousands or more) & Small (tens$\sim$hundreds) \\ \hline
        Network \hspace{1in} connectivity & Low & High \hspace{1in} (oft. fully-connected) \\ \hline
        Network \hspace{1in} synchrony & Asynchronous / partially synchronous & Partially synchronous / synchronous \\ \hline
        Transaction capacity (tps) & Low \hspace{1in} (oft. sub-ten$\sim$tens) & High \hspace{1in} (oft. thousands) \\ \hline
        Application examples & Cryptocurrency, smart contract, public record, DApp & Inter-bank clearing, business contract, supply chain  \\ \hline
    \end{tabular}
    \renewcommand{\arraystretch}{1}
    \label{tab:permissionless-v-permissioned}
\end{table}

Depending on the control of network participation, blockchain networks generally fall into two categories: permissionless and permissioned.
\begin{itemize}
    \item A \textit{permissionless} blockchain allows for free join and leave without any authorization, as long as the node holds a valid pseudonym (account address) and is able to send, receive, and validate transactions and blocks by common rules. 
    Permissionless blockchain is also known as public blockchain for that there is usually one such blockchain network instance on a global scale which is subject to public governance.
    Specifically, anyone can participate in blockchain consensus, though one's voting power is typically proportional to its possession of network resources, such as computation power, token wealth, storage space, etc.
    The operational environment of permissionless blockchain is often assumed to be zero-trust, which often cautions the community against increasing transaction processing capacity or using more efficient consensus schemes \cite{xiao2019distributed}.
    
    \item A \textit{permissioned} blockchain requires participants to be authorized first and then participate in network operation with revealed identity. 
    {\color{\rrcolor}
    The network governance and consensus body can be either the subsidiaries of a single private entity or a consortium of entities \cite{cachin2017blockchain}.
    Compared to permissionless blockchain, the identity-revealing requirement and more effective network governance of permissioned blockchain make it ideal for internal or multi-party business applications. Meanwhile, the limited size of a permissioned blockchain's consensus body} 
    allows for the deployment of more efficient consensus protocols that achieve higher transaction capacity \cite{xiao2019distributed,wust2018you}.
\end{itemize}
Table \ref{tab:permissionless-v-permissioned} summarizes the major differences between permissionless and permissioned blockchain in nine aspects.



Beneath the blockchain peer-to-peer network lies the basic infrastructure of the Internet. Thanks to the transport layer protocols (especially the retransmission mechanism), message delivery is considered guaranteed, while the message delay may vary but most likely will not grow longer as time elapses (weak synchrony) or remains within a certain bound ($\Delta$-synchrony). Therefore we often consider a practical blockchain network partially synchronous, just like most distributed networks overlaying on the Internet. This allows the consensus protocol to take advantage of the timing services of the Internet. For example, in Bitcoin the partial synchrony assumption is echoed by its usage of local timestamps for loose chronological ordering, showing time consciousness. For the blockchain networks that reside on an ad hoc infrastructure not based on the Internet, the message transmission is subject to unexpected network delays, which gives rise to asynchronous consensus protocols such as HoneyBadgerBFT, as we discussed in Section \ref{sec:background}.

\textbf{Transaction~}
A blockchain transaction can be regarded as a public static data record showing the token value redistribution between sender and receiver \cite{wang2018survey}. Take Bitcoin as an example, a transaction transfers token ownership from the sender account to the receiver account(s). It specifies a list of inputs and a list of outputs, with each input claiming a previous unspent transaction output (UXTO) that belongs to the sender, who needs to attach its signature to the inputs to justify the claim. Each output specifies how many tokens go to which receiver and the total token value of the outputs is equal to the UXTOs claimed by the inputs. Therefore, we can always recover the ownership records of any specific token by tracing back the signatures along the chain of transactions. The token balance of an account equals to the summed UXTOs that belong to the account.

\begin{figure}
    \centering
    \includegraphics[width=3.4in]{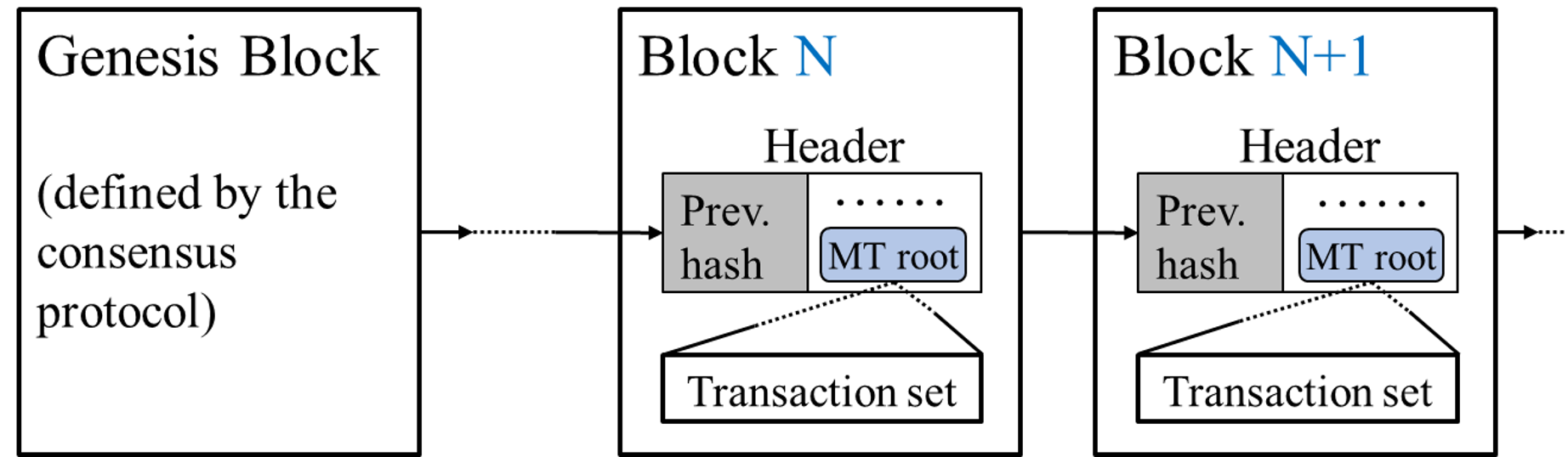}
    \caption{Blockchain data structure. Blocks are sequentially chained together via hash pointers. The Merkle tree root (MT root) is a digest of all transactions included in a block.}
    \label{fig:blockchain-ds}
\end{figure}

\begin{table*}
    \centering
    \caption{Five components of a blockchain consensus protocol.}
    \renewcommand{\arraystretch}{1.2}
    \small
    \begin{tabular} {|p{0.6in}||p{1.4in}| p{1.6in}| p{2.7in}|}
        \hline
        \textbf{Component} & \textbf{Purpose} & \textbf{Counterpart in traditional SMR consensus protocols} & \textbf{Available options} \\ \hline \hline
        
        Block \hspace{1in} proposal & Generating blocks and attaching essential generation proofs (for Sybil attack resistance).  & Clients issuing operation requests and the primary server starting the consensus. & Proof of work (PoW), proof of stake (PoS),  proof of authority (PoA), proof of retrievability (PoR), proof of elapsed time (PoET), round robin, committee-based, etc. \\ \hline
        
        Information propagation & Disseminating blocks and transactions across the network. & Reliable broadcast of operation requests. & Advertisement-based gossiping, block header soliciting, unsolicited block push (broadcast), relay network (for mining pools), etc. \\ \hline
        
        Block \hspace{1in} validation & Checking blocks for generation proofs and validity of enclosed transactions. & Signature check and execution of operation requests. & Proof checking (for proof-of-X block proposal), digital signature \& eligibility checking (for committee-based block proposal), etc. \\ \hline
 
        Block \hspace{1in} finalization & Reaching agreement on the acceptance of validated blocks. & Servers reaching an agreement on current state, executing requests and logging the result. & Longest-chain rule, GHOST rule, BFT and other Byzantine agreements, checkpointing, etc. \\ \hline
        
        Incentive mechanism & Promoting honest participation and creating network tokens. & N/A. & Network token rewards (block rewards, transaction fees), eligibility for issuing new transactions, etc. \\ \hline
    \end{tabular}
    \renewcommand{\arraystretch}{1}
    \label{tab:protocol-components}
\end{table*}

\textbf{Blockchain data structure~}
Blockchain is the underlying data structure for transaction ledger keeping. It is also the consensus target of the network. The basic structure of blockchain is illustrated in Fig. \ref{fig:blockchain-ds}. Every block encloses a set of transactions that should be valid and clear of double spending.
As was pioneered by Bitcoin, the transactions are often organized in a Merkle tree. Merkle tree is a data structure widely used for data storage and efficient data integrity check \cite{merkle1987digital}. In blockchain, every block contains one Merkle tree in which each leaf node is labeled with a transaction hash. The Merkle tree root serves as a digest of the transaction set and is placed in the block header.
The block header also contains a hash of the previous block (except in the genesis block) and other configuration information, which typically includes a timestamp and the blockchain state at block generation.
{\color{\rcolor}The growing chain of blocks and the aforementioned transaction format essentially constitute the blockchain data structure used for the storage, serialization, and validation of new transactions which are continuously injected into the network.}

Aside from recording the transaction history, the blockchain can also record auxiliary information used for other purposes. 
{\color{\rcolor}
The locking and unlocking scripts associated with transaction inputs and outputs can be repurposed for constructing off-chain payment channel (eg. Lightning Network \cite{poon2016bitcoin}) and global state machine that helps build smart contracts, which are the basis of many important applications such as supply chain management and decentralized autonomous organization.
}
The block header may also contain extra fields that facilitate system coordination. For example, Ethereum's proof-of-stake (PoS) scheme Casper FFG \cite{ryan2018casper} utilizes smart contract to implement the staking process; Algorand \cite{gilad2017algorand} attaches a cryptographic proof to each new block to show the block proposer's eligibility to propose. As a result, the blockchain can hold the necessary control information usable by the consensus protocol. We will revisit these protocols in later sections.

\subsection{Consensus Goal}

The goal of a blockchain consensus protocol is to ensure that all participating nodes agree on a common network transaction history, which is serialized in the form of a blockchain. Based on the previous discussion on BFT consensus and the consensus goal abstraction provided in \cite{xiao2019distributed}, we similarly define the following requirements for blockchain consensus:
\begin{itemize}
    \item \textbf{Termination~} At every honest node, a new transaction is either discarded or accepted into the blockchain, within the content of a block.
    
    \item \textbf{Agreement~} Every new transaction and its holding block should be either accepted or discarded by all honest nodes. An accepted block should be assigned the same sequence number by every honest node.
    
    \item \textbf{Validity~} If every node receives a same valid transaction/block, it should be accepted into the blockchain. 
    
    \item \textbf{Integrity~} At every honest node, all accepted transactions should be consistent with each other (no double spending). All accepted blocks should be correctly generated and hash-chained in chronological order.
\end{itemize}

The \emph{termination} and \emph{validity} requirements are similar to their counterparts in classical distributed consensus, as they represent the system's liveness. The \emph{agreement} requirement is enhanced with total ordering, which represents the serialization of blocks and transactions. The \emph{integrity} requirement dictates the correctness of the origin of transactions and blocks, fulfilling the promise of anti-double-spending and ledger tamper-proofing.
These requirements can serve as the design principles of new blockchain protocols. For different application scenarios, they can be tailored or supplemented with more specification.


\subsection{Components of Blockchain Consensus Protocol}

{\color{\rcolor}Based on the discussion on consensus goal and our digest of the blockchain documentation corpus, we identify five key components of a blockchain consensus protocol:
\begin{itemize}
    \item \emph{Block proposal}: Generating blocks and attaching generation proofs.
    \item \emph{Information propagation}: Disseminating blocks and transactions across network.
    \item \emph{Block validation}: Checking blocks for generation proofs and transaction validity.
    \item \emph{Block finalization}: Reaching agreement on the acceptance of validated blocks.
    \item \emph{Incentive mechanism}: Promoting honest participation and creating network tokens.
\end{itemize}
}

For each component we also specify its counterpart in traditional SMR consensus protocols and a list of available options in Table \ref{tab:protocol-components}. The available options are non-exhaustive, as many more are being developed at the time of writing.
{\color{\rcolor}
It is worth noting that the incentive mechanism is unique to blockchain consensus and has no counterpart in traditional SMR consensus protocols. The reason is that traditional SMR protocols are purely designed for transaction processing and serialization within a preexisting network of participants, of which the continuous participation of honest parties is presumed.
{\color{\rrcolor}
Meanwhile, a typical blockchain network allows for voluntary participation and often bears numerous real-world obligations. To this end, a fair and universal incentive mechanism is needed to encourage honest participation, so as to sustain the system's reliable operation. For large-scale permissionless blockchains, a robust incentive mechanism along with the block generation proofs also help demoralize Sybil attackers.
}
}

Though the five components are all vital to successful blockchain consensus, a new blockchain consensus protocol proposal may not cover all of them. For example, the incentive mechanism is indispensable to {\color{\rcolor}permissionless} blockchain networks especially those carrying a financial responsibility;  {\color{\rcolor}however for permissioned blockchains in which participation is sanctioned as a privilege (similar to a traditional distributed computing system), it is not a must-have}. Interestingly, many new public blockchain initiatives have been fixating only on block proposal, while inheriting the other four components from the Nakamoto consensus protocol of Bitcoin. This is likely due to that Bitcoin's PoW-based block proposal attracts the most criticism for its limited scalability and inefficient energy use. 
For this reason, block proposal mechanism can be a good reference angle for a general classification of consensus protocols. In the remaining part we dedicate Section \ref{sec:nakamoto} to Nakamoto consensus and its variations that are built upon PoW, while Section \ref{sec:pos} is dedicated to four genres of PoS-based protocols. Detailed protocol composition is described when it comes to specific features.

\begin{figure*}
    \centering
    \subfigure[]{
        \centering
        \includegraphics[width=1.8in]{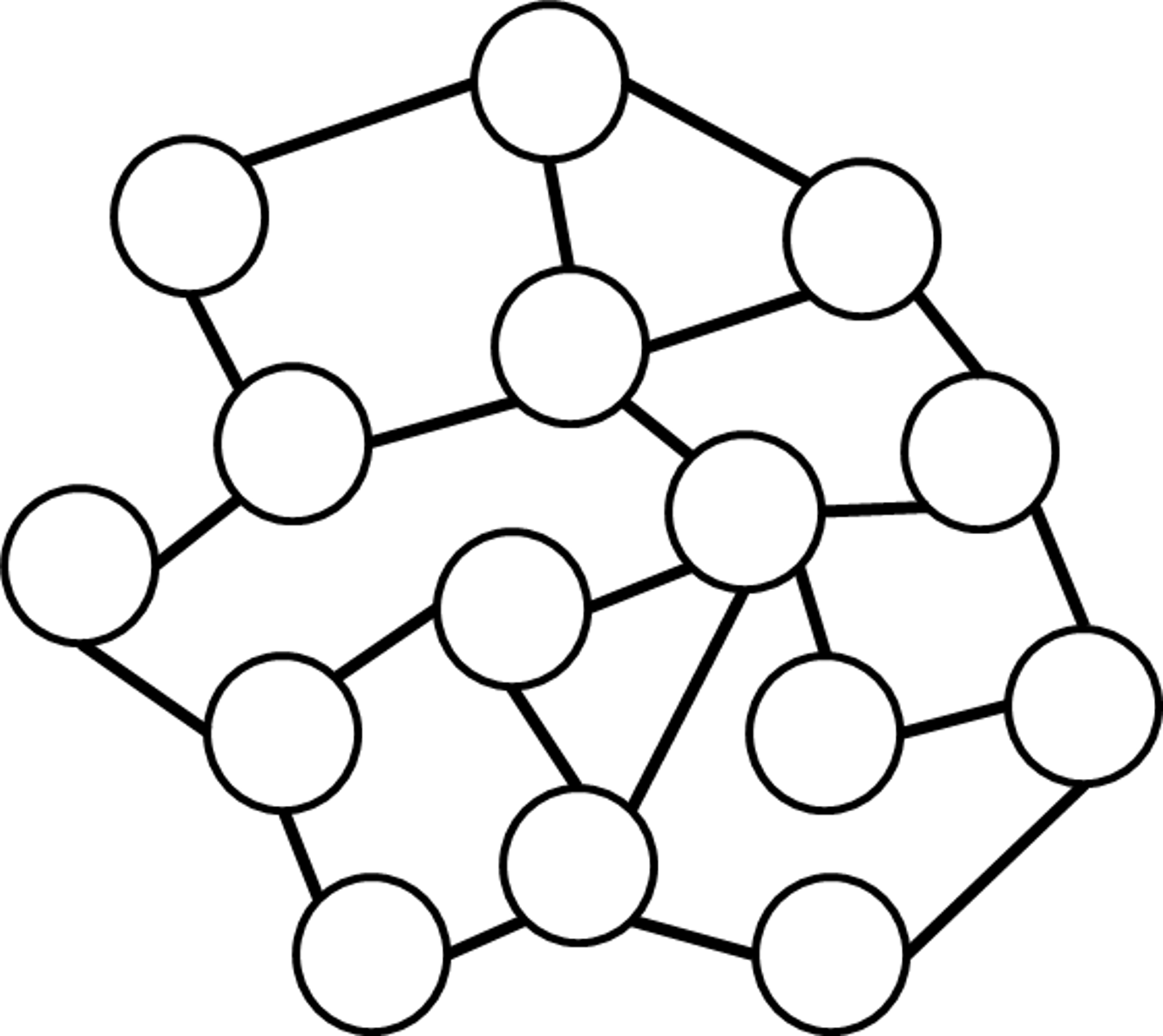}
        \label{fig:block-propagation-a}
    }\hspace{.2in}
    \subfigure[]{
        \centering
        \includegraphics[width=1.8in]{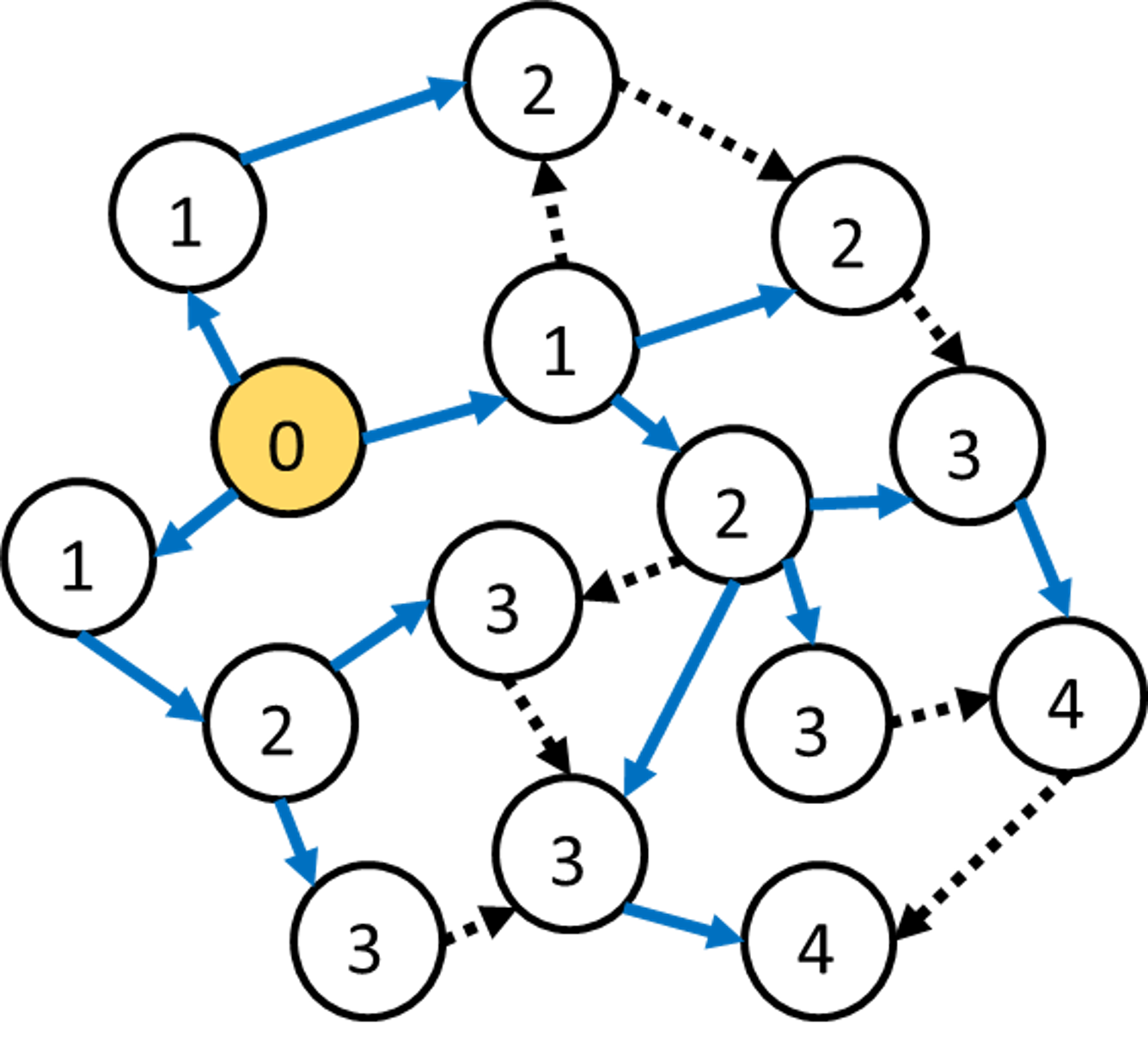}
        \label{fig:block-propagation-b}
    }\hspace{.2in}
    \subfigure[]{
        \centering
        \includegraphics[width=1.8in]{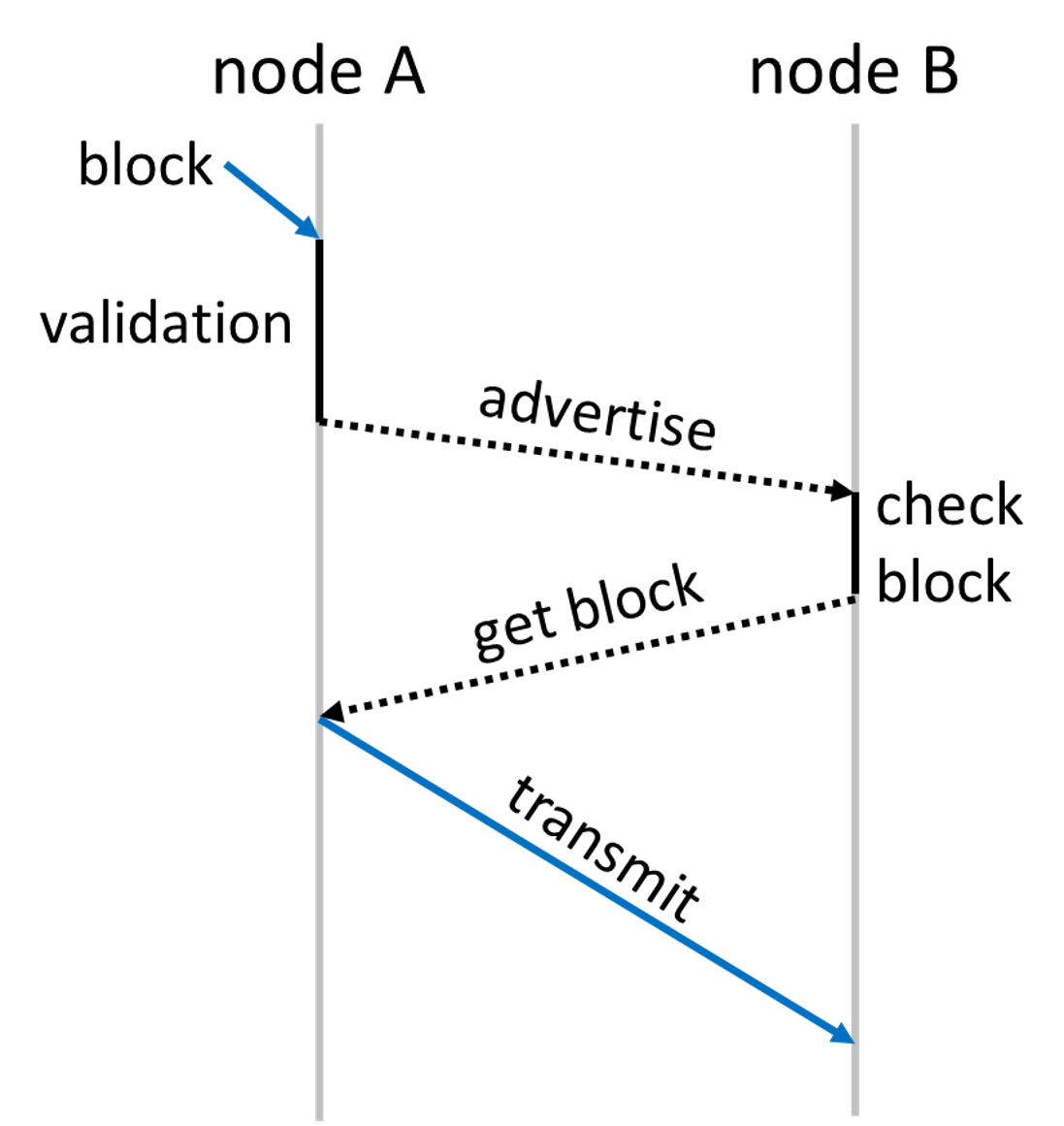}
        \label{fig:block-propagation-c}
    }
    \caption{A toy example of block propagation in the Bitcoin network. (a) The P2P network structure, an undirected graph. (b) The gossiping process. A solid blue arrow represents one-hop block propagation (advertise$\rightarrow$get block$\rightarrow$transmit), while a dotted black arrow represents only advertise. Number denotes the gossiping hop (0 for the block producer). (c) Block propagation in one hop.}
    \label{fig:block-propagation}
\end{figure*}

\section{The Nakamoto Consensus Protocol and Variations}
\label{sec:nakamoto}

The Nakamoto consensus protocol is the key innovation behind Bitcoin \cite{nakamoto2008bitcoin} and many other established cryptocurrency systems such as Ethereum \cite{wood2014ethereum} and Litecoin \cite{litecoinwhitepaper}.
In this section we use Bitcoin as the application background to introduce the Nakamoto consensus protocol and summarize its drawbacks and vulnerabilities. 
We also introduce two well-known improvement proposals and four hybrid PoW-BFT protocols in the later part of this section.




\subsection{Network Setting and Consensus Goal}

In blockchain networks, block or transaction messages are propagated across the P2P network through gossiping. Fig. \ref{fig:block-propagation} shows an example of block propagation in the Bitcoin network. Specifically, the one-hop propagation adopts advertisement-based gossiping, as was first characterized in \cite{decker2013information}. For each new block received and validated, a node advertises it to peers, who will request for this block if it extends their local blockchain. The gossiping process continues until every node in the network has this block. 



Compared to the general consensus goal introduced in Section \ref{sec:blockchain-consensus}, Nakamoto enhances the termination requirement with the probabilistic finality specification:
\begin{itemize}
    \item \textbf{Probabilistic finality~} For any honest node, every new block is either discarded or accepted into its local blockchain. An accepted block may still be discarded but with an exponentially diminishing probability as the blockchain continues to grow.
\end{itemize}


The probabilistic finality property echoes the probabilistic termination property for asynchronous consensus. As we will show later, because of this property the Nakamoto consensus protocol can only achieve eventual double-spending resistance in a decentralized network of pseudonymous participants.

\subsection{The Nakamoto Consensus Protocol}

In correspondence to the five components of a blockchain consensus protocol, the Nakamoto consensus protocol can be summarized by the following rules:
\begin{itemize}
    \item \emph{Proof of Work (PoW)}: Block generation requires finding a preimage to a hash function so the hash result satisfies a difficulty target, which is dynamically adjusted to maintain an average block generation interval.

    \item \emph{Gossiping rule}: Any newly received or locally generated transaction or block should be immediately advertised and broadcast to peers.
    
    \item \emph{Validation rule}: A block or transaction needs to be validated before being broadcast to peers or appended to the blockchain. The validation includes double-spending check on transactions and proof-of-work validity check on block header.
    
    \item \emph{Longest-chain rule}: The longest chain represents network consensus, which should be accepted by any node who sees it. Mining should always extend the longest chain.
    
    \item \emph{Block rewards and transaction fees}: Generator of a block can claim a certain amount of new tokens plus fees collected from all enclosed transactions, in the form of a \emph{coinbase} transaction to itself.
\end{itemize}

The hashing-intensive PoW mechanism is designed for mitigating Sybil attacks. Due to Bitcoin's permissionless and pseudonymous nature, Sybil attackers can obtain new identities or accounts with little effort. Hashing power, however, comes from real hardware investment and cannot be easily forged. The longest-chain rule implies that the stabilized prefix of the longest chain can act as a common reference of the network history, given that no one is authoritative in Bitcoin's decentralized network. 
Block Brewards and transaction fees are used to incentivize miners to participate honestly and inject new coins into circulation.

To better illustrate how these rules harmonize with each other, we present an abstracted version of the Nakamoto protocol in Algorithm \ref{algo:nakamoto-pow}. During block generation, a higher mining difficulty demands more brute-force trials in order to find a fulfilling nonce. To ensure every block is sufficiently propagated before the next block comes out, the mining difficulty is adjusted every 2016 blocks so that the expected block interval remains a constant value (10 minutes in Bitcoin) no matter how the gross hashing power fluctuates. 

\begin{algorithm}
\label{algo:nakamoto-pow}
\caption{Nakamoto consensus protocol general procedure}
\SetAlgoLined
\SetKwProg{Fn}{Function}{:}{end}
\setcounter{AlgoLine}{0}

\tcc{Joining network} 
Join the network by connecting to known peers;\

Start \textit{BlockGen()};\

\tcc{Main loop}
\While{running}
{
    \If{BlockGen() returns block}
    {
        Write \textit{block} into blockchain;\
        
        Reset \textit{BlockGen()} to the current blockchain;\
        
        \tcc{Gossiping rule}
        Broadcast \textit{block} to peers;\
    }
    
    \tcc{Longest-chain/validation rule}
    \If{block received \& is valid \& extends the longest chain}
    {
        Write \textit{block} into blockchain;\
        
        Reset \textit{BlockGen()} to the current blockchain;\
        
        Relay \textit{block} to peers;\
    }
}
\tcc{PoW-based block generation}
\Fn{BlockGen()}
{
    Pack up transactions (including \emph{coinbase});\ 
    
    Prepare a block header context $\mathcal{C}$ containing the transaction Merkle tree root, hash of the last block in the longest chain, timestamp, and other essential information reflecting blockchain status;\
    
    \tcc{PoW hashing puzzle}
    Find a nonce that satisfies the following condition:
    \begin{equation*}
        Hash(\mathcal{C}|\textit{nonce}) < \textit{target}
    \end{equation*}
    wherein more preceding zero bits in \textit{target} indicates a higher mining difficulty;
    
    \KwRet new block;
}
\end{algorithm}

\textbf{Fork resolution~} Ideally, the 10-minute block interval should be enough to ensure the thorough propagation of a new block so that no block of the same height is proposed. However due to the delay during the message propagation and the probabilistic nature of the hashing game, the possibility of two blocks of the same height being propagated concurrently in the network can not be ignored. This situation is called a ``fork", detectable by any node. Correspondingly, the longest-chain rule provides the criterion for fork resolution. Assume a miner receives two valid blocks $B_1^k,B_2^k$ of the same block height $k$ sequentially, then a fork is detected by this miner. It chooses $B_1^k$ (the first arrived) to continue and may encounter the following cases:
\begin{itemize}
    \item Case 1: If receiving or successfully generating a block $B_1^{k+1}$ confirming $B_1^k$, accept $B_1^k$, $B_1^{k+1}$ and orphans $B_2^k$.
    
    \item Case 2: If receiving a block $B_2^{k+1}$ confirming $B_2^k$, switch to $B_2^{k+1}$ and accept $B_2^k$, then orphans $B_1^k$.
    
    \item Case 3: If simultaneously receiving two blocks $B_1^{k+1}$ and $B_2^{k+2}$ confirming respectively $B_1^k$ and $B_2^k$, choose one to follow and continue until case 1 or 2 is met.
\end{itemize}
Wherein to ``orphan'' a block means to deny it into the main chain. Because of the randomized nature of PoW mining, the likelihood of encountering case 3 drops exponentially as time elapses, reflecting the probabilistic finality of Nakamoto consensus.

\textbf{Security analysis~} In contrast to the classical distributed computing system whose fault tolerance capability is characterized by the number of faulty nodes the system can tolerate, fault tolerance of Nakamoto consensus is by characterized by percentage of adversarial hashing power the system can tolerate.
It is proved by Garay et al. \cite{garay2015bitcoin} that if the network synchronizes faster than the PoW-based block proposing rate, an honest majority among the equally-potent (in hashing power and bandwidth) miners can guarantee the consensus on an ever-growing prefix of the blockchain. The prefix represents the probabilistically stable part of the blockchain.
As long as less than $50\%$ of total hashing power is maliciously controlled, the blocks produced by honest miners are timely propagated, the main chain contributed by the honest majority can eventually outgrow any malicious branch.

From the perspective of classical distributed consensus, Nakamoto consensus cleverly circumvents the fundamental 1/3 BFT bound by adopting probabilistic finality. In classical BFT consensus if more than 1/3 of population are malicious, the honest nodes will end up deciding conflicting values, leading to consensus failure. In Nakamoto consensus, however, conflicting decisions are allowed temporarily in the form of blockchain forks, as long as they will be eventually trimmed out by continuing effort of the honest majority. Therefore, the 1/3 BFT bound is not applicable to Nakamoto consensus or other blockchain consensus protocols designed for probabilistic finality.
{\color{\rcolor}
Readers are referred to Abraham et al. \cite{abraham2017blockchain} for an interesting discussion on the correspondence between Nakamoto consensus and classical BFT-SMR framework.
}

As for double-spending resistance, assuming the adversary controls $\alpha$ fraction of the total hashing power and wishes to double-spend an output which is $m$ blocks old, it needs to redo the PoW mining all the way from $m$ blocks behind and grow a malicious chain fast enough to overtake the incumbent main chain. The probability of this adversarial catch-up is $(\frac{\alpha}{1-\alpha})^m$ if $\alpha<50\%$, which drops exponentially as $m$ increases, reflecting probabilistic finality. This probability equals to 1 if $\alpha\geq50\%$. As a result, the 50\% threshold is the safeguard behind Bitcoin's probabilistic finality, as well as resistance to double-spending and transaction history tampering.

\subsection{Drawbacks and Vulnerabilities of Nakamoto Consensus}

\subsubsection{Tight tradeoff between performance and security}

The Nakamoto consensus is widely criticized for its low transaction throughput. For instance, Bitcoin can process up to 7 TPS meanwhile the VISA payment network processes 2500 TPS on average \cite{visa2012visanet}. The limited performance of Nakamoto consensus protocol follows from the security implication of its probabilistic finality and two protocol parameters: block interval and block size. As we discussed previously, the 10-minute block interval ensures every new block is sufficiently propagated before a new block is mined. Reducing the block interval increases the transaction capacity, but will leave new blocks insufficiently propagated and causes more forks incidents, undermining the security of the main chain. Note that although any fork can be resolved given enough time, the higher the fork rate, the larger the portion of honest mining power is wasted, which enables a double-spending attacker to overthrow the main chain with less than 50\% mining power (estimated 49.1\% by Decker et al. \cite{decker2013information} in 2013). On the other hand, increasing the block size (currently 1MB) has the same effect, since larger block sizes lead to higher block transmission delays and  insufficient propagation. According to the measurement and analysis by Croman et al. \cite{croman2016scaling} in 2016, given the current 10-min block interval the maximum block size should not exceed 4MB, which yields a peak throughput of 27 TPS. 

\subsubsection{Energy inefficiency} 

As of November 2019, an average Bitcoin transaction consumes 431 KWh of electricity which can power 21 U.S. households for a day \cite{digiconomist2019bitcoin}. This enormous energy consumption is directly caused by the PoW-based block proposing scheme of Nakamoto consensus. As Bitcoin network's gross mining capacity grows, the Nakamoto consensus protocol has to raise the mining difficulty to maintain the average 10-min block interval, which in turn encourages miners to invest into more mining equipment with higher hashing rates. This vicious cycle shall continue as Bitcoin gains more popularity. In response, the blockchain community has come up with various block proposing schemes such as proof of stake (PoS), proof of authority (PoA), proof of elapsed time (PoET) as energy-saving alternatives to PoW.

\subsubsection{Eclipse attack} 

As was discussed above, the security of Bitcoin network relies on the hashing power and communication capability of honest miners. If a powerful attacker manages to dominate the in/outward communication between a victim miner and the main network (i.e. ``eclipsing"), then the victim will no longer be able to contribute to the extension of the main chain \cite{heilman2015eclipse}. Assume the percentage of hashing power controlled by the eclipse attacker, the eclipsed victims, the remaining honest miners are $\alpha,~\epsilon$, $1-\alpha-\epsilon$, then the attacker's mining power shall be amplified to at least $\frac{\alpha}{1-\epsilon}$. If the attacker decides to exploit the eclipsed victims for growing the malicious chain, its mining power can be further enhanced up to $\alpha+\epsilon$ \cite{gervais2016security}. As a result, a double-spending attack becomes viable for the eclipse attacker if $\alpha+\epsilon>50\%$.

Eclipse attack is in fact an exploit of the weak connectivity of permissionless peer-to-peer network based upon the Internet, which is subject to unpredictable physical bottlenecks and adversarial influences. A general approach to counter eclipse attacks is to secure the communication channels and increase the connectivity and geographical diversity of the peer-to-peer connections.

\subsubsection{Selfish mining} The 50\% fault tolerance of Nakamoto consensus is built upon the assumption that all miners (both honest and malicious) strictly follow the broadcast rule that new blocks are broadcast immediately upon successful generation. If a malicious mining group withholds newly mined blocks and strategically publicizes them to disrupt the propagation of blocks mined by honest miners, they can partially nullify the work of honest miners and amplify their effective mining power. This strategy is known as \textit{selfish mining}. 
It is shown by Eyal et al. \cite{eyal2014majority} that a selfish mining group can generate a disproportionately higher revenue than that from honest mining if the group's mining power surpasses a certain threshold $\theta$, assuming the group has a certain communication capability measured by $\gamma\in[0,1]$, which is the fraction of honest nodes that will follow the malicious chain in case of forks. 
As a result, the selfish mining group can attract new miners and eventually outgrow the honest miners. Notably, this threshold approaches zero if the selfish mining group are able to convince almost all honest miners to follow the malicious chain (i.e. $\gamma\rightarrow1$). It is also shown that by adopting a randomized chain selection strategy at honest miners, which is equivalent to setting $\gamma=0.5$, the threshold can be raised up to $25\%$. A later work by Sapirshtein et al. \cite{sapirshtein2016optimal} shows that an optimized selfish mining strategy can further enhance the selfish mining pool's effective mining power fraction from $\alpha$ to the upper bound $\frac{\alpha}{1-\alpha}$ (achievable when $\gamma=1$).

Selfish mining attack and eclipse attack have only happened to smaller blockchains such as Monacoin \cite{monacoin2018selfish}, but never to Bitcoin or other mainstream blockchains. This is probably due to two reasons. First, miners in established public blockchain networks actually care about the system's longevity and reputation, which can positively affect the exchange rate of the cryptocurrency and thus their mining revenue. Second, established blockchains tend to be better connected (reflected by the existence of dedicated mining pools and relay networks), which allows for an effective detection of any selfish mining and eclipse attack behavior.

\subsubsection{Mining pools and centralization risk}
According to the incentive mechanism of Nakamoto consensus, the mining revenue of a miner is proportional to its computing power.
Since bitcoins can be traded for fiat currencies at exchanges, higher-earning miners have the financial advantage to purchase more efficient mining hardware, which consumes less joules per hash operation. Furthermore, higher-earning miners are often backed by large organizations that can direct huge capital into the mining business. As a result, small individual miners are either forced out of the game, or alternatively join in mining pools for stabler income. All members in a mining pool are registered with a coordinator and work to extend a common chain, while transaction validation, packaging and block proposal can be performed independently.  
{\color{\rrcolor}
To incentivize pool participation, block rewards are redistributed among the pool through a reliable remuneration scheme so that every pool member routinely gets a fair share of the pool's mining rewards according to its registered computing power.
}

In fact, joining in a mining pool has become the dominant way of participation in major PoW-based blockchains.
The measurement study by Gencer et al. \cite{gencer2018decentralization} in 2018 shows that throughout a one-year observation period, over 50\% of the gross mining power was controlled by eight mining pools in Bitcoin and five mining pools in Ethereum.
Moreover, the empirical study by Kondor et al. \cite{kondor2014rich} in 2014 shows that the wealth distribution among Bitcoin addresses has been converging to a stable exponential distribution, and the wealth accumulation of node is positively related to its ability to attract new connections, which is another advantage of established large miners.


\subsection{Improvements to the Nakamoto Consensus Protocol}

\subsubsection{GHOST Rule}

The greedy heaviest-observed subtree (GHOST) {\color{\rcolor}block finalization rule} was proposed by Sompolinsky et al. \cite{sompolinsky2015secure} for Bitcoin in 2015. According to the longest-chain rule, all unconfirmed blocks in a fork shall be orphaned, resulting in a waste of honest mining power which could otherwise have been used to contribute to the longest-chain's security. The longest-chain rule also limits the transaction capacity {\color{\rcolor}since the tight tradeoff between performance and security mandates that the block interval should be sufficiently long}. The GHOST rule is an alternative to the longest-chain rule that the orphaned blocks also contribute to the main chain security, effectively reducing the impacts of forks, which allows for a shorter block interval and thus higher transaction capacity. GHOST requires that given a tree of blocks with the genesis block being the root, the longest chain within the heaviest subtree shall be used as the main chain. Similar to the Nakamoto consensus, the probabilistic finality of the heaviest subtree up to the current block height will hold as long as more than 50\% of mining power are honest. 

The simulation result in \cite{sompolinsky2015secure} shows that given the same block interval, applying GHOST rule leads to a slightly lower transaction throughput than that with the longest-chain rule when block interval is fixed, but near-perfectly prevents the security degradation when the block interval decreases, allowing for a higher transaction throughput.
A variation of GHOST is implemented in the Ethereum blockchain, wherein the ``uncle blocks'' (i.e. blocks with a valid proof of work but orphaned out of the main chain) may get rewarded for their redundant mining effort. As a result, Ethereum adopts a much shorter block interval (10-15 seconds) and achieves up to 25 TPS throughput, in contrast to Bitcoin's 10-minute block interval and 7 TPS throughput.


\subsubsection{Bitcoin-NG}

\begin{figure}
    \centering
    \includegraphics[width=3.4in]{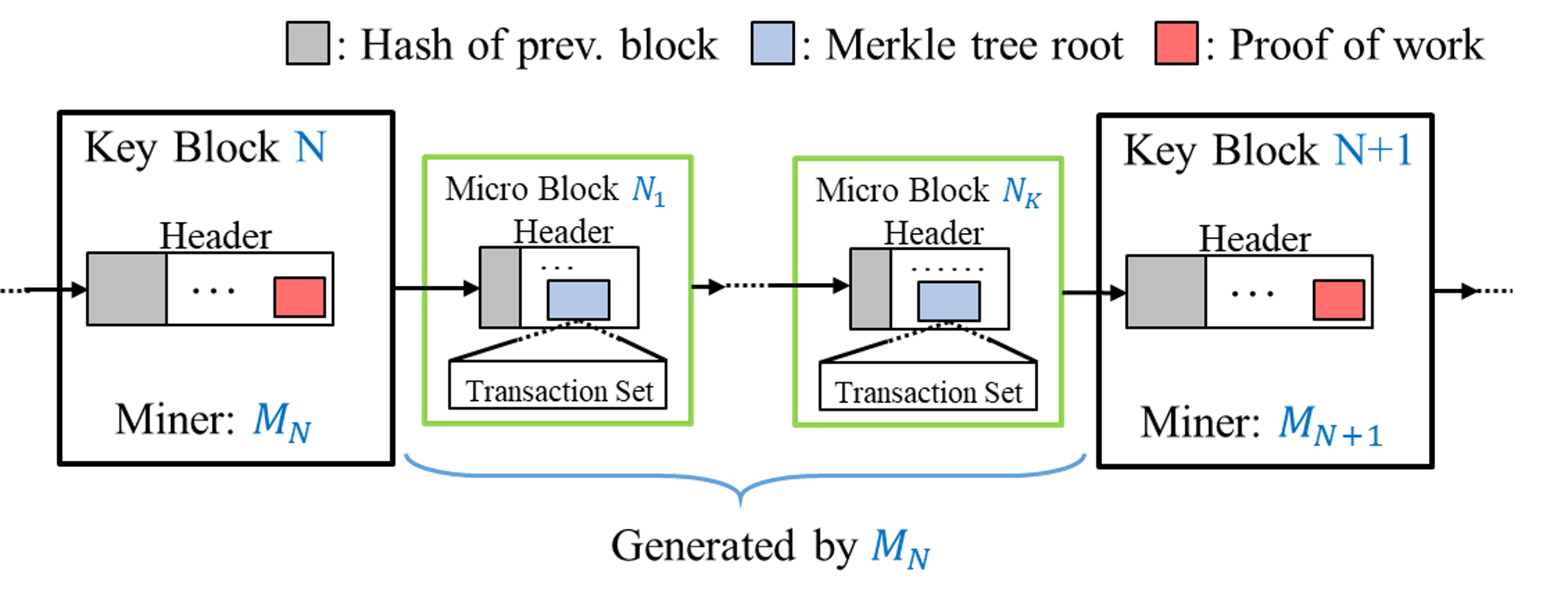}
    \caption{Bitcoin-NG blockchain data structure. Key blocks track miners and are generated through PoW. Micro blocks curate transactions and are generated by the most recent key block miner.}
    \label{fig:bitcoin-ng-dr}
\end{figure}

Bitcoin-NG was proposed by Eyal et al. \cite{eyal2016bitcoin} in 2016 to scale up Bitcoin's transaction capacity. A variant of Bitcoin-NG called Waves-NG \cite{wavengdocs} is currently used in Wave Platform, a blockchain initiative. The key insight of Bitcoin-NG is decoupling block generation into two planes: \emph{leader-election} and \emph{transaction serialization}, which respectively correspond to two types of blocks: \emph{key blocks} and \emph{micro blocks}. The key blocks resemble Bitcoin's blocks, which contain a solution to a hash puzzle representing the proof of work and have an average block interval of 10 minutes, except for the actual transactions which are included in the micro blocks. Once a key block is mined, all subsequent micro blocks shall be generated by the current key block miner until the generation of the next key block. The generation of micro blocks is deterministic and does not contain proof of work. As a result, the micro block frequency is in the control of the key block miner (up to a maximum) to accommodate as many transactions as possible. The blockchain data structure of Bitcoin-NG is shown in Fig. \ref{fig:bitcoin-ng-dr}. 

The longest-chain rule is still applied to finalize and resolve forks of key blocks. As for the micro blocks, since they are batch-generated by key block miners, Bitcoin-NG relies on a combination of a heaviest-chain extension rule and a longest-chain extension rule to finalize and resolve forks of micro blocks.
To encourage honest participation and discourage the current key block miner from double-spending and other malfeasance, 60\% of the transaction fees collected from micro blocks by the current key block miner are redistributed to the miner of the next key block. 

As for Bitcoin-NG's performance, since key blocks do not carry transactions, the transaction throughput entirely depends on the micro block size and frequency.
The micro block frequency needs to be controlled, for an excessive amount of them may exhaust the network bandwidth and cause frequent key block forks.
Hypothetically, if we kept the key block frequency at 10 minute and micro block size at 1MB, set the micro block frequency to 12 seconds (the minimum practical block interval under current Bitcoin network condition \cite{li2018scaling}), Bitcoin-NG would achieve up to 200 TPS throughput.

On the downside, due to the determinism in micro block generation, the key block miner may become a target of denial-of-service or corruption attacks. A compromised key block miner may enclose transactions selectively or finalize contradicting transactions, the inconsistency caused by which can cost the network more than one key block cycle to remedy.



\subsection{Hybrid PoW-BFT Consensus Protocols}

The limited transaction capacity and tight tradeoff between performance and security of Nakamoto consensus are much warranted by its probabilistic finality and decentralized ideal. In contrast, BFT consensus assumes fixed participants with revealed identities and achieves deterministic finality, allowing much shorter block intervals and thus much higher transaction throughput. In response, hybrid PoW-BFT protocols have been proposed to get the best of two worlds. Here we introduce four popular proposals.

\subsubsection{PeerConsensus} Proposed by Decker et al. \cite{decker2016bitcoin} in 2014, PeerConsensus uses a PoW-based blockchain to throttle and certify new identities joining the network, while being agnostic to any application built upon it. The number of identities a player may control is proportional to its share of computation power, which provides Sybil resistance. With the identities established by the blockchain, the application can employ an efficient BFT protocol such as PBFT and SGMP \cite{reiter1996secure} for committing transactions. The transaction fees collected are distributed to all identities equally. As a result, PeerConsensus effectively decouples participation management from transaction processing, allowing the latter to scale up throughput. On the downside, since the transaction history is not recorded in blockchain, PeerConsensus cannot control the malleability of transactions.

\subsubsection{SCP} The scalable consensus protocol (SCP), proposed by
Luu et al. \cite{luu2015scp} in 2015, incorporates BFT and sharding into blockchain consensus. The key idea of SCP is to partition the network into sub-committees (i.e. shards) with a PoW mechanism, so that each sub-committee controls a limited amount of computation power and the number of sub-committees is proportional to the network's gross computation power. This is aimed to limit the size of a sub-committee, which operates independently and curates a local blockchain using a BFT consensus protocol. A dedicated finalization committee is responsible for combining the outputs of all sub-committees into the global blockchain. A block in the global chain stores the hash and transaction Merkle tree root of every block proposed by every sub-committee. To ensure consensus safety, SCP requires each sub-committee as well as the whole network to maintain a two-thirds majority of honest computation power. However, the use of sharding and a dedicated finalization committee assumes the preexistence of network coordination, which to some extent counters the decentralized ideal of public blockchains. 

Notably, using sharding to scale up blockchain transaction throughput has been extensively studied in the developer communities. Interested readers are referred to \cite{sharding-ethereum} for a development history and the state-of-the-arts of blockchain sharding.

\subsubsection{ByzCoin} ByzCoin was proposed by Kogias et al. \cite{kogias2016enhancing} in 2016 as a blockchain consensus protocol that leverages PoW for consensus group membership management and BFT for transaction finalization. 
{\color{\rcolor}
ByzCoin takes inspiration from Bitcoin-NG's ledger structure but features a subtle difference. Instead of a linear structure,
}
ByzCoin's ledger consists of two parallel blockchains: a keyblock chain and a microblock chain. Keyblocks are mined via PoW as in Nakamoto consensus and used for maintaining a consensus group from recent keyblock miners according to a \textit{sliding-share-window} mechanism.
{\color{\rcolor}
Specifically, when a miner finds a new keyblock, it is credited one \textit{share} in the consensus group and the share window moves one share forward. Only miners with shares in the window can participate in the subsequent consensus. Old shares expire once being left out of the window and so does the share owner's eligibility of consensus participation. The window length is subject to designer's choice on the consensus group size as well as the overall consensus participation fairness.
}

A microblock is produced by the current consensus group via an adapted PBFT protocol based on collective signing (CoSi) \cite{syta2016keeping}. Compared to the original PBFT, the CoSi-based PBFT reduces the communication complexity from $O(N^2)$ to $O(N\log N)$ and thus scales better to large consensus groups. As for transaction finalization, the current keyblock miner packs up new transactions into a microblock and acts as the leader to trigger the CoSi-based PBFT. In the end, the microblock will be finalized and contain the collective signature of the consensus group and the hash pointer to the preceding keyblock, which also contains the collective signature.

ByzCoin configures that the sliding-share-window mechanism replaces one member of the consensus group at a new keyblock. This yields a slightly tighter fault-tolerance threshold than that of classical BFT consensus: $N\geq3f+2$ is needed anytime, where $N$ is the consensus group size and $f$ the Byzantine population. Meanwhile, ByzCoin's PoW-based keyblock chain is still susceptible to forks. A fork can split the consensus group and potentially make the PBFT consensus stall, which can further aggravated by the presence of selfish miners. In response, ByzCoin relies on a deterministic prioritization function tweaked with high output entropy to resolve forks timely and reduce the impact of selfish miners.

\subsubsection{Pass and Shi's hybrid consensus} As a concurrent work to ByzCoin, the hybrid consensus protocol proposed by Pass and Shi \cite{pass2017hybrid} adopts a sliding-window idea similar to ByzCoin's but less susceptible to forks. That is, assuming the window size $\lambda$, the consensus group is populated by the last $\lambda$ miners of the ``stable part" of the blockchain, which is the main chain truncated off $\Theta(\lambda)$ blocks. This protocol keeps the PBFT consensus off-chain; only the consensus epoch number and a digest of the transaction log are attached to the new block. Moreover, this protocol advocates using FruitChain \cite{pass2017fruitchains} as the underlying PoW blockchain, which was proposed by the same authors and allegedly achieves better ledger tamper-resistance than Nakamoto's blockchain.

\textbf{A short summary~}
Due to the scalability concern that BFT protocol's communication overhead would be overwhelmingly high if the consensus group grew out of control, the above hybrid PoW-BFT protocols share a common trait that the PoW mechanism is used for maintaining a stable consensus group for each BFT protocol instance. Since the PoW-based participation control does not involve actual authorization and is open to any one with computation power, we consider it a form of \textit{soft permission} control for public blockchains. Moreover, novel signature schemes such as CoSi \cite{syta2016keeping} and aggregated signature gossip \cite{long2019scalable} can help reduce communication complexity in a sparsely-connected peer-to-peer network and allow for a larger number of consensus participants.

We stress that there are more ways to hybridize PoW and BFT in addition to the above proposals. Moreover, independently proposed cryptographic techniques are often complementary to the hybrid design. This observation also applies to general hybrid PoX-BFT schemes.


\section{Proof-of-Stake Based Consensus Protocols}
\label{sec:pos}

Proof-of-Stake (PoS) originates from the Bitcoin community as an energy efficient alternative to PoW mining. 
In the simplest terms, a stake refers to the coins or network tokens owned by a participant that can be invested in the blockchain consensus process.
From the security point of view, PoS leverages token ownership for Sybil attack mitigation.
Compared to a PoW miner whose chance to propose a block is proportional to its brute-force computation power, the chance to propose a block for a PoS miner is proportional to its stake value.
From the economics point of view, PoS moves a miner's opportunity cost from outside the system (waste of computation power and electricity) to inside the system (loss of capital and investment gain) \cite{poelstra2014distributed}. Because of the lack of real mining, we often refer to a PoS miner as a \textit{validator}, \textit{minter}, or a \textit{stakeholder} for PoS's close resemblance to investing in capital markets.

\begin{figure}
    \centering
    \includegraphics[width=3.3in]{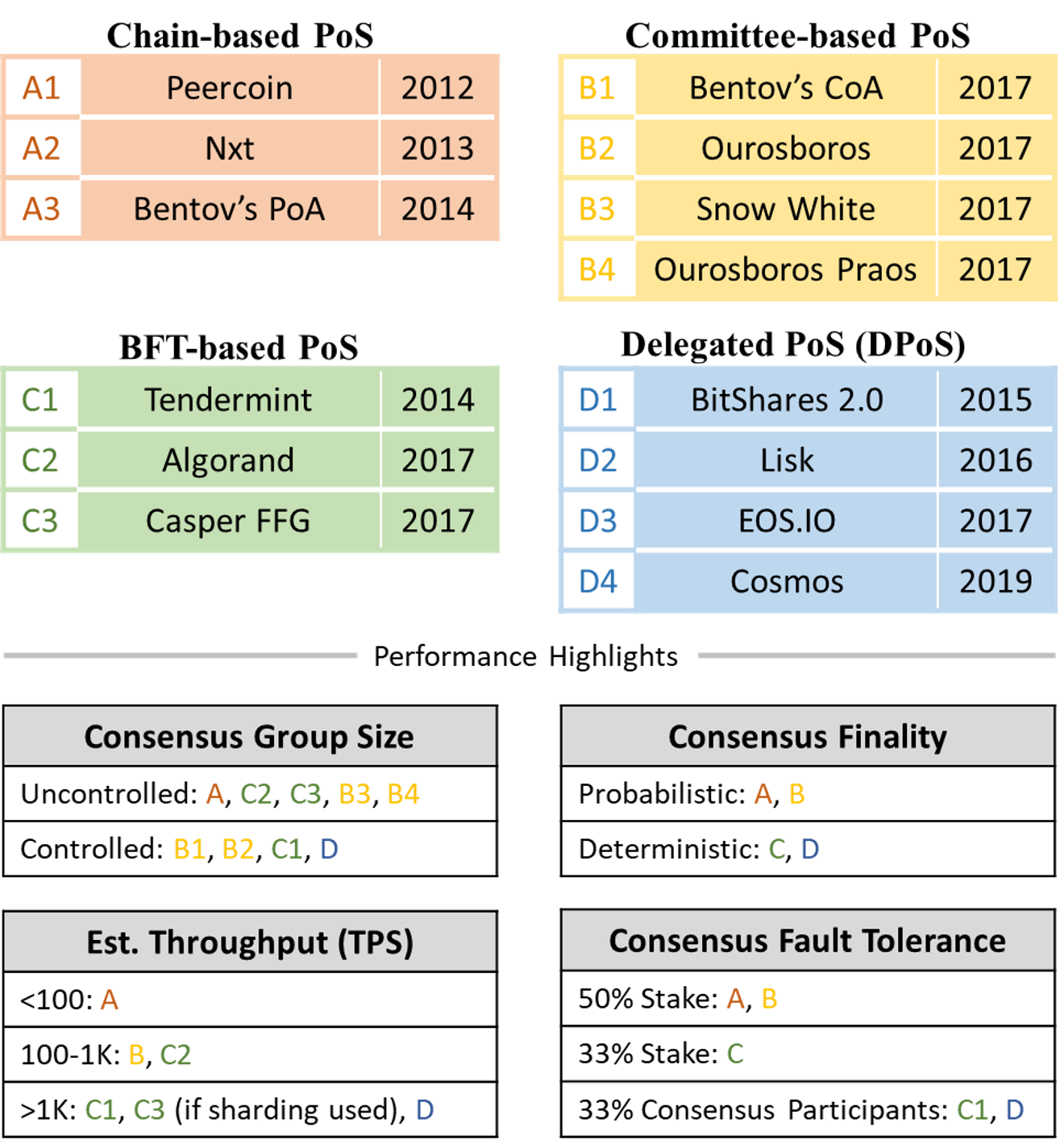}
    \caption{Popular PoS blockchain initiatives classified under four classes and performance highlights.}
    \label{fig:pos-classes}
\end{figure}

We identify four classes of PoS protocols: \textit{chain-based PoS}, \textit{committee-based PoS}, \textit{BFT-based PoS}, and \textit{delegated PoS} (DPoS). Chain-based PoS inherits many of the components of the Nakamoto consensus protocol such as information propagation, block validation, and block finalization (i.e. longest-chain rule), except that the block generation mechanism is replaced with PoS. Committee-based PoS leverages a multiparty computation (MPC) scheme to determine a committee to orderly generate blocks. BFT-based PoS combines staking with BFT consensus which guarantees deterministic finality of blocks. DPoS employs a social voting mechanism that elects a fixed-size group of delegates for transaction validation and blockchain consensus on behalf of the voters. Popular examples for each PoS class are listed in Fig. \ref{fig:pos-classes}.

\subsection{Chain-based PoS}
\label{subsec:chainbased-pos}

Chain-based PoS is an early PoS scheme proposed by Bitcoin developers as an alternative block generation mechanism to PoW. It is within the framework of Nakamoto consensus in that the gossiping-style message passing, block validation rule, longest-chain rule, and probabilistic finality are preserved. Early full-fledged chain-based PoS blockchain systems include Peercoin and Nxt.

The general procedure of a chain-based PoS minter can be summarized by Algorithm \ref{algo:chainbased-pos}.
Unlike PoW, PoS does not hinge on wasteful hashing to generate blocks. A minter can solve the hashing puzzle only once for a clock tick.
Since the hashing puzzle difficulty decreases with the minter's stake value, the expected number of hashing attempts for a minter to solve the puzzle can be significantly reduced if her stake value is high.
Therefore, PoS avoids the brute-force hashing competition that would occur had PoW been used, thus achieving a significant reduction of energy usage.

\begin{algorithm}
\label{algo:chainbased-pos}
\caption{Chain-based PoS general procedure (Peercoin, Nxt)}
\SetAlgoLined
\SetKwProg{Fn}{Function}{:}{end}
\setcounter{AlgoLine}{0}

Join the network by connecting to known peers;\

Deposit in the stake pool;\

Start \textit{BlockGen()};\

\tcc{Main loop}\
\While{running}
{
    (Same with Nakamoto's protocol except that block validation should include PoS check.)
}

\tcc{PoS-based block generation}
\Fn{BlockGen()}
{
    Pack up transactions and prepare a block header context $\mathcal{C}$ containing the transaction Merkle tree root and other essential blockchain information;\
    
    \tcc{PoS hashing puzzle}
    Set up a clock (whose tick interval is a constant) and check for the following condition per clock tick:
    \begin{equation*}
        Hash(\mathcal{C}|\textit{clock\_time}) < \textit{target} \times \textit{stake\_value}
    \end{equation*}
    wherein more preceding zero bits in \textit{target} indicates a higher mining difficulty per unit of stake value;
    
    \KwRet new block;
}
\end{algorithm}


\subsubsection{Peercoin and Nxt} Both Peercoin \cite{king2012ppcoin} and Nxt \cite{Nxt2014whitepaper} generally follow Algorithm \ref{algo:chainbased-pos}. Their major difference lies in how the stake is valued. Stake value is initially proportional to stake quantity. To ensure the profitability of small stakeholders, a stake valuation scheme can be used to adjust the value of an unused stake as time passes.
Peercoin uses the \textit{coin age} metric for stake valuation, which lets the value of a stake appreciate linearly with time since the deposit. At the end of a block cycle, the value of the winner's stake returns to its base value. To avoid stakeholders from locking in a future block by deliberately waiting long, stake appreciation only continues for 90 days and stays flat since then. As a result, the chances of small stakeholders to generate a block are supplemented with time value that encourages them to stay participated even if they have not generated a block for a long time.


In comparison, Nxt does not appreciate stake value continuously across block cycles like what Peercoin does. This is because the latter's coin age metric may lower the attack cost---attackers can just invest a small amount of stakes and keep attempting to generate blocks until they succeed, which is especially unfair to big stakeholders who perform honestly. Instead, Nxt requires the stake value appreciate only within one block cycle and be reset to the base value once the block cycle ends. Without coin age, Nxt addresses the monopolization problem from the incentive angle. First, stakes in Nxt are not actively managed by stakeholders; they are essentially the account balances---the more tokens held in its account, the higher the chance the stakeholder will win the right to generate a block. Second, total token supply is determined at the beginning and block rewards only come from transaction fees, which effectively aligns a stakeholder's revenue with its validation work. As a result, all stakeholders have the incentive to honestly validate transactions since it is the only way to accumulate wealth.

\subsubsection{Bentov's PoA} Compared to Peercoin and Nxt, Bentov's proof of activity (PoA) \cite{bentov2014proof} 
is a hybrid PoW-PoS adaptation of the Bitcoin protocol that utilizes an algorithm called \textit{follow-the-satoshi} (FTS) to involve staking. 
FTS works as follows: 1) Use a pseudo-random function (PRF) to locate an atomic piece of token (eg. satoshi in Bitcoin, wei in Ethereum) in the token universe; 2) If the atomic piece belongs to stakeholder $A$, then output $A$. With the input being a sequence of random seeds, FTS outputs a pseudo-random sequence of stakeholders such that the chance for any stakeholder to be in it is proportional to the volume of tokens owned by the stakeholder.

Bentov's PoA works as follows. At the beginning of block cycle $k$, an empty block header $EB_k$ is generated according to the PoW rule and propagated across the network. After receiving $EB_k$, a stakeholder computes the $N$-tuple vector seed $S$ as follows:
\begin{equation*}
    S_j=hash\Big(hash(EB_k)|hash(B_{k-1})|SF_j\Big) ~~\text{for}~j=1,...,N
\end{equation*}
$B_{k-1}$ is the previous block, $SF$ is a $N$-tuple of fixed suffix values. $N$ is a predefined value that should not be too large. Then $S$ is used as the input for FTS to compute the pseudo-random sequence of stakeholders $pSeq$, which is also a $N$-tuple. Every stakeholder in $pSeq$ needs to sign the block and broadcast the signature; the last stakeholder in $pSeq$ wraps up the block by including transactions and the $N$ signatures and broadcasts the final block $B_k$ to the network. All stakeholders in $pSeq$ shares the reward of $B_k$ with the PoW miner of $EB_k$.

To avoid name conflict with proof of authority, we also refer to Bentov's PoA as PoAct.

\textbf{Security analysis~} Chain-based PoS can tolerate up to 50\% of all stakes being maliciously controlled. And since every token can be staked, the fault tolerance further generalizes to 50\% of all tokens in the network. If colluding attackers control more than 50\% of stakes, they can grow their malicious chain faster than the others and carry out a double-spending attack, which is analogous to the 51\% attack in PoW blockchains. However, from the economic perspective, PoS attackers have lower incentives to do so because of the capital loss risk. As staking is recorded in the form of transaction scripts, the blockchain users can retrieve the staking records from which the consensus protocol can legally issue punishment to violators, such as nullifying stakes and disbarring the violators from participating in the future staking process.

\subsection{Committee-based PoS}
\label{subsec:committee-pos}

Chain-based PoS still relies on the hashing puzzle to generate blocks. As an alternative mechanism, committee-based PoS adopts a more orderly regime: determining a committee of stakeholders based on their stakes and allowing the committee to generate blocks in turns. A secure \textit{multiparty computation} (MPC) scheme is often used to derive such a committee in the distributed network. MPC is a genre of distributed computing in which multiple parties beginning with individual inputs shall output the same result \cite{rabin1989verifiable}. The MPC process in the committee-based PoS essentially realizes the functionality that takes in the current blockchain state which includes the stake values from all stakeholders, and outputs a pseudo-random sequence of stakeholders (we call it the \textit{leader sequence}) which will subsequently populate the block-proposing committee. This leader sequence should be the same for all stakeholders and those with higher stake values may take up more spots in the sequence. A general procedure for a stakeholder of committee-based PoS is shown in Algorithm \ref{algo:committee-pos}. In this algorithm, the \emph{CommitteeElect()} functionality can also be implemented in a privacy-preserving way with verifiable random function (VRF) \cite{micali1999verifiable} in that only the stakeholder itself knows if it gets elected into the committee.

\begin{algorithm}
\label{algo:committee-pos}
\caption{Committee-based PoS general procedure}
\SetAlgoLined
\SetKwProg{Fn}{Function}{:}{end}
\setcounter{AlgoLine}{0}

\tcc{Joining network and staking} 
Join the network by connecting to known peers;\

Deposit in the stake pool;\


\tcc{Main loop}
\While{running}
{   
    \tcc{Committee election}
    \If{new block cycle}{
        Participate in \textit{CommitteeElect()};\ 
        
        Check \textit{BlockGenSeq} for my turns;\
    }
    
    \tcc{Block proposing \& broadcast}
    \If{my turn to generate block}{
        Collect transactions and generate \textit{block};\
        
        Write \textit{block} to blockchain;\
        
        Broadcast \textit{block} to the network;\
    }
    
    \tcc{Longest-chain\&validation rule}
    \If{block is received \& is valid \& extends the longest chain}
    {
        Write \textit{block} into blockchain;
        
        Relay blocks to other committee members;
    }
    
}

\tcc{PoS-based committee election}
\Fn{CommitteeElect()}
{
    Fetch the current blockchain state and the stake information of all participants; use them as the MPC input;\
    
    Participate in the MPC that produces \textit{BlockGenSeq}, a pseudo-random sequence of block generation opportunities;\
    
    \KwRet \textit{BlockGenSeq};
}
\end{algorithm}

Well-known committee-based PoS schemes include Bentov's chain of activity (CoA), Ouroboros, Snow White, and Ouroboros Praos. These protocols and Algorand (see Section \ref{subsec:bft-pos}) were developed concurrently by academics around 2017 and share many common traits.

\subsubsection{Bentov's CoA}
Bentov's CoA \cite{bentov2016cryptocurrencies} was proposed in 2016 partially based on Bentov's PoA. It follows the main routine in Algorithm \ref{algo:committee-pos} in that each nominated stakeholder gets to generate its own block. CoA first leverages a MPC process to generate a string $S$ of $N$ random bytes. Then $S$ is fed to a FTS algorithm that outputs a pseudo-random sequence \textit{BlockGenSeq}. All parties should output the same \textit{BlockGenSeq}, which is then used to coordinate the generation of the next $N$ blocks.

\subsubsection{Ouroboros} 
\begin{figure}
    \centering
    \includegraphics[width=3.4in]{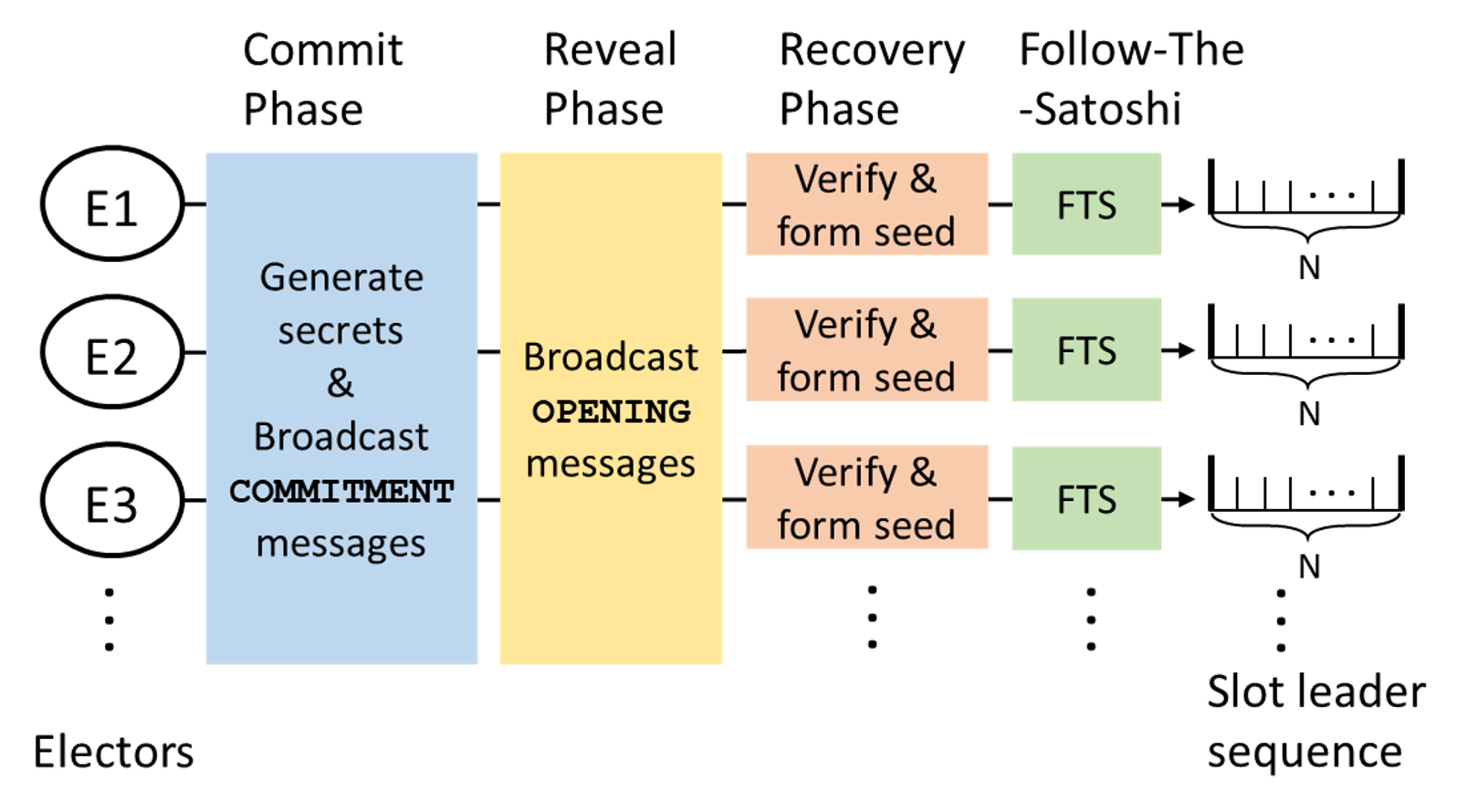}
    \caption{PVSS-based slot leader sequence generation in Ouroboros. Secrets generated by electors at the beginning of the commit phase eventually form the random seed string. All electors shall go through FTS and obtain the same slot leader sequence at the end.}
    \label{fig:Ouroboros-mpc}
\end{figure}

Ouroboros was developed by Kiayias et al. \cite{kiayias2017ouroboros} in 2017 and has been used as the consensus protocol for cryptocurrency Cardano. 
Ouroboros divides the physical time into fixed-time epochs and each epoch is subdivided into $N$ slots, each can be used by only one slot leader to generate a block for the network. For each epoch, stakeholders with enough stake can become electors, who will collectively elect slot leaders (i.e. the committee) for the next epoch through a MPC procedure known as publicly verifiable secret sharing (PVSS), as is shown in Fig. \ref{fig:Ouroboros-mpc}. In the commit phase, elector $E_i$ broadcasts a commitment message that includes a random secret. In the reveal phase, $E_i$ broadcasts an opening message that reveals the previously sent secret. In the recovery phase every elector verifies that commitments and openings match and then form a seed string with the revealed secrets. All electors have the same seed string and shall obtain the same slot leader sequence after executing FTS.
As we shall see next, Ouroboros' one-slot-one-leader arrangement entails stringent network synchrony, and the PVSS-based leader selection may expose the elected leaders to targeted attacks.

\subsubsection{Ouroboro Praos} Ouroboros Praos was proposed by David et al. \cite{david2018ouroboros} in 2017 to address two security concerns of Ouroboros. First, Ouroboros requires stringent network synchrony for slot leaders to use their allocated slots precisely, which is vulnerable to desynchronization attacks. In comparison, Ouroboros Praos is designed for partially synchronous networks wherein a maximum delay of $\Delta$ slots is allowed for message delivery, albeit $\Delta$ is unknown to electors. This is achieved by allocating empty slots that help electors to re-synchronize to the network and allowing a slot to have multiple slot leaders.
Second, Ouroboros is susceptible to adaptive corruption against slot leaders. Since the leader sequence for the next epoch is known to all network participants, an attacker may target and try to corrupt slot leaders ahead of their block proposing. In comparison, Ouroboros Praos employs a locally executed verifiable random function (VRF) \cite{micali1999verifiable} that allows only the elector herself to know her block proposing slots for the next epoch, which is verified by the corresponding VRF proofs. Compared to Ouroboros' PVSS-based leader selection, the VRF scheme saves much of the communication overhead at the cost of local cryptographic computation. Similar schemes are also used in contemporary protocols such as Snow White and Algorand.

Because of its reduced synchrony requirement and the privacy-preserving nature of the VRF scheme, Ouroboros Praos does not limit the size of consensus participants and allows for a flexible committee.
Ouroboros Praos also adopts key-evolving signatures (KES) to counter posterior corruption and provide forward security, which we will elaborate in Section \ref{subsec:pos-vuln}.




\subsubsection{Snow White} Snow White was developed by Daian et al. \cite{daian2017snow} in 2017. It
is a PoS protocol specifically designed to accommodate the sporadic participation model in which nodes can switch online/offline arbitrarily. Similar to the committee-based PoS schemes above, Snow White employs a MPC procedure to decide the block proposing committee, each of which is issued an eligibility ticket privately. For each epoch, every stakeholder takes the current blockchain (including staking information) as input and output the committee for the next epoch. Specially, Snow White executes a modified version of the sleepy consensus protocol \cite{pass2017sleepy}, an asynchronous consensus protocol that ensures the consensus safety in case of sporadic participation and committee reconfiguration. Compared to contemporary schemes (including Ouroboros, Praos and Algorand), this feature enables Snow White to work under harsh network conditions with frequent disconnections and volatile message delays. Moreover, Snow White uses a checkpointing scheme to finalize earlier history that protects the blockchain from posterior corruption attack \cite{bentov2014proof} and adaptive key selection attack.

\textbf{Security analysis} In spite of having an orderly block proposing scheme, committee-based PoS still adheres to the longest chain rule for probabilistic finality. So long as fewer than 50\% stakes are held by the malicious party, the honest parties can safely maintain the longest chain. 
Meanwhile, the expansion of the committee may lead to deterioration in network connectivity which can result in a significant drop in protocol performance and desynchronization of block proposal. The round-based committee election process with a predetermined round duration (eg. Ouroboros' fixed timeslot) also faces scalability problems, as large committee sizes may lead to never ending consensus cycles due to the excessive communication overhead.
To mitigate such risks, the duration of a communication round can be extended sufficiently to ensure that all broadcast messages are delivered before the participants proceed to the next round.
This however leads to longer transaction conformation latency and lower throughput. A more straightforward approach is limiting the committee size by imposing a minimum stake requirement for the committee members. For example, Cardano, the cryptocurrency platform that deploys Ouroboros, mandates that every committee member should own no less than 2\% of total tokens in circulation. This effectively limits the committee size to 50, safeguarding an efficient consensus process.

\subsection{BFT-based PoS}
\label{subsec:bft-pos}

Chain-based PoS and committee-based PoS largely follow the Nakamoto consensus framework in that  the longest-chain rule is still used to provide probabilistic finality of blocks. In comparison, BFT-based PoS (or hybrid PoS-BFT) incorporates an extra layer of BFT consensus that provides fast and deterministic block finalization. Algorithm \ref{algo:bftbased-pos0} shows the general procedure of BFT-based PoS at every participant.
Block proposing can be done by any PoS mechanism (round-robin, committee-based, etc.) as long as it injects a stable flow of new blocks into the BFT consensus layer. 

Aside from the general procedure, a checkpointing mechanism can be used to seal the finality of the blockchain (not shown in Algorithm \ref{algo:bftbased-pos0}). As a result, the longest-chain rule can be safely replaced by the most-recent-stable-checkpoint rule for determining the stable main chain.
Popular BFT-based PoS blockchain protocols include Tendermint, Algorand, and Casper FFG. DPoS protocols such as EOSIO also use BFT consensus for block finalization within delegates.

\begin{algorithm}
\label{algo:bftbased-pos0}
\caption{BFT-based PoS general procedure}
\SetAlgoLined
\SetKwProg{Fn}{Function}{:}{end}
\setcounter{AlgoLine}{0}

Join the network by connecting to known peers;\

Start \emph{BlockGen()};\

\tcc{Main loop}
\While{running}{
    \tcc{Block proposing \& broadcast}
    \If{BlockGen() returns block}{
        
        Add \textit{block} to its \textit{tempBlockSet};\
        
        Broadcast \textit{block} to the network;
    }
    
    \tcc{Block validation}
    \If{block is received \& is valid}{
        Add \textit{block} to its \textit{tempBlockSet};\
        
        Relay \textit{block} to the network;\
    }
    
    \tcc{BFT consensus layer}
    \If{new consensus epoch}{
        Perform \textit{BlockFinBFT()} on \textit{tempBlockSet};
        
        Write the winning block to blockchain;
        
        Clear \textit{tempBlockSet};\
    }
}

\tcc{PoS-based block generation}
\Fn{BlockGen()}
{
    Elect a block proposer, whose success rate is proportional to stake value;\
    
    Propose block;\
    
    \KwRet block;\
}

\tcc{BFT-based block finalization}
\Fn{BlockFinBFT()}
{

    Participate in a BFT consensus that finalizes one winning block out of \textit{tempBlockSet};    
    
    \KwRet the winning block;
}
    
\end{algorithm}

\subsubsection{Tendermint} 
Tendermint was developed by Kwon et al. \cite{kwon2014tendermint,buchman2016tendermint} in 2014 and currently used in Cosmos Hub network \cite{cosmoshub}. It is the first public blockchain project to incorporate a BFT consensus layer and takes inspiration from the DLS protocol \cite{dwork1988consensus} and PBFT \cite{castro1999practical}.
Tendermint works in consensus cycles. Each cycle involves a multi-round BFT consensus process to finalize one block. Each round consists of three phases: Propose, Prevote, Precommit. Specially, in the Propose phase a validator is designated by a deterministic algorithm as the block proposer in a round-robin fashion such that validators are chosen with frequency proportional to the value of their deposited stakes. A validator continues iterating the three-phase rounds until one block receives more than 2/3 of Precommits. The validator then broadcasts Commit votes for the block and listens for other validators' Commit votes. When a block receives more than 2/3 of Commit votes, it will be finalized in blockchain. As long as more than 2/3 of validators of each round are honest, Tendermint can achieve consensus safety. On the other hand, because Tendermint selects block proposers deterministically, the future block proposers are susceptible to targeted attacks (eg. DDoS). Tendermint addresses this risk by deploying sentry nodes which act as proxies of block proposers and never reveal the IP addresses of the latter \cite{tendermintsentry}.
{\color{\rcolor}
Notably, since Tendermint decouples the PoS mechanism from the BFT layer, a validator's stake value does not add weight to its consensus votes. For this reason, Tendermint is often considered an early effort on applying BFT consensus to blockchain.

Tendermint's also features an equal-sharing-style incentive mechanism instead of winner-takes-all. For every block height, the block reward is distributed among the block proposer and validators from whom the proposer received Commit votes. However, fairness may be impaired if Commit votes are not delivered in time before the next cycle, as is demonstrated by Amoussou-Guenou et al. \cite{amoussou2018correctness}. 
}

\subsubsection{Algorand} Algorand is a cryptocurrency system developed by Gilad et al. \cite{gilad2017algorand} at MIT CSAIL in 2017. It employs committee-based PoS for block proposing and Byzantine agreement for block finalization.
{\color{\rcolor}
First, similar to Ouroboros Praos' block proposer election mechanism, the election of Algorand's block proposing committee is done by a VRF scheme called \emph{cryptographic sortition} which sorts candidates according to the amount of coins they own. 
}
Only those with rankings above a threshold are admitted into the committee for the next block cycle. Every individual user can check privately if it is in the committee. At each user $i$, cryptographic sortition also outputs an eligibility proof signed by the user's private key $\sigma_{sk_i}^{ep}$ showing that it is truly a committee member. $\sigma_{sk_i}^{ep}$ is broadcast to the network along with the new block proposed by user $i$. Upon receiving the block, other users can verify the proof via the user $i$'s public key $pk_i$.

On top of the cryptographic sortition-based block proposing scheme, Algorand relies on a Byzantine agreement protocol called $BA\star$ for block finalization. $BA\star$ reduces the consensus problem to binary Byzantine agreement: either agreeing on a proposed block or an empty block. In the ideal case where strong network synchrony is assumed, the committee follows $BA\star$ to exchange votes on proposed blocks so that they will decide a final block, or an empty block if no blocks pass the eligibility proof check. In a weakly synchronous network where block propagation and message exchange among committee members can suffer from uncertain delays, $BA\star$ outputs tentative blocks if none of the proposed blocks can be finalized, which results in a fork. To resolve the forks of tentative blocks, Algorand periodically runs a recovery protocol to accept a tentative block if there is any. Specially, the recovery protocol needs to be invoked by a synchronized committee. Therefore, according to the authors \cite{gilad2017algorand}, weak synchronony is sufficient for consensus safety during $BA\star$'s routine operation while strong synchrony is required for consensus liveness when kicking off the recovery protocol for resolving forks.

\begin{algorithm}
\label{algo:bftbased-pos}
\caption{Casper FFG}
\SetAlgoLined
\SetKwProg{Fn}{Function}{:}{end}
\setcounter{AlgoLine}{0}


Deposit in the stake pool;\



\tcc{Main loop}
\While{running}{
        
            
    
    (Block proposing and block validation are the same as in Algorithm \ref{algo:bftbased-pos0}, except that blocks are attached to \textit{BlockTree} rather than stored in a temporary set.)
    
        
    
    
    
    \tcc{BFT consensus layer}
    \If{new consensus epoch}{
        
        Identify valid checkpoint blocks and attach them to \textit{CheckPointTree};\
        
        Participate in \textit{CheckPointVote()} w.r.t. \textit{CheckPointTree}, which returns $CP_s,CP_t$;\
        
        Mark $CP_s$ \emph{finalized} and $CP_t$ \emph{justified};
    }
}

\tcc{Staked checkpoint voting}
\Fn{CheckpointVote()}
{
    Broadcast a vote for a source-target checkpoint pair in \textit{CheckPointTree};\
    
    Check received votes against the slashing rules and then evaluate them by signer's deposited stake;\

    \If{pair $\langle CP_s, CP_t \rangle$'s votes cover more than 2/3 of total deposited stakes}{
        \KwRet $CP_s, CP_t$;\
    } 
}
    
\end{algorithm}

\subsubsection{Casper FFG} Casper FFG was first envisioned by Zamfir \cite{zamfir2015introducing} in 2015 and formally presented by Buterin \cite{buterin2017casper} in 2017.
It is a light-weight PoS consensus layer built on top of Ethereum's current PoW-based block proposing mechanism (Ethash). Casper FFG slightly deviates from Algorithm \ref{algo:bftbased-pos0} for that it directly incorporates PoS into block finalization.
Algorithm \ref{algo:bftbased-pos} shows the general procedure of Casper FFG for each validator. Newly generated and received blocks are attached to the \emph{BlockTree}, which is similar to the tree data structure used by the GHOST rule. The actual consensus subject, however, is the \emph{CheckPointTree}, which is a subtree of \emph{BlockTree}.
Specifically, for every consensus epoch (100 in \emph{BlockTree}'s height or 1 in \emph{CheckPointTree}'s height), every validator broadcasts to peers a vote for a block as the checkpoint. The height of the block in \emph{BlockTree} must be divisible by 100. The vote consists of a \emph{justified} source checkpoint $CP_s$ and its height $h(s)$, a target checkpoint $CP_t$ and its height $h(t)$ ($h(s)<h(t)$), and the validator's signature $S$. All votes are broadcast to the network and are weighted by the signer's stake value. If the source-target checkpoint pair $\langle CP_s, CP_t \rangle$ is voted by validators who possess more than 2/3 of total deposited stakes, then $CP_t$ is justified and $CP_s$ is \emph{finalized}. All blocks between $CP_s$ and $CP_t$ are finalized as well. Casper FFG relies on two so called \emph{Casper Commandments} for ensuring consensus safety: 1) validator must not cast two distinct votes for the same checkpoint height, and 2) validator must not cast a new vote whose source-target span is within that of its existing vote. Violators are subject to slashing rules including forfeiting stakes and temporarily banning from staking. Since every vote is signed with the validator's private key and received by peer validators, Casper FFG can conveniently detect violators and enforce the slashing rules. 

The current smart contract implementation of Casper FFG is documented in EIP 1011 \cite{ryan2018casper}. A stakeholder becomes a participating validator by depositing a stake in the dedicated smart contract, which encodes Casper FFG and can be accessed via Ethereum transactions.
Notably, Casper FFG is the preamble project of Casper Correct-by-Construction (Casper CBC), the PoS protocol family that will be used by Ethereum 2.0 to complete the transition to pure PoS \cite{ethereumcbcfaq}. 

To further improve performance and scalability, Ethereum 2.0 also plans to combine PoS with sharding \cite{ethereumshardingfaq}. In a nutshell, all Ethereum 2.0 participants are divided into shards. Each shard runs a blockchain instance via a consensus scheme not limiting to PoS. On the top level, the main chain, known as the ``beacon chain'', will be maintained by a group of known validators via a Casper CBC protocol. Each validator is randomly assigned to a shard as the shard manager and periodically commits a digest of the shard chain to the main chain. The parallelism of sharding and the energy efficiency of PoS can theoretically scale up both transaction throughput and network size. Nonetheless, the sharding scheme is still an ongoing work and faces several challenges before being harmonized with Casper CBC. They include the increased take-over risk related to small shard size, the difficulty in coordinating inter-shard communication and token transfer.

\textbf{Security analysis} BFT-based PoS's consensus fault tolerance varies among the three above-mentioned implementations. In Tendermint, although block proposers are determined based on PoS, all validators have the equal weight in the consensus process. Therefore Tendermint tolerates up to 1/3 of Byzantine validators. In comparison, Algorand and Casper FFG tolerate up to 1/3 of maliciously-possessed stakes. In Algorand, if an attacker owns more than 1/3 of total tokens, then chance is high that more than 1/3 of the elected committee members is compromised by the attacker, leading to consensus failure of $BA\star$. In Casper FFG, if an attacker owns more than 1/3 of total deposited stakes and dominates the communication within the network, the compromised validators can vote on conflicting checkpoints without getting punished. Since a typical BFT consensus protocol can incorporate a checkpointing mechanism to ensure deterministic finality of blocks, costless simulation attacks can be naturally avoided (to introduce in Section \ref{subsec:pos-vuln}). 



\subsection{Delegated PoS (DPoS)}

\begin{figure}
    \centering
    \includegraphics[width=2.8in]{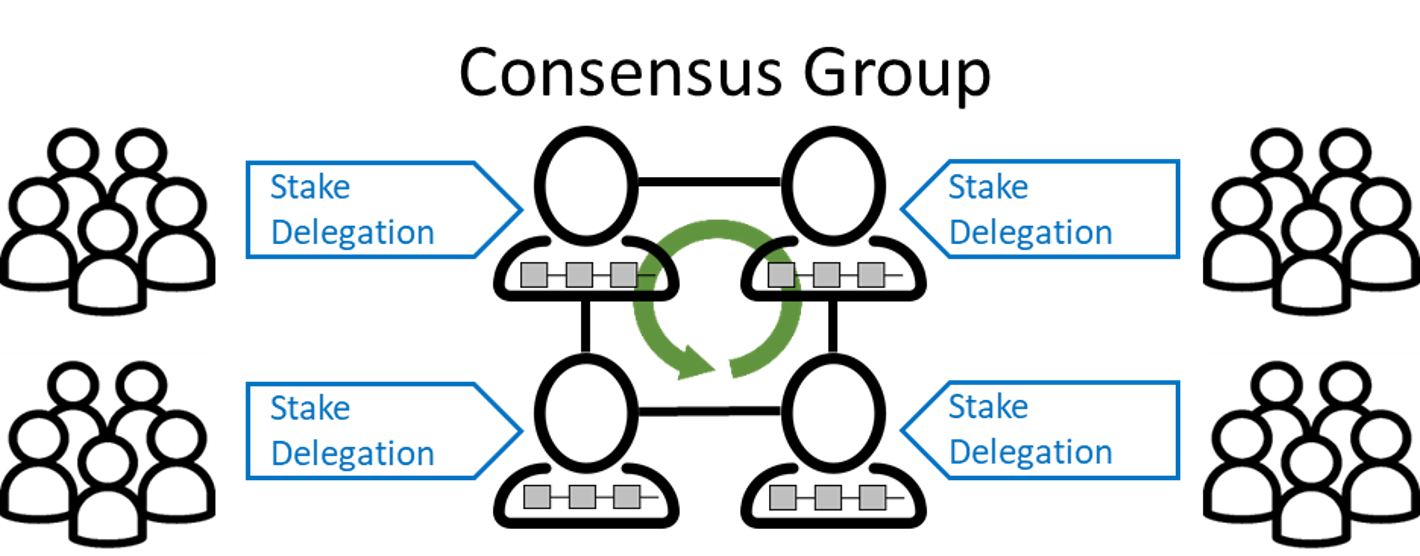}
    \caption{Illustration of the delegation process in DPoS. Small stakeholders vote for delegates with their stakes and the most voted delegates form the consensus group.}
    \label{fig:delegation}
\end{figure}

DPoS can be seen as a democratic form of committee-based PoS in that the committee (consensus group) is chosen via public stake delegation. It is currently used by BitShares 2.0 (2015) \cite{Bitshares2015docs}, Lisk (2016) \cite{lisk2016docs}, EOSIO (2017) \cite{eos2018whitepaper}, and {Cosmos \cite{cosmoshub}}.
DPoS was designed to control the size of the consensus group so that the messaging overhead of the consensus protocol remains manageable. Members of the consensus group are also called \emph{delegates}. The election of delegates is called the \emph{delegation} process, and a general example is shown in Fig. \ref{fig:delegation}. In the actual case, the delegation process and the soliciting of votes may involve outside incentives. And the delegation process may turn out to be an interesting socioeconomic phenomenon.

Generally, an aspirating delegate needs to attract enough votes from normal token holders. This is often accomplished by offering a popular application and building up reputation through propaganda campaigns. 
By casting a vote to a delegate via a blockchain transaction, a token holder entrusts the delegate with its own stake. As a result, the delegate harvests the stake voting power from her voters and acts as their proxy in the consensus process. A token holder can switch vote to another delegate via another delegation transaction.
Take EOSIO for example, the EOSIO protocol mandates that anyone can be a delegate and solicit votes, but only those who ascend to top 21 can join the consensus group, among whom the right of block proposal is equally shared.
EOSIO employs a pipelined PBFT-style consensus scheme to finalize the proposed blocks across the 21 delegates \cite{eosio2018dposbft}. 
Specially, the physical time is divided into slots and the 21 delegates take turn to propose a block in a round-robin fashion. At each slot when a delegate proposes a block, the consensus scheme goes through pre-commitment and commitment phases and decides on the fate of the proposed block.
Because of the small consensus group size and orderly PBFT-style procedure, every new valid transaction can be near-instantly propagated to the consensus group and finalized in the blockchain.

To enforce transaction validation and consensus safety, DPoS's incentive mechanism is designed to encourage honest delegation and consensus participation. Every delegate receives daily vote-reward proportional to the votes she has. Once ascending to the consensus group, delegates also receive block rewards for validation work.

\textbf{Security analysis~} Assuming BFT is used by the consensus group for block finalization, which is recommended since the group size is limited, DPoS can tolerate 1/3 of delegates being malicious. For example, EOSIO can tolerate at most 6 out of 21 delegates being malicious. In the real world they may not wish to misbehave or collude at all, since all delegates have revealed their identities to voters and would be scrutinized for any misconduct.

\subsection{Vulnerabilities of PoS}
\label{subsec:pos-vuln}

Although heralded as the most promising mechanism to replace PoW, PoS still faces several vulnerabilities.

\subsubsection{Costless simulation}
Costless simulation is a major vulnerability of non-BFT-based PoS schemes, especially chain-based PoS in which PoS is used to simulate the would-be PoW process. Costless simulation literally means any player can simulate any segment of blockchain history at the cost of no real work but speculation, as PoS does not incur intensive computation while the blockchain records all staking history. This may give attackers shortcuts to fabricate an alternative blockchain. The four subsequent vulnerabilities, namely \textit{nothing-at-stake}, \textit{posterior corruption attack}, \textit{long-range attack}, and \textit{stake-grinding attack} are all based on costless simulation.

\subsubsection{Nothing-at-stake}
Nothing-at-stake is the first identified costless simulation problem that affects chain-based PoS. It is also known as ``multi-bet" or ``rational forking" problem. Unlike a PoW miner, a PoS minter needs little extra effort to validate transactions and generate blocks on multiple competing chains simultaneously. This ``multi-bet" strategy makes economical sense to PoS nodes because by doing so they can avoid the opportunity cost of sticking to any single chain. Consequently if a significantly fraction of nodes perform the ``multi-bet" strategy, an attacker holding far less than 50\% of tokens can mount a successful double spending attack \cite{martinez2018understanding}. 
Nothing-at-stake problem can be practically solved by penalizing whoever multi-bets, such as forfeiting part of or all their stakes. However the penalties could still be reversed if the attacker eventually succeeds in growing a malicious chain.

\subsubsection{Posterior corruption}
Dubbed by Bentov \cite{bentov2014proof} as ``bribing attack" in 2014, posterior corruption is another attack utilizing costless simulation against PoS. The key enabler of posterior corruption is the public availability of staking history on the blockchain, which includes stakeholder addresses and staking amounts.
An attacker can attempt to corrupt the stakeholders who once possessed substantial stakes but little at present by promising them rewards after growing an alternative chain with altered transaction history (we call it a ``malicious chain''). When there are enough stakeholders corrupted, the colluding group (attacker and corrupted once-rich stakeholders) could own a significant portion of tokens (possibly more than 50\%) at some point in history, from which they are able to grow an malicious chain that will eventually surpass the current main chain.
Since posterior corruption is only possible because the private/public keys are fixed for blockchain participants, key-evolving cryptography (KEC) \cite{franklin2006survey} can be applied so that the past signatures cannot be forged by the future private keys. Ouroboros Praos \cite{david2018ouroboros} currently adopts KEC for this purpose.
Alternatively, as is in Snow White \cite{daian2017snow} and Casper FFG \cite{buterin2017casper}, checkpointing can be used to finalize the ledger and eliminate the possibility of posterior corruption.


\subsubsection{Long-range attack}
Coined by the Ethereum founder Vitalik Buterin \cite{buterin2014long}, long-range attack can be viewed as the ultimate form of costless simulation. It foresees that a small group of colluding attackers can regrow a longer valid chain that starts not long after the genesis block. Because there were likely only a few stakeholders and a lack of competition at the nascent stage of the blockchain, the attackers can grow the malicious chain very fast and redo all the PoS blocks (i.e. by costless simulation) while claiming all the historical block rewards. If the blockchain network is not incentivized by block rewards, the attackers can still deploy a similar long-range scheme called \textit{stake-bleeding attack} \cite{gavzi2018stake} as long as transaction fees exist. That is, the attackers can accumulate significant amount of wealth by collecting most of historical transaction fees through the malicious chain. Through either scheme, once the malicious chain overtakes the main chain, it is released to public and becomes the new main chain. 

While zero transaction fees can counter stake-bleeding attacks, general long-range attacks (and costless simulation as a whole) can be resolved by checkpointing, a more radical measure. Checkpointing is widely used in BFT protocols to ensure the finality of system agreements and safely discard older records. For permissionless blockchains (the main venue for chain-based PoS), however, checkpointing can undermine decentralization, as the finality of checkpoints always requires the endorsement from certain authoritative entities.


\subsubsection{Stake-grinding attack}
Generally, the block generation competition in Nakamoto-style blockchain with proof-of-X block proposal is a pseudo-random process that a higher X (eg. work, stake) yields a higher probability of winning the competition. 
However, unlike PoW in which pseudo-randomness is guaranteed by the brute-force use of a cryptographic hash function, PoS's pseudo-randomness is influenced by extra blockchain information---the staking history. Malicious PoS minters may take advantage of costless simulation and other staking-related mechanisms to bias the randomness of PoS in their own favor, thus achieving higher winning probabilities compared to their stake amounts \cite{kiayias2017ouroboros}. For example, in chain-based PoS blockchains such Peercoin, at a certain historical point of the blockchain, attackers may iterate through different block headers and increase the probability of generating a valid block with their stakes \cite{ethereum2018grinding}.
For committee-based and BFT-based PoS schemes that decouple the election of block proposers from block generation, stake-grinding attacks can be mitigated by ensuring the PoS pseudo-randomness with a secure block proposer election scheme that involves minimal local information. Examples include Ouroboros, Ouroboros Praos, and Algorand.

\subsubsection{Centralization risk}
PoS faces a similar wealth centralization risk as with PoW. In PoS the minters can lawfully reinvest their profits into staking perpetually, which allows the one with a large sum of unused tokens become wealthier and eventually reach a monopoly status. When a player owns more than 50\% of tokens in circulation, the consensus process will be dominated by this player and the system integrity will not be guaranteed. Take Ethereum's Casper FFG for example, the proposed PoS scheme is built upon the current PoW system, of which the cryptocurrency ethers can be directly used for staking. This gives initial advantages to those who have already accumulated huge wealth during Ethereum's PoW operation.
Potential countermeasures against monopolization in PoS mainly come from the incentive mechanism and economic perspective. In addition to the stake valuation scheme that improves the winning chances of small stakeholders (see Section \ref{subsec:chainbased-pos}), we can use off-chain factors to complicate the staking process (EOSIO for example) and impose taxation on the blocks generated by large stakeholders, to name a few.

\section{Other Emerging Blockchain Consensus Mechanisms and Protocols}
\label{sec:other-promising}

The majority of contemporary blockchain consensus protocols make use of established schemes including PoW, PoS, and BFT. There are also numerous other promising proposals for specific application scenarios, many of which are still conceptual and cover just one or two of the five blockchain consensus protocol components. In this section we present five of such proposals, namely proof of authority (PoA), proof of elapsed time (PoET), proof of TEE-stake (PoTS), proof of retrievability (PoR), Ripple consensus protocol/algorithm (RCPA).
{\color{\rcolor}
A comprehensive review of two types of DAG-based protocols, namely blockDAG and txDAG, are also provided.
}

\subsection{Proof of Authority (PoA)}
\label{subsec:PoA}


Proof of authority was coined by Ethereum co-founder Gavin Wood as an alternative to PoW and PoS. It is currently deployed in Ethereum's Rinkeby (2017) and Kovan (2017) testnet, and POA Network (2018) \cite{poa2018whitepaper}. To avoid name conflicts, in later comparisons we refer to Bentov's PoA as PoAct, and proof of authority as PoA. In a nutshell, PoA is a special case of PoS in that a validator stakes with its identity instead of monetary tokens. To qualify as a PoA validator in the consensus group, a participant needs go through a mandatory certification process to build up its authority. It generally involves having the unique identity verified, demonstrating the ability to contribute consistently to the consensus, and making all certification documents publicly available. The consensus group should be stable and small in size, and publicly scrutinized so that users can entrust the consensus group for reliable transaction processing and blockchain curation. If a validator shows incompetence in such tasks or misbehaves, it will be discredited by users and peer validators.

\textbf{Security analysis~} The fault tolerance of a PoA blockchain depends on the consensus protocol used by the consensus group. Besides BFT protocols with 1/3 fault tolerance threshold, Nakamoto-style protocols such as Parity's AuthorityRound (AuRa) \cite{aura2017doc} can also be used to tolerate up to 50\% of colluding validators. 
Featured by its small but trusted consensus group, PoA is a good example of trading decentralization for security and performance. To prevent validators from collusion, they are required to operate independently and constantly monitored by users. PoA is currently used in Ethereum's Rinkeby/Kovan testnet and POA Network \cite{poa2018whitepaper}.

\subsection{Proof of Elapsed Time (PoET)}
PoET was proposed by Intel as an alternative mining mechanism in 2016 and subsequently used in the Hyperledger Sawtooth family \cite{HYPerledgersawtooth}. Instead of undergoing the hashing-intensive mining, PoET simulates the time that would be consumed by PoW mining. That is, every node randomly backs off for an exponentially distributed period of time before announcing its block. To ensure that the local time truly elapses, PoET requires the back-off mechanism to be carried out in a trusted execution environment (TEE), which is an isolated memory area that provides integrity and confidentiality to the program running inside, against a compromised hosting platform. Specially, the program enclosed in a TEE is called an ``enclave". Intel SGX \cite{costan2016intel} and Arm TrustZone \cite{arm2018trustzone} are the two major TEE solutions. Among other utilities, TEE provides an integrity proof of enclave program through remote attestation, which essentially helps the network establish trust on consensus participants. 

Taking Hyperledger Sawtooth for example, the PoET protocol works in two phases for every participating node $i$:
\begin{enumerate}
    \item \textit{TEE setup}: Node $i$ first obtains the PoET protocol program from a trusted source and instantiates the program on its SGX machine wherein the random back-off routine runs inside an enclave. The trusted program generates a signing key pair $\langle PK_i,SK_i \rangle$ for node $i$ and starts the attestation process which results in sending an attestation report to the network. The attestation report contains the public key $PK_i$ and the enclave measurement signed by the Intel Enhanced Privacy Identification (EPID) private key inside the enclave. Other nodes in the network will validate the node $i$'s hardware authenticity through Intel Attestation Service (IAS) and validate the attestation report before accepting node $i$.
    
    \item \textit{Participating in consensus}: The consensus process is similar to Nakamoto except for block generation and validation. For each block cycle, node $i$ waits for a period dictated by the random back-off routine running in the enclave before producing a new block. The enclave then generates a certificate of back-off completion signed by node $i$'s private key $SK_i$ which is broadcast to the network along with the new block. Upon receipt of the new block, other nodes validate the block content and the back-off completion certificate with node $i$'s public key $PK_i$. A block shall be appended to the blockchain if it passes validation and the finalization rule.
\end{enumerate}

While PoET was initially proposed to substitute the PoW for block proposal without touching on the longest-chain rule, the usage of hardware-assisted public key cryptography actually enables the use of more efficient BFT algorithms for block finalization. The hybrid PoET-BFT idea is currently used by Sawtooth PBFT \cite{sawtoothpbft2018}, a sub-project of Hyperledger Sawtooth.

\textbf{Security analysis~} PoET can tolerate up to 50\% TEE nodes being malicious. Since every TEE node runs the same random back-off routine in its enclave, a player can shorten its expected back-off time by running multiple TEE nodes, which is susceptible to Sybil attacks. If more than 50\% of TEE devices collude or are controlled by an attacker, they can ultimately win the block race and keep extending the malicious chain. Therefore, PoET is most suited for permissioned blockchains, where every participant is authenticated and runs one TEE node.

For Hyperledger Sawtooth, Intel is the sole TEE hardware vendor and attestation service provider, which poses a single point of risk to the network. The validity of remote attestation depends on the integrity of SGX implementation and the availability of IAS. Recent attacks such as \textit{cache attacks} \cite{brasser2017software}, \textit{Foreshadow} \cite{van2018foreshadow} and \textit{Foreshadow-NG} \cite{weisse2018foreshadow} have demonstrated the ability to extract the EPID private key from hardware exploiting side channels and speculative execution, posing a security threat at the hardware level.



\subsection{Proof of TEE-Stake (PoTS)}

PoTS is another protocol harmonizing TEE and blockchain consensus. It was first proposed by Li et al. \cite{li2017securing} in 2016 and formalized by Andreina et al. \cite{andreinapots} in 2019. A PoTS node $i$ follows the same setup procedure as in PoET to bootstrap a TEE enclave, generate the signing key pair $\langle SK_i,PK_i \rangle$, and attest the setup to the network. Instead of simulating the would-be elapsed time of PoW mining, the enclave program of PoTS is akin to Algorand's cryptographic sortition scheme that randomly selects a committee according to the stake distribution. Every node in the committee is eligible to propose a new block for the coming block cycle. To prove the block proposal eligibility to the network, the block also includes the eligibility proof signature $\sigma_{SK_i}^{ep}$, which is produced by the enclave program of the block generating node $i$. Once other nodes receive this block, they validate the block content as well as signature $\sigma_{SK_i}^{ep}$ using node $i$'s public key $PK_i$. The longest-chain rule is then used to determine whether to accept this block into the blockchain. 

\textbf{Security analysis~} PoTS can tolerate up to 50\% of all stake value at TEE nodes being maliciously controlled, the same as the fault tolerance of chain-based or committee based PoS. Compared to PoET, the incorporation of staking gives PoTS higher robustness against Sybil attacks, which implies its applicability to permissionless blockchains. Furthermore, the security offered by public key cryptography and TEE-certified execution of committee selection helps PoTS counter the stake-bleeding and stake-grinding attack. The single point of risk in TEE hardware vendor still exists.


\subsection{Proof of Retrievability (PoR)}

PoR was originally proposed by Juels et al. \cite{juels2007pors} in 2007 as a cryptographic building block for a semi-trusted distributed archiving system. The core feature of PoR is to allow a file owner to check if its online files or file fragments are securely stored and retrievable through a challenge-response protocol. The retrievability of a target file $F$ at a remote node $n_i$ can prove $n_i$ indeed spends the required amount of storage resources on $F$. Because of the space requirement behind retrievability, PoR is also known as proof of space. 

In the role of a consensus protocol, PoR was first used by the cryptocurrency Permacoin, proposed by Miller et al. \cite{miller2014permacoin} in 2014. It was designed as a mining-free alternative to PoW. First, a central dealer publishes a target dataset $F$ and computes the digest of $F$ (the Merkle hash tree root of all segments of $F$). Then each participant stores some random segments of $F$ per its storage capability, and computes the digest of these segments. For every block cycle, the dealer initiates a lottery game with a random puzzle. Then every participant derives a lottery ticket consisting of a fixed number of PoR challenges from its locally stored segments, public key, and the puzzle. Participants with more segments stored have higher probability of winning the lottery and thus being eligible to generate a block. All PoR challenges are stored in the new block and verified by the whole network. Permacoin also implements a signature scheme to discourage participants from outsourcing the storage task. Aside from PoR, Permacoin inherits Bitcoin for other consensus components.

Compared to PoW, PoR has two economical advantages. First, file storage consumes far less energy than brute-force mining, and storage space as a resource can be recycled. Second, PoR can be repurposed for meaningful storage tasks. For example the target dataset can be some extremely large but useful public dataset. In fact, the latter advantage is not seen in any other proof-of-X schemes. 

\textbf{Security analysis~} Since the block winning rate of a participant is proportional to its local storage space, PoR can tolerate up to 50\% of gross storage being held up by the malicious party. Although this still can trigger an arm race of storage resources, it downplays the efficacy of ASICs and encourages a wider variety of mining participants. Meanwhile, the 50\% threshold depends on the job of the randomness of the central dealer. To ensure the diversity of lottery tickets across all participants and increase the randomness of the lottery, the target dataset should be large enough so that participants stores almost non-overlapping segments. This assumption can be undermined if the dealer chooses a dataset not large enough or deliberately distributes overlapped segments.




\subsection{Ripple Consensus Protocol/Algorithm (RCPA)}

RCPA was proposed by Schwartz et al. \cite{schwartz2014ripple} in 2014 as the underlying protocol for Ripple, a global payment and gross settlement network operated by the Ripple company. Unlike public blockchains such as Bitcoin and Ethereum, Ripple treats individual transactions as the ledger's atomic items, similar to a transaction log.

In Ripple network, only validator nodes can participate in consensus. We call a validator node ``node'' for simplicity. Nodes collect transactions from clients and propose them to peer nodes for consensus. In initialization, every node establishes a unique node list (UNL) which identifies the nodes it can trust and directly exchange messages with. A UNL relationship is reciprocal. We call a group of nodes that are fully connected by UNL relationships a UNL clique.

\begin{algorithm}[!h]
\label{algo:ripple}
\caption{RPCA (validator node)}
\setcounter{AlgoLine}{0}

Joining Ripple network as a validator;\

\tcc{Main loop}
\For{new epoch}{
Collect valid transactions (new or leftover from previous epochs) $\Rightarrow$ $CandidateSet$;\

\For{$r=1\to MaxRound$}{
Broadcast $CandidateSet$ to UNL peers;

After receiving transactions from UNL peers, add them to $CandidateSet$ and broadcast a vote on the veracity (yes/no) of every transaction;

After receiving votes from UNL peers, discard transactions from $CandidateSet$ whose yes-votes fall short of a threshold $TH_r$ ($TH_{MaxRound}=80\%$);
}

The remaining transactions in $CandidateSet$ are accepted into the ledger; 
}
\end{algorithm}

The operation of RCPA at one node is shown in Algorithm \ref{algo:ripple}. RCPA runs in epochs, each of which finalizes a certain set of transactions into the ledger. Every epoch involves multiple rounds
of transaction filtering. The ledger is marked ``closed'' after the final round. Specially, the yes-vote threshold of the final round $TH_{MaxRound}=80\%$, which is designed to cope with the network connectivity assumption that any two UNL cliques $UNL_i,UNL_j$ must be at least 20\% overlapped, i.e. sharing at least $\frac{1}{5}\max(|UNL_i|,|UNL_j|)$ inter-clique UNL relationships. 
In an ideal environment, as long as $MaxRound$ is large enough, all valid transactions proposed by non-faulty nodes should eventually surpass the 80\% yes-vote threshold.

\textbf{Security analysis~} RCPA imitates a relaxed DLS protocol with an artificial BFT bound of 1/5. It further requires no more than 1/5 of nodes are faulty in every UNL clique in order to ensure overall network consensus. 
Compared to PBFT that achieves 1/3 Byzantine fault tolerance with $O(N^2)$ message complexity in a fully connected network, RPCA's 1/5 fault tolerance bound trades for a lower connectivity requirement and thus lower message complexity per block cycle, which is $O(MK^2)=O(NK)$ where $K$ is the clique size and $M=N/K$ is the number of cliques. Therefore Ripple {demonstrates the possibility of} trading fault tolerance for better performance when a certain level of trust is assumed. 
On the down side, RCPA's multi-round broadcast scheme among a UNL clique and the quick convergence of votes require high synchrony among clique members. This impairs Ripple's decentralization capability in practical settings.

Ripple's current customers are primarily established corporations and financial institutions. Possible reasons include the above-mentioned synchrony requirement and that the 1/5 fault tolerance can be too restrictive for low-trust environments. Interestingly, Stellar \cite{mazieres2015stellar}, originally a fork project of the Ripple, 
{\color{\rcolor}
has shifted to federated Byzantine quorum systems (FBQS) for a loose synchrony requirement on participants and achieves the 1/3 BFT bound \cite{cachin2019asymmetric}.
}

\subsection{BlockDAG-based Consensus Protocols}

{\color{\rcolor}
Starting from 2016 there have been increasing interests in non-linear ledger structures for the aim of better performance, among which directed acyclic graph (DAG) has received the most attention.
Consensus schemes with DAG ledger structure mark a significant divergence from Nakamoto's blockchain design. Their key insight is that transaction throughput should not be limited by a restrictive consensus object, such as a linearly growing chain of blocks with fixed time intervals. Instead, the influx of transactions should drive the ledger expansion. There are two types of DAG ledgers: block-based DAG (blockDAG) and transaction-based DAG (txDAG). In this subsection we focus on blockDAG and defer the latter type to the next subsection. 

In a blockDAG, every vertex contains a collection of transactions which is similar to the block concept in blockchain. What sets blockDAG apart from blockchain is that every block can be hash-pointed to multiple parent blocks. This leads to a situation that every new block can be appended to the DAG with considerable flexibility on how many and which parents to point to. This parent-selecting conundrum is commonly regarded as the major challenge for DAG-based consensus schemes and pertains to the system's transaction processing capability and security against Sybil and double-spending attacks. Next we introduce two concurrently proposed blockDAG schemes, SPECTRE and PHANTOM, with a focus on the parent-selection mechanism.

\subsubsection{SPECTRE} Proposed by Sompolinsky et al. \cite{sompolinsky2016spectre} in 2016, SPECTRE is one of the first well-documented blockDAG proposals. SPECTRE requires any node who wants to mine a new block to find all blocks of zero in-degree (i.e. ``tips'') in the DAG and hash-point the new block header to these tips before starting the PoW mining for the new block. The node broadcasts the newly mined block to the network. Every node initiates a recursive voting procedure to determine the order of any two blocks in the current DAG. This recursive procedure eventually results in a pairwise ordering over the blockDAG. Every new block should be incremental to the past pairwise ordering effort. This pairwise ordering scheme essentially allows SPECTRE to decide between two conflicting blocks (i.e. containing transactions that spend the same UXTO).

\subsubsection{PHANTOM} PHANTOM \cite{sompolinsky2018phantom} is a concurrent blockDAG proposal of SPECTRE by the same authors. One notable weakness of SPECTRE is that the pairwise ordering of blocks may not extend to a fully linear ordering. This leads to SPECTRE's weak liveness (i.e. only supports naturally chronological transactions) and an increased risk of balancing attacks in that conflicts may not be solved. In contrast, PHANTOM realizes fully linear ordering of transactions and blocks in the DAG via the following algorithm. Every node searches the blockDAG for the largest $k$-cluster of blocks. $k$ denotes the node degree in the cluster, and is a predefined security parameter that rarely $k$ or more honest blocks are created simultaneously. The $k$-cluster is regarded as honest and all blocks within are linearly ordered. The transactions covered by the cluster are then validated in the new order. 
Notably, the largest $0$-cluster case is equivalent to the longest-chain rule of Nakamoto consensus. The authors also argues that PHANTOM can be utilized alongside SPECTRE for more flexible consensus performance, as the two ordering schemes are complimentary to each other. 

\textbf{Security analysis~} Though with an innovative blockDAG ledger structure and a carefully designed ordering scheme for finality, SPECTRE and PHANTOM inherit other protocol components from Nakamoto's, including PoW-based block proposal, gossip-based information propagation, validity check on PoW and transactions. Therefore, the two blockDAG based schemes can tolerate up to 50\% of maliciously controlled computing power. On the performance side, blockDAG-based schemes can theoretically support arbitrary throughput capacity, only to be capped by network bandwidth and nodes' processing speed. On the downside, the increased parent-selection flexibility may expose more attack surfaces to adaptive attackers (such as the balancing attack against SPECTRE), which is still an ongoing research topic.
Alternatively, BlockDAG can be designed to incorporate a specific blockchain as a main chain, reflecting a similar idea as the GHOST rule. Conflux, a recent blockDAG scheme proposed by Li et al. \cite{li2018scaling}, resorts to the GHOST rule for the finalization of a pivot chain, which is used as the reference for partitioning the blockDAG into chronological order. This scheme yields high transaction throughput but also higher confirmation latency.
}

\subsection{\color{\rcolor}TxDAG-based Consensus Protocols}

{\color{\rcolor}
Compared to blockDAG, txDAG further breaks the shackles of block. Since different blocks in a blockDAG can still hold overlapped transactions, it takes extra bandwidth and processing to solve conflicts. In txDAG, every vertex represents a unique transaction and diverging branches from a vertex shall hold disjoint transactions. This conveniently avoids the conflict resolution effort as in blockDAG and effectively frees up more processing power at each node and allows for higher transaction throughput, only to be capped by network bandwidth. On the other hand, the blockless design of txDAG also leads to more challenges especially regarding to the convergence and effective management of ledger. Next we introduce three txDAG schemes: IOTA Tangle, Byteball, and Nano.
}

\subsubsection{IOTA Tangle}
\begin{figure}
    \centering
    \includegraphics[width=3in]{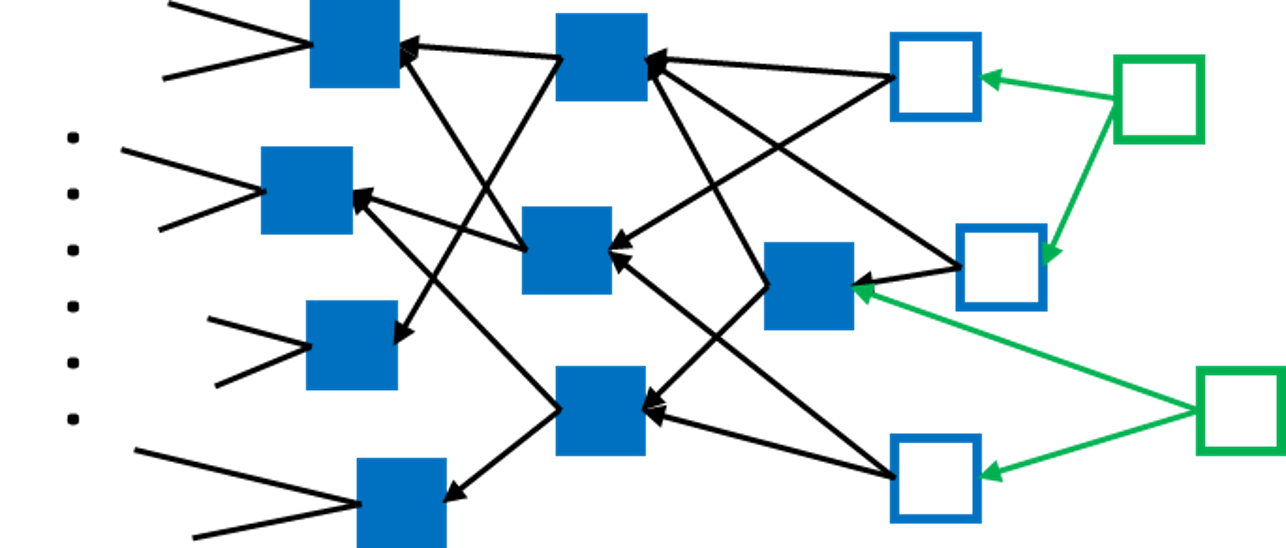}
    \caption{An example of IOTA Tangle's DAG structure. Every new transaction (hollow green) needs to approve two unapproved transactions (``tips", hollow blue) and becomes a new tip. If only one tip is available, choose an approved transaction for the other. The arrow direction symbolizes a user's tip selection behavior.}
    \label{fig:iota-tangle-dag}
\end{figure}

IOTA is a blockchain initiative designed for machine-to-machine micro payments in the IoT setting. Tangle, its underlying consensus protocol, was formalized by Popov \cite{popov2016tangle} in 2016. 
An example of Tangle's DAG is shown in Fig. \ref{fig:iota-tangle-dag}. To append a new transaction to the DAG, a user needs to approve (i.e. validate) two unapproved transactions (tips) of the DAG in order to submit a new transaction. If there is only one tip available, the user chooses an approved transaction instead. The user attaches the hashes of the two chosen transactions to the new transaction and works on a PoW puzzle. The proof is broadcast along with the new transaction. Therefore, every vertex in the DAG has an out-degree of two and contains a PoW proof.
The two-tip rule also serves as the incentive mechanism for honest participation. For this reason, IOTA Tangle does not reward block miners with cryptocurrency nor charge transaction fees.

When multiple conflicting tips (i.e. double-spending) are detected by a user, only one can be valid. In this case the user resorts to a tip-selection scheme to choose one that yields the highest acceptance probability in the long term. Other conflicting tips are considered invalid and then orphaned. Tangle currently employs a Markov-chain Monte Carlo (MCMC) based scheme to estimate the long-term acceptance probabilities of conflicting tips.

When there are more than two non-conflicting tips available (i.e. multiple valid choices), the user chooses two tips according to a weight based strategy, which works as follows. Every transaction is assigned an initial weight, which is proportional (discrete values) to the PoW effort spent on this transaction. The user chooses two tips with the highest weights. If the user only possesses a subgraph of the DAG, it performs two weighted random walks from the beginning of the subgraph to decide two tip choices \cite{iotatangle-tipselection}. The weights of a tip transaction and its preceding transactions along the lineage of approvals in the DAG shall accumulate by the weight of the approving transaction. As a result, as time passes the DAG ledger will look like a cascade of transactions that advances toward the direction of the highest accumulative weight. The MCMC-based scheme, the highest-accumulated-weight strategy, and the weighted random walk algorithm can be regarded as the DAG version of longest-chain rule, and essentially contribute to Tangle's tamper-resistance and probabilistic finality.

Theoretically, Tangle's DAG data structure can boost IOTA's transaction capacity to millions per second, or unbounded. This is due to the DAG's native support for branches: when multiple valid tips are present, a new transaction just chooses two to approve, while leaving the remaining tips to other new transactions. As a result, the more new transactions are generated, the more likely a tip gets approved, which can accommodate more new transactions.
In practice, Tangle's transaction capacity is still capped by link-/physical-layer communication bandwidth.

{\color{\rcolor}

\subsubsection{Byteball}
Byteball was proposed by Churyumov \cite{churyumov2016byteball} in 2016 and currently used in the Obyte cryptocurrency. Byteball features three major differences from Tangle: arbitrary tip number, a main chain (MC) for tip selection, a fixed witness group for DAG consensus and finalization. Every vertex $v$ is assigned a main chain index (MCI), which refers to the height of the closest MC vertex that has a directed path to $v$. Whenever a node wants to attach a new transaction to a tip, the tip should be chosen based on its validity and MCI. Meanwhile, the MC is determined by the collective effort of the witness group and the network. Byteball's witness group consisting of 12 reputable entities decides the next vertex of MC. The participation of witnesses in the blockDAG helps assign a witness level to any candidate vertex. The one with the higher witness index is attached to MC.

To mitigate Sybil attacks and spams, Byteball collects a storage fee for every byte of transaction content (both header and payload) recorded in the txDAG. These fees, also known as commissions, are paid to witnesses for their honest contribution. Therefore, the witnesses are essentially the consensus participants that work towards the Byteball's operation safety. Their revealed identities and small population contributes to deterministic finality and efficient consensus, but also an increased risk of targeted DoS attack.

\subsubsection{Nano}
Nano was proposed by LeMahieu in 2014, formally presented in 2018 \cite{lemahieu2018nano}, and is currently used by the namesake cryptocurrency. Compared to Tangle and Byteball's txDAG ledger, Nano has a more refined ledger structure which is built upon parallel blockchains. Nano's txDAG can be characterized as a block-lattice structure. Every node runs a local blockchain that is not entangled with the others. Notably, we call it ``blockchain'' for the consistency with the whitepaper \cite{lemahieu2018nano}, though every block is actually a single transaction. A normal transaction is fulfilled by two transactions: a \emph{send} transaction at the sender's blockchain and a \emph{receive} transaction at the receiver's blockchain. PoW is also attached to transactions for anti-spamming. When an account is detected double-spending a transaction received by another node (i.e. a fork), the receiver initiates a voting procedure wherein each vote is weighted by the voter's account balance (i.e. stake). 

A key insight of Nano's block-lattice based txDAG is that forks and other faulty transactions only affect the accounts referenced in the said transactions. In other words, the resolution of forks does not hamper the processing of transactions from unaffected accounts. This property allows Nano to provide stable and fast transaction confirmation. On the downside, the stake-weighted voting scheme requires high synchrony among nodes and may not scale well in network size.
}

\textbf{Security analysis~} 
For Tangle, since PoW is enforced for transaction generation, probabilistic finality is guaranteed as long as the honest nodes control more than 50\% of gross computing power. Byteball's consensus safety relies on the majority consensus of the 12 witness. Since Nano resorts to a stake-weighted scheme for fork resolution, its fault tolerance is capped by 50\% of tokens.
In terms of network decentralization, we consider Tangle the most desirable design among the three txDAG schemes because it does not require a dedicated identified group or a synchronized voting mechanism to ensure consensus finality.
Nonetheless, the current version of Tangle's MCMC-based tip-selection is susceptible to parasite chain attack, as is documented in \cite{popov2016tangle}. An attacker can grow a parallel chain from some early point all the way to the current DAG height, with intermittent connections to the DAG. An attacker may need significantly less than 50\% of network computing power to succeed, as the malicious chain may carry far fewer transactions than the main DAG. To address this issue, Tangle currently relies on a centralized checkpointing scheme to periodically confirm past transactions. The IOTA community plans to improve the MCMC-based tip-selection scheme in hope of obtaining the 50\% fault tolerance in a truly decentralized fashion. 

Snowflake-Avalanche \cite{rocket2018snowflake}, a consensus protocol family proposed by Cornell Team Rocket, provides an alternative to Tangle's tip selection problem. Specially, the txDAG-based Avalanche protocol builds upon a hierarchy of leaderless BFT schemes, namely Snowflake and Snowball, for conflict resolution during parent transaction selection. Interested readers are referred to the whitepaper \cite{rocket2018snowflake} for more details.





\section{Comparison and Discussion}
\label{sec:comparison}

\begin{table*}
    \caption{A Comparison of Blockchain Consensus Protocols}
    \small
    \renewcommand{\arraystretch}{1.1}
    \begin{tabular}{
        |>{\raggedright}p{.9in}|
        |>{\raggedright}p{.67in}
        |>{\raggedright}p{.7in}
        |>{\raggedright}p{.7in}
        |>{\raggedright}p{.75in}
        |>{\raggedright}p{.75in}|
        |>{\raggedright}p{.57in}
        |p{.65in}|
        }
        \hline
        \textbf{Consensus \hspace{.5in} Protocol} & \multicolumn{5}{c||}{\textbf{Five-Component Framework}} & \textbf{Fault \hspace{.5in} Tolerance} & \textbf{Throughput} (TPS)  \\ \cline{2-6}
        
         (Examples/ Realizations) & \textbf{Block Proposal} & \textbf{Block Validation$^*$} & \textbf{Information Propagation} & \textbf{Block \hspace{.5in} Finalization} & \textbf{Incentive Mechanism} & & \textbf{\color{\rcolor}[Confirmati- on Latency]} \\ \hline \hline
         
         \textcolor{\rcolor}{\textbf{BFT-SMR}} \hspace{1in} (PBFT, HoneyBadgerBFT)
         & Client operation request & Signature check & Broadcast & Mutual agreement on the same state & (N/A) & 33\% servers & Thousands \hspace{1in} {\color{\rcolor}[1s-1m]} \\ \hline
         
         \textbf{Nakamoto} \hspace{.5in} (Bitcoin, Litecoin) & PoW & PoW check & Gossiping & Longest-chain rule$^\dagger$ & Block reward and transaction fee & 50\% computing power & Sub-ten \hspace{1in} {\color{\rcolor}[10m-60m]} \\ \hline
         
         \textbf{Nakamoto-GHOST} \hspace{.5in} (Ethereum) & PoW (Ethash) & PoW check & Gossiping \hspace{1in}via secure channels & A variation of GHOST rule$^\dagger$ & Block reward and transaction fee & 50\% computing power & Tens \hspace{1in} {\color{\rcolor}[6m-10m]} \\ \hline
         
         \textbf{Bitcoin-NG} \hspace{.5in} (Waves-NG) & PoW for key blocks & PoW check for key blocks & Gossiping & Longest-chain rule$^\dagger$ & Block reward and transaction fee & 50\% computing power & Hundreds \hspace{1in} {\color{\rcolor}[10m]} \\ \hline
         
         \textbf{Hybrid PoW-BFT} (PeerConsens, SCP, ByzCoin) & PoW-based BFT committee election & PoW check & Broadcast among BFT committee & Longest-chain rule$^\dagger$ for chain (PBFT for transactions) & $\odot$ & 33\% computing power & Hundreds \hspace{1in} {\color{\rcolor}[10s-1m]} \\ \hline
         
         \textbf{Chain-based PoS} (Peercoin, Nxt, PoAct) & PoS & PoS check & Gossiping & Longest-chain rule$^\dagger$ & Block reward and transaction fee & 50\% deposited stake value & Tens \hspace{1in} {\color{\rcolor}[10-60m]} \\ \hline
         
         \textbf{Committee-bas- ed PoS} (Ourobo- ros, Praos, CoA, Snow White) & PoS-based committee election & Proposer eligibility check & Broadcast among committee & Longest-chain rule$^\dagger$ & Block reward & 50\% token wealth & Hundreds \hspace{1in} {\color{\rcolor}[1m-10m]} \\ \hline
         
         \textbf{BFT-based PoS ---Tendermint} (Cosmos Hub) & PoS-based round robin & Proposer eligibility check & Broadcast among validators & BFT (adapted DLS) & Block reward & 33\% token wealth & Thousands \hspace{1in} {\color{\rcolor}[1s-3s]} \\ \hline
         
         \textbf{BFT-based PoS ---Algorand} & PoS-based committee election & Proposer eligibility check & Broadcast among committee & BFT (adapted Byzantine agreement) & Block reward & 33\% token wealth & Hundreds \hspace{1in} {\color{\rcolor}[20s-1m]} \\ \hline
         
         \textbf{BFT-based PoS ---Casper FFG} (Ethereum 2.0) & PoW (Ethash) & PoW \& Checkpoint tree check  & Broadcast among validators & BFT (with staked votes) & Block reward for miners and validators & 33\% deposited stake value & Thousands (if sharding used) \\ \hline
         
         \textbf{DPoS} \hspace{.5in} (EOSIO, Lisk, BitShares) & PoS with stake delegation & Delegate eligibility check & Broadcast among delegates & BFT (suggested, not limited to) & Block reward and vote reward & 33\% delegates & Thousands \hspace{1in} {\color{\rcolor}[$<$1s]} \\ \hline
         
         \textbf{PoA} \hspace{.5in} (Rinkby, Kovan, POA Network) & PoA & Block propo- ser identity check & $\odot$ \hspace{1in} Broadcast$^\ddagger$ & $\odot$ \hspace{1in} HoneyBadger- BFT$^\ddagger$ & $\odot$ \hspace{1in} Transaction fee$^\ddagger$ & 50\% TEEs (33\% if BFT used) & Tens to hundreds \hspace{1in} {\color{\rcolor}[10s-1m]} \\ \hline
         
         \textbf{PoET} (Hyperledger Sawtooth family) & PoET within TEE & TEE certificate check & $\odot$ \hspace{1in} Broadcast$^\ddagger$ & $\odot$ \hspace{1in} PBFT$^\ddagger$ & $\odot$ & 50\% IDs (33\% if BFT used) & Tens to thousands \hspace{1in} {\color{\rcolor}[10s-1m]}  \\ \hline
         
         \textbf{PoTS} & PoTS-based committe election & Proposer eli- gibility\&TEE cert. check & $\odot$ & $\odot$ & $\odot$ & 50\% token wealth & (Unavailable) \\ \hline
         
         \textbf{PoR} \hspace{1in} (Permacoin) & PoR & File retrievability check & $\odot$ & $\odot$ & $\odot$ & 50\% storage space & (Unavailable) \\ \hline
         
         \textbf{RPCA} \hspace{.5in} (Ripple) & Any server can propose transactions & UNL membership check & Broadcast to UNL peers & Accepting $>80\%$ voted transactions & Transaction fee & 20\% nodes in each UNL & Thousands \hspace{1in} {\color{\rcolor}[$<$1s]} \\ \hline
         
        
        
        

    \end{tabular}
    \vspace{.08in}
    
    $^*$Block validation also includes validating all transactions inside the block, which is omitted here for space saving.
    
    $\dagger$: With probabilistic finality.
    $\ddagger$: Unspecified in the protocol white paper, but found in one or more blockchain realizations.
    $\odot$: Unspecified in the protocol white paper, but the Bitcoin counterpart is usable.
    
    \label{tab:comparison}
\end{table*}

\begin{table*}
    \caption{A Comparison of DAG-based Consensus Protocols}
    \small
    \renewcommand{\arraystretch}{1.1}
    \begin{tabular}{
        |>{\raggedright}p{.9in}|
        |>{\raggedright}p{.67in}
        |>{\raggedright}p{.7in}
        |>{\raggedright}p{.7in}
        |>{\raggedright}p{.75in}
        |>{\raggedright}p{.75in}|
        |>{\raggedright}p{.57in}
        |p{.65in}|
        }
        \hline
        \textbf{Consensus \hspace{.5in} Protocol} & \multicolumn{5}{c||}{\textbf{Five-Component Framework}} & \textbf{Fault \hspace{.5in} Tolerance} & \textbf{Throughput} (TPS)  \\ \cline{2-6}
        
         (Blockchain Realizations) & \textbf{Block/TX Proposal} & \textbf{Block/TX Validation} & \textbf{Information Propagation} & \textbf{Block/TX \hspace{.5in} Finalization} & \textbf{Incentive Mechanism} & & \textbf{\color{\rcolor}[Confirmati- on Latency]} \\ \hline \hline
         
         \textcolor{\rcolor}{\textbf{BlockDAG ---SPECTRE ---PHANTOM}} 
         & PoW & PoW check & Gossiping & Pairwise ordering$^\dagger$/ $k$-cluster-blocks ordering$^\dagger$ & $\odot$ & 50\% computing power & Thousands or more \hspace{1in} {\color{\rcolor}[10s-2m]} \\ \hline
         
         
         \textcolor{\rcolor}{\textbf{BlockDAG ---Conflux}} 
         & PoW & PoW check & Gossiping & GHOST rule to finalize pivot chain$^\dagger$ & $\odot$ & 50\% computing power & Thousands \hspace{1in} {\color{\rcolor}[5m-20m]} \\ \hline
         
         \textcolor{\rcolor}{\textbf{TxDAG ---Tangle}} \hspace{1in} (IOTA) 
         & Approving two tips \& PoW & Tip approval check \& PoW check & Gossiping & Highest cumulative weight rule$^\dagger$ & Eligibility for issuing new transactions & 50\% computing power & Thousands or more {\color{\rcolor}[1m-10m]}\\ \hline
         
         \textcolor{\rcolor}{\textbf{TxDAG ---Byteball}} (Obyte) 
         & Approving tips per main chain index & Tip approval check & Gossiping & Witnesses consensus on main chain & Commissions collected from storage fees & 50\% (or 6) witnesseses & Thousands or more {\color{\rcolor}[30s-10m]}\\ \hline
         
         \textcolor{\rcolor}{\textbf{TxDAG \hspace{1in}---Nano}} \hspace{1in} (Nano) 
         & Cross-chain sender-rece- iver \& PoW & Sender account \& PoW check & Gossiping & stake-weighted voting$^\dagger$ & (Unspecified) & 50\% token wealth & Thousands or more {\color{\rcolor}[1s-10s]}\\ \hline
         
         \textcolor{\rcolor}{\textbf{TxDAG ---Avalanche}} 
         & Approving one parent & Parent approval check  & Gossiping & Confidence ordering to solve conflict$^\dagger$ & (Unspecified) & 50\% participants & Thousands {\color{\rcolor}[1s-5s]}\\ \hline

    \end{tabular}
    \vspace{.08in}
    
    $\dagger$: With probabilistic finality.
    $\odot$: Unspecified in the protocol white paper, but the Bitcoin counterpart is usable.
    
    \label{tab:comparison-dag}
\end{table*}


Table \ref{tab:comparison}, \ref{tab:comparison-dag} compare the protocols that we have discussed so far from three aspects: five-component framework composition, fault tolerance, and transaction processing capability. The latter includes maximum throughput and transaction confirmation latency.
For some consensus protocols, not all the five components are specified in the white paper. Protocols designed for permissioned blockchains (eg. BFT-SMR, PoET, PoTS) do not need to specify incentive mechanism; protocols that were initially proposed to substitute PoW (eg. PoA, PoET, PoR) inherit other components from Nakamoto's protocol by default.

The fault tolerance capability of a consensus protocol largely depends on the block finalization mechanism. For example, the 50\% bound (of computing power or stake value) applies to all protocols that inherit the probabilistic finality property from Bitcoin; protocols with BFT-style block finalization typically achieve 33\% fault tolerance and guarantee deterministic finality. {\color{\rcolor}In hybrid schemes, the 33\% fault tolerance is usually coupled with the block proposal mechanism or a preexisting identity scheme. In Hybrid PoW-BFT and BFT-based PoS, each node's voting weight in the BFT consensus stage is augmented by its block proposing capability. This is achieved by either linking a player's opportunity of participation in BFT consensus to the PoX process (eg. Hybrid PoW-BFT, Tendermint), or directly weighting a player's vote in the BFT consensus by the player's stake value (eg. Algorand, Casper FFG). In other hybrid schemes such as PoA-BFT and PoET-BFT, the PoX process is purposed for identity management and thus the fault tolerance threshold is 33\% of identities. DAG-based protocols generally achieves probabilistic finality (except Byteball) and 50\% fault tolerance of either computing power, consensus participants, or token wealth.
}

{\color{\rcolor}
The statistics on transaction processing capability were fetched from either the simulation result in protocol white paper or documented experiment. The throughput metric measures the maximum TPS a protocol can handle, and is only precise to the scale of magnitude, i.e. sub-ten, tens, hundreds, or thousands. The confirmation latency metric estimates time for a submitted transaction to be finalized in the blockchain. We observe that protocols with probabilistic finality tend to have higher confirmation latency (typically more than 1 minute), and high throughput is often accompanied by low conformation latency.

\begin{figure}
    \centering
    \includegraphics[width=3.3in]{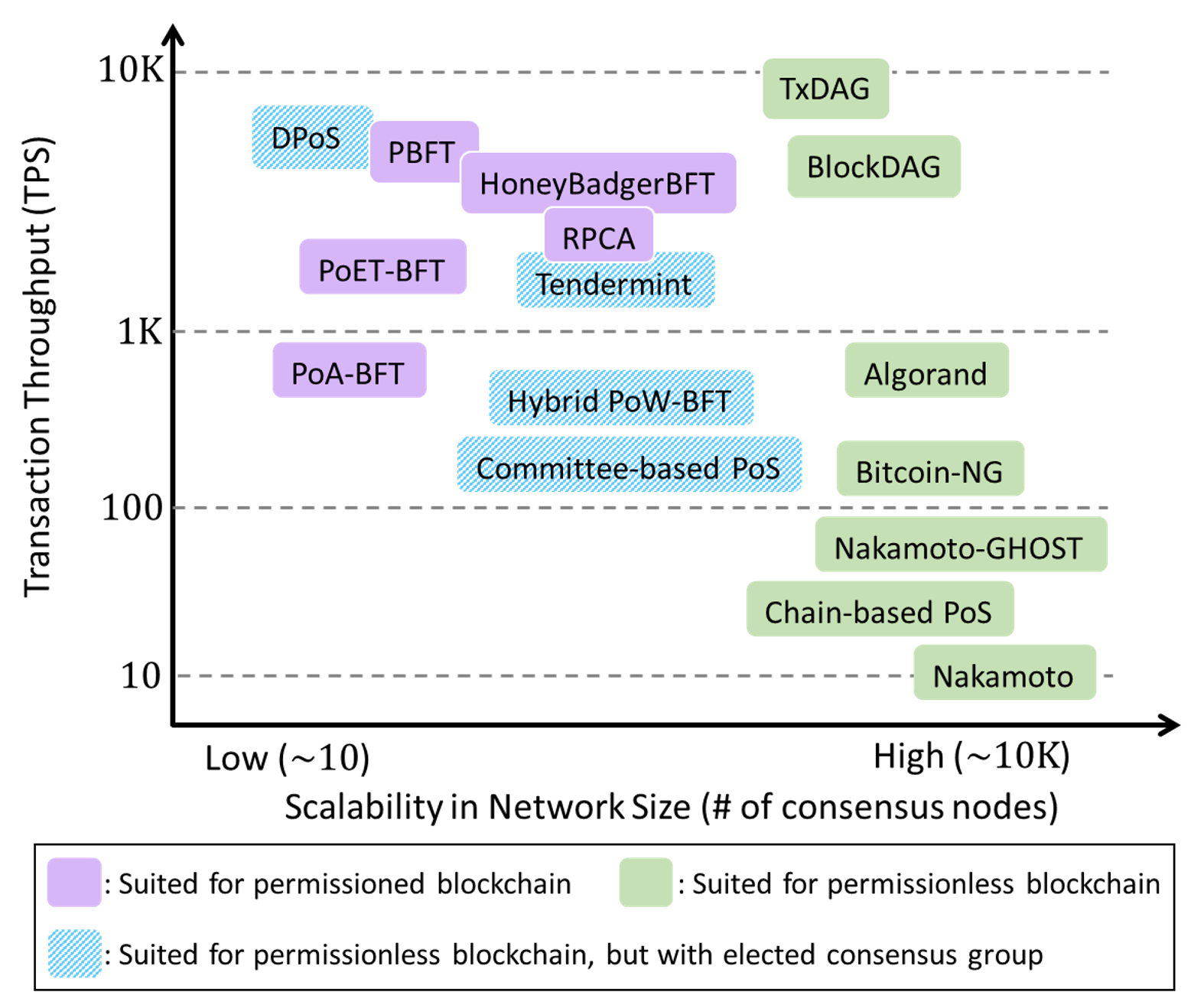}
    \caption{Comparing different blockchain consensus protocols on transaction throughput and scalability in network size. Note that for comparison purposes, we consider DAG-based schemes as wide-sense blockchain protocols.}
    \label{fig:quant-compare}
\end{figure}

We are also interested in another performance metric---scalability in network size, which measures how well a consensus protocol can maintain its transaction capacity when network size grows. 
Given the difficulty of performing large-scale experiments on actual blockchain networks, the scalability results are either fetched from the simulation study in protocol white paper or based on our speculation on the current network status. It is worth noting that transaction throughput and scalability in network size are often recognized as two integral parts in scalability evaluation of a blockchain system \cite{vukolic2015quest,croman2016scaling,natoli2019deconstructing}.

Fig. \ref{fig:quant-compare} illustrates performance of consensus protocols with respect to transaction throughput and scalability in network size along with suitability for permissioned or permissionless blockchains. 
For protocols with BFT-style block finalization mechanism to achieve thousands of TPS, the network size should be small, typically around several hundred. Meanwhile, protocols that work for large-scale public networks are predominantly permissionless and employ a block finalization mechanism that only achieves probabilistic finality. Their transaction throughput is usually capped by a hundred TPS for security reasons. Generally, protocols illustrated farther to the top right in Fig. \ref{fig:quant-compare} tend to achieve better overall performance and scalability. However the security implication of such superiority, exemplified by Bitcoin-NG and DAG-based protocols, is subject to constant scepticism and attracts ongoing research effort.

As for the suitability for permissioned or permissionless blockchains, we stress that protocols that need a stable consensus group with participation control and identification are suitable for permissioned blockchains. These protocols include PBFT, HoneyBadgerBFT, PoET-BFT, PoA-BFT, and RPCA. In comparison, protocols including Nakamoto, Nakamoto-GHOST, chain-based PoS, Bitcoin-NG, Algorand, and Tangle are specifically designed for large-scale permissionless blockchains. Meanwhile DPoS, Tendermint, hybrid PoW-BFT, committee-based PoS are designed for permissionless scenarios but rely on an election mechanism for maintaining a stable consensus group, of which the identities may be revealed for public supervision. DAG-based protocols are designed with the aim of scaling up in participant count and thus suited for permissionless networks.
}

{\color{\rcolor}
\section{On Designing Blockchain Consensus Protocol}
\label{sec:tutorial}

The rich variety of blockchain initiatives often confounds novice protocol designers.
In this section we offer a succinct tutorial that hopefully helps protocol designers form reasonable objectives and avoid pitfalls.
}

\subsection{The Paradigm Shift in Protocol Design}

{\color{\rcolor}
In the early days of blockchain development, consensus protocol designs were often coupled with designer's ingenuity and heuristics. Anticipating that the increasing number of participants would lead to more block contentions and blockchain forks in Bitcoin, Nakamoto designated the longest-chain rule for block finalization and cautiously fixed the average block interval to ten minutes for consensus security \cite{nakamoto2008bitcoin}. Litecoin \cite{litecoinwhitepaper} and Ethereum \cite{wood2014ethereum} adopted shorter block intervals for faster transaction settlement. Observing the energy inefficiency of Bitcoin and the financial value of unused tokens, Peercoin pioneered the PoS mechanism which could substitute PoW with far less energy consumption \cite{king2012ppcoin}. Yet its stake valuation scheme seemed to be heuristically designed. These early-day initiatives also commonly lacked sufficient security analysis, which were often scrutinized by researchers and practitioners \cite{garay2015bitcoin,eyal2014majority,Nxt2014whitepaper}.

Later blockchain consensus protocols, especially those proposed after 2016, have started to take inspirations from the established research of distributed computing, cryptography, and trusted computing. For example, HoneyBadgerBFT \cite{miller2016honey}, BEAT \cite{duan2018beat}, Algorand \cite{gilad2017algorand} extend reliable broadcast and Byzantine agreement to blockchain scenarios. Committee-based PoS protocols, such as Ouroboros \cite{kiayias2017ouroboros}, Snow White \cite{daian2017snow}, Bentov's CoA \cite{bentov2016cryptocurrencies}, make use of MPC for the management of consensus committee and block proposal. PoET \cite{HYPerledgersawtooth} and PoTS \cite{andreinapots} leverage trusted execution environment (TEE) for secure block proposal and mining substitution.
Furthermore, some proposals incorporate innovative cryptographic primitives for enhancing privacy of network participants. Zero-knowledge proof (ZKP) and ring signatures are used by Zerocash \cite{sasson2014zerocash} and Monero \cite{noether2015ring} to construct privacy-preserving transactions that hide essential transaction values and sender-receiver addresses; Algorand and Ouroboros Praos \cite{david2018ouroboros} use cryptographic sortition and VRF for the PoS-based election of consensus committee without revealing participant's stake value. The privacy of consensus participants in large-scale permissionless blockchains is still a fresh research topic.

Besides increased sophistication in design, new protocol proposals are also often accompanied by formal security analysis, via security frameworks such as the \textit{random oracle model} \cite{canetti2004random} and \textit{universal composibility} \cite{canetti2001universally}. In summary, This paradigm shift has brought the design and analysis of blockchain consensus protocols into formal scientific research, echoing Cachin et al. \cite{cachin2017blockchain} that blockchain design should follow the rigor established by prevailing wisdom.
}


\begin{figure}
    \centering
    \includegraphics[width=3.3in]{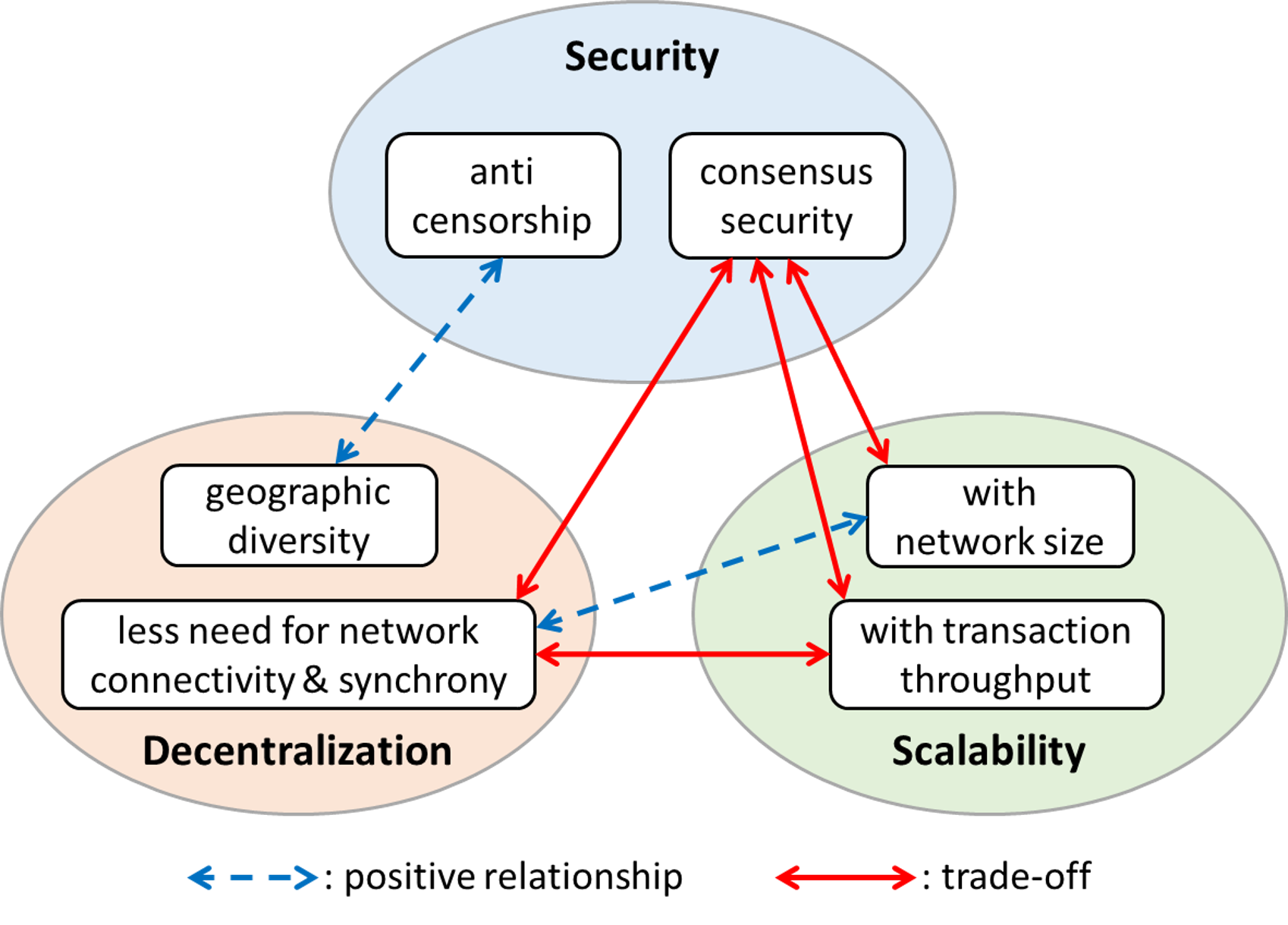}
    \caption{Illustration of the security-decentralization-scalability trilemma.}
    \label{fig:trilemma}
\end{figure}

\subsection{The Security--Decentralization--Scalability Trilemma}

Observing the existence of various blockchain consensus protocols, we stress that the design of consensus protocol should seek a balance between three objectives: security, decentralization, and scalability. 
\begin{itemize}
    \item \textit{Security} refers to the blockchain's consensus security in the presence of malicious players and anti-censorship capability. The two concepts correspond to the safety and liveness property in classical distributed consensus. 
    \item \textit{Decentralization} refers to the decentralization profile of the blockchain network, which is normally associated with geographic diversity, connectivity pattern, and synchrony of networked nodes.
    \item \textit{Scalability} means two-fold: the system needs to remain secure and efficient with rising transaction throughput as well as larger network size. Scalability is often considered as a broader concept of system performance.
\end{itemize}
Fig. \ref{fig:trilemma} sketches the relationships and trade-offs between the three objectives.

For most blockchain consensus protocols especially those tailored for financial application, security is always the top priority. Meanwhile, security level can be practically traded for system's scalability. Lower security requirement gives rise to more scalable protocol design. For classical BFT protocols, lower fault tolerance threshold (the ``$f$'' parameter) leads to fewer consensus rounds \cite{lamport1982byzantine,bracha1984asynchronous} or less communication per round \cite{castro1999practical}, effectively alleviating message complexity caused by large network sizes. For many public blockchain networks that use Nakamoto consensus, the probabilistic finality and mandatory multi-block confirmation give rise to permissionless access and higher scalability in network size. Meanwhile, a shorter block interval yields higher transaction throughput but also leads to higher chance of 51\%-attack \cite{decker2013information}. For DAG-based protocols, the security implications associated with their scalability with both network size and throughput are not well studied and remain an ongoing research.

A protocol designer should also anticipate the security implications of a blockchain network's decentralization profile. A more decentralized network tends to be less synchronized, owing to the increased diversity of geographic locations and sparsity of connections. On one hand, the increased geographic diversity makes censorship by one suppressing regime more challenging, which effectively enhances the system's liveness. On the other hand, the reduced synchrony implies that inter-player communications are less bounded by delay requirements. This contributes to the heterogeneity of network connections and allows better connected players to gain unfair advantages, which potentially impairs consensus security of the network. 
Take protocols with the gossiping rule and longest-chain rule for example, the heterogeneity of network connections gives well-connected nodes a communication advantage of disseminating new blocks faster in the network than the less-connected ones, effectively enhancing the former's winning chance in blockchain fork races. 
As a result, in a sparsely connected network but with one well-connected adversary, the adversary may need significantly less than 50\% of mining power to commit a double-spending attack.


Last but not least, a protocol designer should be aware that the tradeoff between decentralization and scalability depends on the practical needs.
For example, Bitcoin was designed to operate in a weakly synchronized network and there is no enforceable criterion for message delivery and timeout. As a result Nakamoto consensus protocol needs to rely on best-effort mining and long block intervals to achieve the security against double spending, at the price of low transaction throughput. In comparison, BFT-based protocols are designed for permissioned networks where every participant operates with revealed identity and the overall network is small and well-connected and with enforceable time-out mechanisms. This enables the network to achieve higher transaction throughput. On the flip side, the high dependence on network connectivity and synchony for throughput brings high messaging complexity and network management overhead. When the network grows in size and inevitably becomes sparse, message relaying can be a major burden for each participant which potentially leads to jammed communication channel and stalled consensus process. Therefore, we argue that under a certain security premise, increased need for network connectivity and synchrony leads to higher transaction throughput but lower scalability with network size.



\subsection{Protocol Composability}

The discussions above demonstrate that there is never a one-fits-all blockchain consensus protocol. The existence of hybrid consensus protocols highlights the possibility to cherry-pick individual protocol components to fulfill specific application needs.
Sawtooth \cite{HYPerledgersawtooth}, a Hyperledger project featuring the utilization of trusted hardware for PoET-based block proposal, has the flexibility of plugging in different block finalization schemes---Raft for crash fault tolerance and PBFT for Byzantine fault tolerance. Similarly, POA Network \cite{poa2018whitepaper} uses PoA for block proposal and has opted for HoneyBadgerBFT for block finalization to achieve best performance under asynchronous network conditions. 

Besides block proposal and finalization, the choice of information propagation mechanism largely depends on the underlying network topology. In a public blockchain network where gossiping is the default information propagation method, broadcast can be used among a small set of identified nodes for better efficiency \cite{croman2016scaling}.
The block validation mechanism is associated with the block proposal mechanism and takes the current blockchain as input. The incentive mechanism is aimed for promoting honest execution of the former tasks. It needs to account for off-chain factors as well, such as the long-term system sustainability, financial stability, and subtle security issues \cite{buterin2014long,gavzi2018stake,ethereum2018grinding}. 

The composability of blockchain consensus protocol also demonstrates the feasibility of progressive protocol improvement, in that one individual protocol component is updated at a time instead of overhauling the entire protocol. 
This is particularly important for large-scale established blockchain systems in which any protocol change requires network-wide consensus and significant joint effort.















\section{Conclusion}
\label{sec:conclude}

In this survey we provided a summary of classical fault-tolerance consensus research, a five-component framework for a general blockchain consensus protocol, and a comprehensive review of blockchain consensus protocols that
have gained great popularity and potential.
We analyzed these protocols with respect to fault tolerance, performance, vulnerabilities and highlighted their use cases. Notably, many of these protocols are still under development and are subject to major changes at the time of writing.
We hope the five-component framework, classification methodology, protocol abstractions, {\color{\rcolor}performance analyses, and discussion on protocol design} provided in this survey can help researchers and developers grasp the fundamentals of blockchain consensus protocols and facilitate future protocol design.



\section*{ACKNOWLEDGMENT}
The authors would like to thank Dr. Christian Cachin, Dr. Yonggang Wen, and the anonymous reviewers for their constructive comments on this paper.
This work was supported in part by US National Science Foundation under grants CNS-1916902, CNS-1916926.



\bibliography{main}
\bibliographystyle{IEEEtran}

\end{document}